\documentclass[aps,prb,reprint,superscriptaddress]{revtex4-2}
\usepackage{graphicx}
\usepackage{amsmath}
\usepackage{braket}
\usepackage{booktabs}
\usepackage{longtable}
\usepackage{xcolor}
\usepackage{float}  

\begin{document}

\title{From Atomic Defects to Integrated Photonics: A Perspective on Solid-State Quantum Light Sources}
\author{Anuj Kumar Singh}
\author{Parul Sharma}
\author{Kishor Kumar Mandal}
 \affiliation{Laboratory of Optics of Quantum Materials, Department of Physics, Indian Institute of Technology Bombay (IITB), Mumbai -- 400076, India}
\author{Lekshmi Eswaramoorthy}
 \affiliation{Laboratory of Optics of Quantum Materials, Department of Physics, Indian Institute of Technology Bombay (IITB), Mumbai -- 400076, India}
\affiliation{IITB-Monash Research Academy, Indian Institute of Technology Bombay, Mumbai -- 400076, India}
\author{Anshuman Kumar}
\email{anshuman.kumar@iitb.ac.in}
\affiliation{Laboratory of Optics of Quantum Materials, Department of Physics, Indian Institute of Technology Bombay (IITB), Mumbai -- 400076, India}
\affiliation{Centre of Excellence in Quantum Information, Computation, Science and Technology (QuICST), Indian Institute of Technology Bombay, Mumbai -- 400076, India}

\begin{abstract}
Single-photon emitters (SPEs) constitute a foundational resource for quantum technologies, including secure communication, photonic quantum computing, and emerging quantum network architectures. A wide range of quantum materials, from atom-like point defects in bulk crystals to excitonic states in low-dimensional semiconductors, now provide bright, coherent, and scalable sources of non-classical light. Meanwhile, advances in photonic integration have enabled efficient routing, filtering, and on-chip manipulation of these emitters. From this perspective, we survey and discuss the technological landscape in which solid-state emitters interface with quantum sensing, quantum communication, quantum computation, and emerging photonic AI platforms.
Further, we discuss the materials landscape underpinning modern single-photon sources from the zero-dimensional, one-dimensional, two-dimensional and three-dimensional materials. Lastly, we highlight key integration pathways for these single-photon emitters into scalable quantum photonic systems.

\end{abstract}
\maketitle
\section{Introduction}

Quantum information science is structured around three principal domains: quantum communication \cite{gisin2002quantum}, photonic quantum computing \cite{obrien2009photonic}, and quantum sensing \cite{degen2017quantum}. In quantum communication, quantum key distribution (QKD) represents a central yet specific application, utilizing quantum states to establish secret \cite{bennett1992,zhang2025}. The broader field also encompasses quantum teleportation, entanglement distribution, and the development of quantum networks\cite{laneve2025,hou2025,xu2025}. Quantum sensing includes quantum imaging as a subfield, which employs quantum-enhanced measurements to acquire spatially resolved information, distinguishing it from other quantum sensors such as magnetometers or atomic clocks that provide scalar or temporal outputs\cite{samantaray2023}. Quantum computing comprises various computational models, including gate-based, adiabatic or annealing, measurement-based, and topological approaches \cite{farhi2001,raussendorf2001,kitaev2003}. It also involves distinct algorithmic families, such as Shor-type, Grover-type, variational, and simulation algorithms, as well as foundational layers like quantum error correction and fault tolerance\cite{georgescu2014}. These elements are implemented across a range of physical platforms, including superconducting circuits, trapped ions, neutral atoms, spin qubits and single photon emitters as qubit\cite{kjaergaard2020,saffman2010,blatt2008}.

Realization of such photonic quantum systems requires fundamental resources, including compatible interfacing of qubit sources with ultra-low-loss photonic elements, to enable strong interactions and efficient routing of quantum signals at the high-density circuit level to facilitate scaling of quantum photonic integrated circuits (QPICs) \cite{moody20222022, Maring2024, Elshaari2020, Aharonovich2016, Kim2020, Mandal2024, zelaya2025chip, kumar2023photonic, singh2023low, kumar2023universal, singh2025plasmonic}. Emitters that produce single photons with high purity, brightness, and indistinguishability on demand are required by these technologies\cite{somaschi2016,he2013}. Materials used for quantum emitters must have long coherence durations, high radiative quantum efficiency, and little spectrum diffusion in order to satisfy these strict requirements.

The phenomenon of photon indistinguishability, central to many quantum interference protocols, is dominated by the coherence properties of the emitter \cite{Santori2002}. Furthermore, emitter brightness and photon purity must be maintained under practical conditions including room-temperature operation and integration with photonic circuits \cite{Aharonovich2016}. These conditions place severe restrictions on the physical properties of the host material, such as band structure, phonon interactions and dielectric medium.

Therefore, one of the main challenges in the engineering of quantum emitters is material selection. The main materials which produce the single photon emitter are from bulk materials like nitrogen-vacancy (NV) centers, Silicon Nitride, III-V quantum dots, and two-dimensional materials  \cite{doherty2013nitrogen} \cite{rogers2014multiple},\cite{chakraborty2015voltage, tran2016quantum}. In this perspective article, we examine the range of quantum materials being researched for quantum emitter applications, emphasizing their operational regimes, performance metrics, and applicability for scalable quantum systems.

\section{Quantum Applications}
\label{sec:Quantum application}

\begin{figure*}[htbp]
    \centering
    \includegraphics[width= 0.98\linewidth]{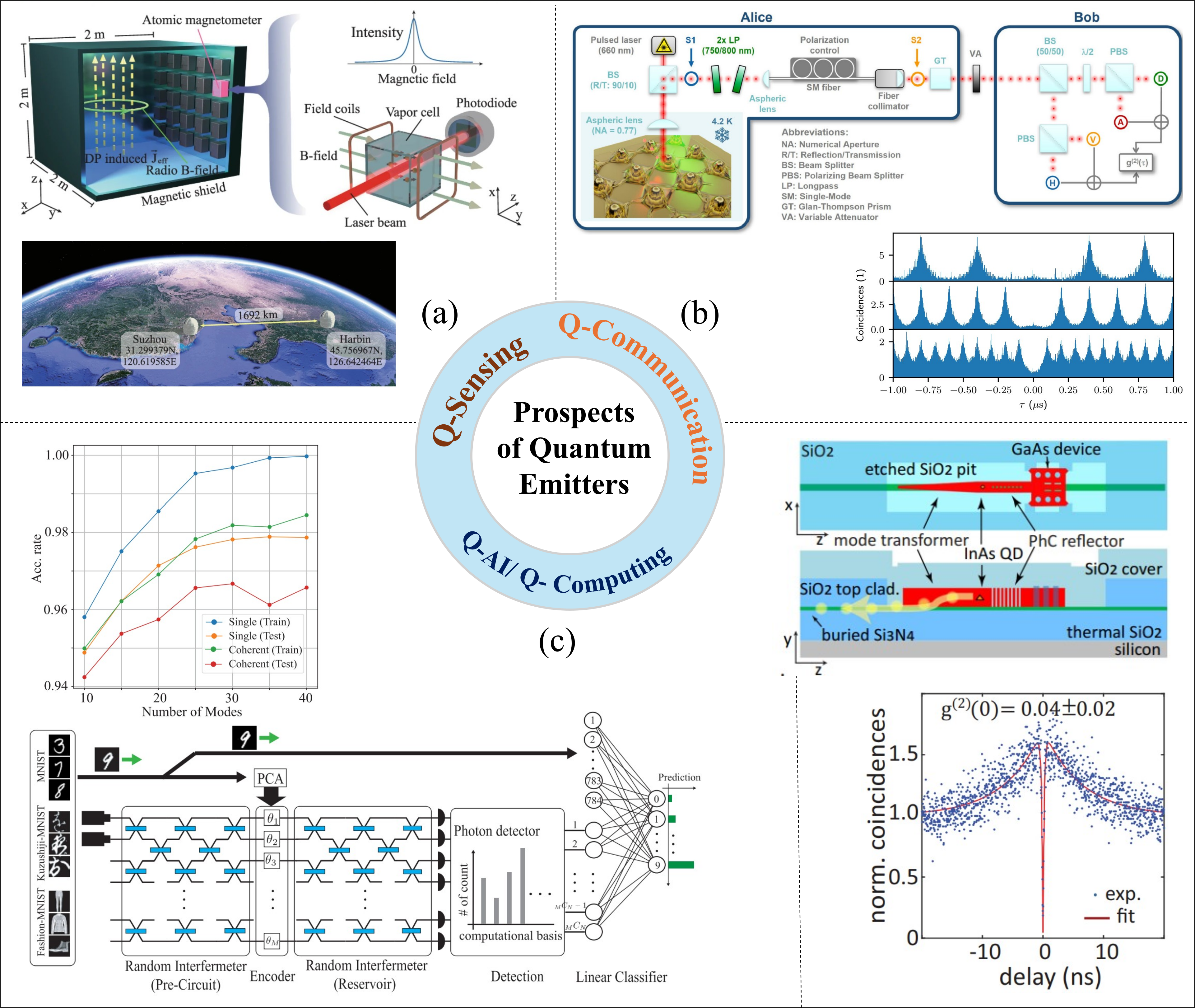}
    \caption{
\textbf{Outlook and applications of integrated single-photon emitter technology.} Schematic overview of emerging directions where integrated single-photon emitters underpin quantum sensing, communication, and computing.  
\textbf{(a) Quantum sensing} High-brightness, high-purity single-photon emission engineered via cavity quantum electrodynamics, illustrated in the context of a long-baseline quantum sensor network of 15 GPS-synchronized atomic magnetometers in two multi-layer mu-metal shield rooms in Suzhou and Harbin (separation $\sim$1700 km), where wall-mounted zero-field magnetometers detect dark-photon–induced radio-frequency magnetic fields with femtotesla sensitivity.  
\textbf{(b) Quantum-secured communication} Integrated single-photon sources for on-chip quantum key distribution, where a strain-engineered WSe$_2$ monolayer in a closed-cycle cryostat (Alice) is pumped by a pulsed diode laser, spectrally filtered, polarization-controlled, and fiber-coupled to send BB84 polarization states (H, V, D, A) through a variable-loss channel to a four-state polarization analyzer (Bob); spectra isolating an 807~nm emission line and second-order autocorrelation histograms up to 10~MHz confirm low-noise, high-purity operation suitable for high-speed quantum links. 
\textbf{(c) Photonic quantum computing and quantum AI} Integration of indistinguishable quantum-dot emitters in reconfigurable photonic circuits, exemplified by a GaAs/InAs QD single-photon source on an ultra-low-loss Si$_3$N$_4$ waveguide feeding a 50:50 MMI splitter and exhibiting high-purity $g^{(2)}(\tau)$ under resonant excitation (right panel). Other hand a quantum optical reservoir computing (QORC) architecture that encodes classical data onto multimode single-photon resource states via random interferometric networks and linear optics, and shows enhanced classification performance when driven by single photons compared to coherent light (left panel).
All figures are adapted with permission from: (a) ref. \cite{Jiang2024}; (b) ref. \cite{Gao2023}; (c) ref. \cite{Sakurai:25, Chanana2022ULLW}.}
    \label{fig:future}
\end{figure*}

The rapid advancement of integrated single-photon emitter (SPE) technologies represents a significant milestone toward scalable quantum photonic platforms. Specifically, the on-chip integration of solid-state emitters, including quantum dots, color centers, and defect-based emitters in two-dimensional materials, with low-loss photonic circuits has substantially enhanced source brightness, purity, and indistinguishability, while supporting operation at technologically relevant wavelengths\cite{Pelucchi2022PotentialIntegrated}. These developments are establishing SPEs as practical components for large-scale photonic quantum computing architectures, long-distance quantum communication networks, and high-precision quantum sensing protocols\cite{Giordani2023IntegratedPhotonicQT,Ramakrishnan2023IntegratedPlatformsReview}. Furthermore, the compatibility of many SPE platforms with established semiconductor processing enables wafer-scale fabrication and heterogeneous integration with detectors and control electronics, which is critical for achieving scalable quantum technologies.

\subsection{Quantum sensing}

Quantum sensors use coherent spin dynamics to detect weak external fields with sensitivities that can go beyond classical limits\cite{Degen2017QSensing}. Spin-exchange-relaxation-free (SERF) alkali-vapour magnetometers, for example, turn tiny magnetic fields into optical signals using optically pumped atomic spins, reaching sub-femtotesla noise levels in shielded settings\cite{Kominis2003SERF,Sheng2013Multipass}. Figure~1(a) shows a long-baseline array of these magnetometers placed in separate shield rooms and synchronized with GPS timing\cite{Jiang2024DPDM}. An ultralight dark-photon field, mixed with the photon, creates an oscillating current in the shield walls, which then produces a monochromatic radio-frequency magnetic field. The field's amplitude depends on the size of the cavity\cite{Chaudhuri2015Radio,Caputo2021DarkPhotonHandbook}. By cross-correlating the outputs from all SERF sensors, the experiment reaches a sensitivity of a few $\mathrm{fT}/\sqrt{\mathrm{Hz}}$ across the audio band and sets new terrestrial limits on the dark-photon kinetic-mixing parameter in the $10^{-15}$–$10^{-12}$~eV mass range\cite{Jiang2024DPDM}.
Figure~1(a) shows how a single SERF magnetometer works: circularly polarized resonant light pumps the alkali atoms into a polarized steady state. Transverse magnetic fields then cause Larmor precession, which rotates the spin polarization and changes the transmitted intensity at a photodiode\cite{Kominis2003SERF,Degen2017QSensing}. These same techniques are used in quantum biosensing platforms, such as optically pumped magnetometers and nitrogen-vacancy (NV) centres in diamond. These devices can detect pico- to femtotesla biomagnetic fields from brain and heart activity when placed close to the body inside multi-layer shield rooms\cite{Boto2018WearableMEG,Aslam2023BioQSens}. Large shielded biomedical facilities with dense magnetometer arrays serve two purposes: they support high-performance biosensing and can also act as nodes in global quantum sensor networks. In return, methods from dark-matter searches, like optimized cross-correlation, array calibration, and common-mode noise rejection, can help improve quantum-enabled biomedical diagnostics by making them more precise and reliable.

\subsection{Quantum communication}

Quantum communication protocols use quantum states of light to enable tasks such as secret key exchange, entanglement distribution, and the development of a quantum internet\cite{Kimble2008QI}. Most current systems rely on weak coherent pulses with decoy-state protocols to partially address the Poissonian photon-number statistics and vulnerability to photon-number-splitting attacks\cite{Wang2005Decoy,Ma2005Decoy}. A long-term solution is to use true single-photon emitters, whose sub-Poissonian statistics naturally suppress multi-photon events\cite{Aharonovich2016SPE}. Atomically thin transition-metal dichalcogenides (TMDCs) are promising for this purpose because they offer straightforward, low-cost fabrication, spectral tunability across all three telecom windows, and compatibility with large-scale photonic integration.\cite{Mak2016TMDCReview,Turunen2022Review}

Figure~1(a) shows a typical BB84 quantum key distribution (QKD) setup using a strain-engineered WSe$_2$ monolayer single-photon source.\cite{Gao2023TMDCQKD} Localized excitons in the monolayer are excited by a pulsed laser inside a cryostat, and their emission is spectrally filtered with low-loss long-pass filters before being coupled into a single-mode fiber. Polarization encoding of the four BB84 states $\{\ket{H},\ket{V},\ket{D},\ket{A}\}$ is achieved using fiber-based polarization control and a high-extinction polarizer on Alice’s side. The encoded single-photon pulses travel through a free-space or fiber link, where variable attenuation simulates channel loss, and are analyzed by Bob with a passive-basis polarization decoder consisting of non-polarizing and polarizing beam splitters and four single-photon detectors. This architecture represents a broader class of TMDC-based transmitters that can interface with existing QKD receivers and operate in wavelength regimes suitable for both terrestrial and satellite free-space links.\cite{Ursin2007FSO,Liao2017Satellite}

Figure~1(c) captures the essential non-classicality of the WSe$_2$ source by showing the second-order intensity autocorrelation $g^{(2)}(\tau)$ for different clock rates. In all cases, the strong suppression of coincidences around zero delay, with raw values of $g^{(2)}(0)\ll 0.5$, confirms dominantly single-photon emission into the quantum channel even close to saturation.\cite{Gao2023TMDCQKD} From the detected count rates and the calibrated transmission of Bob’s receiver, one can infer the mean photon number per pulse $\mu$ and hence the multiphoton probability that enters security analyses such as the GLLP framework.\cite{Gottesman2004GLLP,Waks2002SubPoisson} Together with the experimentally measured quantum bit error rate (QBER), which remains below the percent level when background and polarization misalignment are well controlled, these parameters determine the achievable secret-key rate and maximum tolerable channel loss.\cite{Takeoka2014RateLoss,Pirandola2017Repeaterless,Kupko2020Tools,Vajner2022QDReview} When benchmarked against semiconductor quantum dots and color centers in diamond,\cite{Waks2002Turnstile,Leifgen2014NVQKD,Takemoto2015QD120km} atomically thin TMDC emitters already reach competitive performance, and realistic improvements such as Purcell-enhanced extraction in microcavities and integration with low-dark-count detectors are expected to push TMDC-based QKD toward the rate–loss frontier required for global-scale quantum communication networks.\cite{Luo2018Cavity,Iff2021Purcell,Turunen2022Review}

\subsection{Photonic Quantum Computing}
In photonic quantum computing, the main challenge is maintaining a high flux of quality photons through increasingly complex circuits. As devices advance toward quantum simulation, optical neural networks, and fault-tolerant computation, propagation and insertion losses become the primary barriers to scaling
\cite{Sparrow2018,Steinbrecher2019,Choi2019,Rudolph2017,Wang2019Boson,Brod2019,Deshpande2022}. Ultra-low-loss Si$_3$N$_4$ waveguides address this issue by offering wafer-scale CMOS compatibility, broadband transparency from visible to infrared, and propagation losses near $10^{-2}\,\mathrm{dB/m}$ at telecom wavelengths. These features make them a strong foundation for large quantum photonic processors\cite{Liu2021,Blumenthal2020,Blumenthal2018}.

The hybrid platform described in Ref.~\cite{Chanana2022ULLW} addresses a key challenge by integrating deterministic single-photon emitters directly onto ultra-low-loss circuits. Previous integrated sources using spontaneous parametric down-conversion or four-wave mixing face a trade-off: increasing photon generation probability also raises the risk of unwanted multi-photon events, so higher brightness often reduces state purity, even with advanced multiplexing\cite{Kaneda2019}. Solid-state quantum emitters such as InAs quantum dots avoid this issue and can, in principle, produce near-transform-limited photons at high repetition rates \cite{Lodahl2018}. The device shown in Fig.~1(c) (right)implements this approach by placing a GaAs nanowaveguide with InAs quantum dots onto a buried Si$_3$N$_4$ ultra-low-loss waveguide, followed by an adiabatic mode transformer and a 50:50 multimode interference coupler\cite{Chanana2022ULLW,Schnauber2019}. This setup enables triggered single-photon emission with $g^{(2)}(0)\!<\!0.1$ directly into a waveguide network with propagation losses around $1\,\mathrm{dB/m}$ at $\sim 930\,\mathrm{nm}$. It combines high-purity single-photon generation with a circuit platform that supports long on-chip paths and dense linear-optical processing\cite{Chanana2022ULLW}.

The resonance-fluorescence measurements in Fig.~1(c) (right) further demonstrate the relevance of this approach for quantum computing. Under resonant driving, the integrated quantum dot shows strong antibunching with $g^{(2)}(0)\approx 0.04$ and clear evidence of coherent control, including a Mollow triplet and Rabi oscillations in Fourier-transform spectroscopy\cite{Chanana2022ULLW,Flagg2009,Ulhaq2012,Liu2018}. Along with previous demonstrations of waveguide-coupled resonance fluorescence and large-scale hybrid integration of artificial atoms,\cite{Makhonin2014,Reithmaier2015,Wan2020,Elshaari2020,Kim2020} these results point to quantum photonic processors where on-demand, nearly ideal single photons are injected into ultra-low-loss, reconfigurable networks with long delay lines, time-multiplexing stages, and integrated detectors. This combination meets the requirements of measurement-based photonic quantum computing, Gaussian boson sampling, and quantum repeater architectures, and offers a practical path to scaling photonic quantum information processing on chip\cite{GimenoSegovia2017,Dietrich2016}.

\subsection{Quantum-Driven Artificial Intelligence (Quantum AI)}
Photonic platforms that use boson sampling offer a promising way to explore quantum versions of artificial intelligence and machine learning. In boson sampling, $N$ identical single photons move through an $M$-mode linear-optical interferometer, and the output detection statistics depend on matrix permanents, which are believed to be very hard to compute with classical computers in the worst case\cite{Aaronson2011BosonSampling,Arute2019Supremacy,Zhong2020PhotonicAdvantage,Madsen2022GaussianAdvantage}. In the quantum optical reservoir computing (QORC) model,\cite{Sakurai2025QORC} these complex interference patterns are used as a high-dimensional feature map for learning tasks. As shown in Fig.~1 (c)(left), classical data such as images are first compressed, for example by principal component analysis (PCA), into a smaller set of real parameters $\{\theta_i\}$ that set the phases in a random interferometer with an $N$-photon input. The resulting multi-photon output is then sampled by coincidence detection, and these probabilities are sent to a classical linear classifier. There is no need for quantum backpropagation: the interferometer provides a fixed, hardware-based quantum feature map, and all trainable weights are in the final classical layer\cite{Nakajima2019QRC,Fujii2020QRC}.

Fig~1(c) (left)shows that this boson-sampling reservoir works as a quantum random-feature model. As the number of optical modes $M$ increases, and with it the size of the coincidence space ${M \choose N}$, the classification accuracy on datasets also improves. It even outperforms a purely linear support-vector classifier using the same PCA features\cite{Sakurai2025QORC}. When QORC is compared to classical random Fourier features (RFF),\cite{Rahimi2007RFF,Belkin2019DoubleDescent} its performance is similar to that of an RBF-kernel SVM, but it uses a physically generated feature map instead of a large random matrix. In this way, the boson-sampling interferometer creates a complex kernel on the input angles $\{\theta_i\}$, with the difficult structure of multi-photon interference producing a rich, nonlinear decision boundary in the space of input images.\cite{Huh2015BosonMolecules,Xiong2025QELM}

This approach also fits well with the wider field of quantum reservoir and extreme learning machines,\cite{Ghosh2019QReservoir,Domingo2022NoisyQRC,Suprano2024QELM,DeLorenzis2025QELMImage} where quantum dynamics provide a fixed, complex transformation and only the readout layer is trained. An important finding from QORC is that single-photon inputs perform much better than coherent-state inputs in the same interferometer, even when both are selected to have the same total photon number.\cite{Nakajima2021CoherentRC,Ma2023IntegratedRC,Sakurai2025QORC} This shows that true many-boson interference, not just classical wave interference, is what creates the larger and more expressive feature space used by the quantum model. These results point to a practical way forward for “quantum AI” on photonic chips: hybrid systems where boson-sampling interferometers generate quantum features or act as reservoirs, closely linked with classical linear classifiers or shallow neural networks that work in the expanded feature space. These designs could become specialized quantum accelerators for pattern recognition, time-series prediction, and other AI tasks where kernel methods are already strong.


\section{Materials for Single photon emitter}

\begin{figure*}[htbp]
    \centering
    \includegraphics[width=0.98\textwidth]{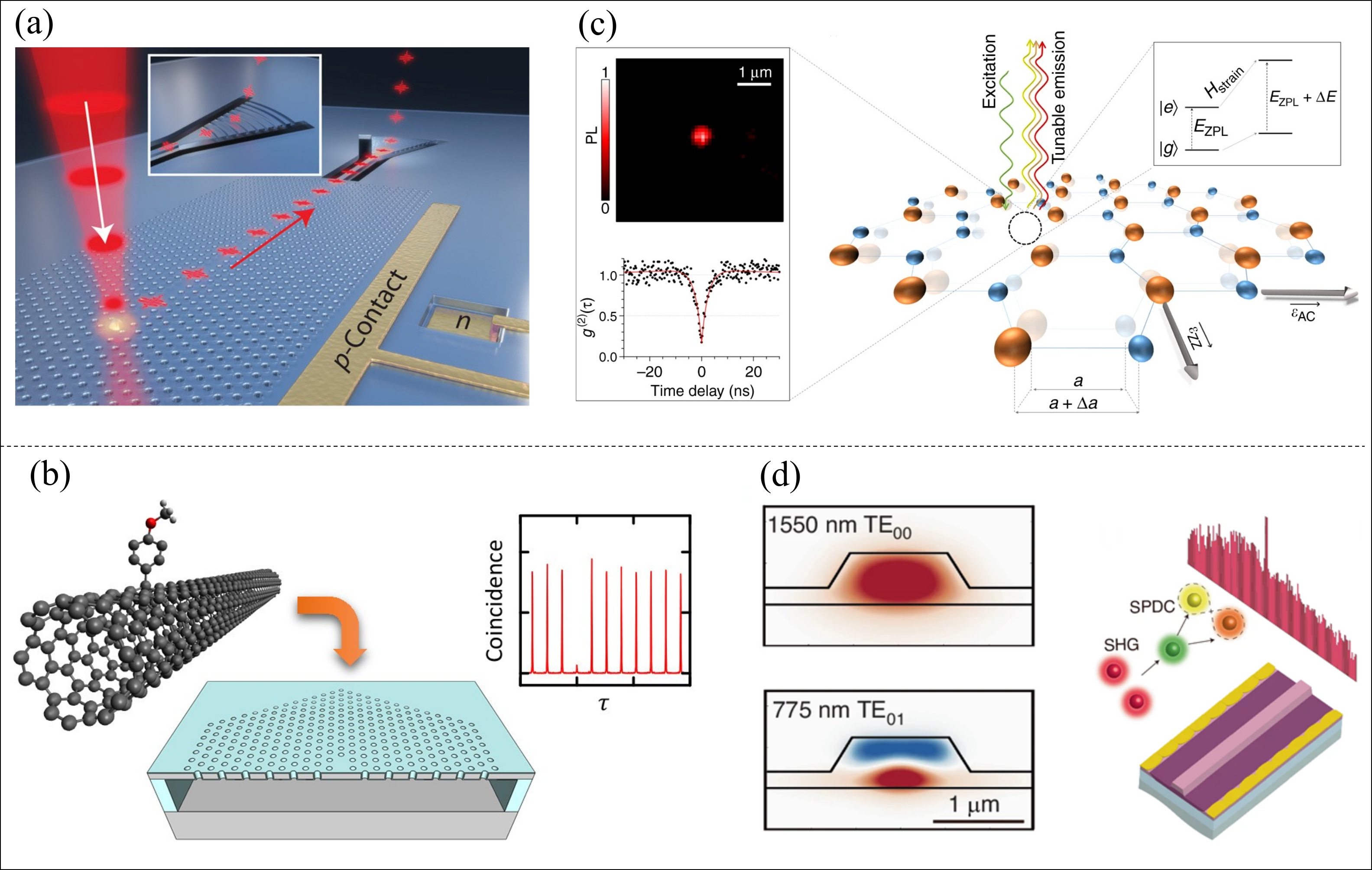}
    \caption{\textbf{Materials landscape of integrated quantum light sources across dimensions} (a) Electrically controlled InAs quantum dot embedded in a GaAs nanophotonic waveguide with adiabatic coupling to an on-chip output, exemplifying a \textbf{zero-dimensional} emitter engineered into a bright, nearly deterministic single-photon source via a large $\beta$-factor and charge stability. (b) Doped single-walled carbon nanotube (SWCNT) hosting $sp^{3}$-defect–localized excitons coupled to a silicon photonic crystal cavity, representing \textbf{one-dimensional} host that provides telecom-band, room-temperature single-photon emission enhanced by cavity quantum electrodynamics. (c) Atom-like defect centres in hexagonal boron nitride (hBN), with (insets) an atomic model, confocal map and antibunching trace, illustrating a \textbf{two-dimensional} van der Waals host that supports room-temperature single-photon emission with transition energy tunable by strain and local environment. (d) Layer-poled thin-film lithium niobate (LN) waveguide supporting cascaded SHG and spontaneous parametric down-conversion (SPDC), highlighting a \textbf{three-dimensional}  ferroelectric platform nanostructured and poled to generate bright photon pairs across multiple telecom bands on an integrated chip (refer to Table 1.1 for a quantitative comparison of operating temperature, radiative lifetime, and coherence time of quantum materials). Figure panels are adapted with permission from: (a) ref. \cite{Uppu2020SciAdv}; (b) ref. \cite{Ishii2018NanoLett}; (c) ref. \cite{Grosso2017NatCommun}, (d) ref. \cite{Shi2024LPLN} }
    \label{fig:intro_figure}
\end{figure*}

Bulk materials are well-established platforms for quantum emitters, offering long coherence times and high optical quality. Diamond is the most widely studied, with nitrogen-vacancy (NV) centers that work at room temperature and have millisecond spin coherence\cite{doherty2013nitrogen}. Silicon-vacancy (SiV) and germanium-vacancy (GeV) centers in diamond need cryogenic temperatures but offer better optical linewidths and more consistent photons\cite{rogers2014multiple, Sipahigil2014}. Silicon carbide (SiC) is another strong candidate because it is compatible with CMOS technology and has spin-active colour centres,\cite{castelletto2020silicon}. Rare-earth doped crystals, like Y$_2$SiO$_5$:Er$^{3+}$, also stand out for their very long coherence times and optical transitions in the telecom band, making them suitable for quantum memory \cite{Zhong2015}.  
On the other hand, new materials discovered through experiments in the 21st century, known as two-dimensional materials, are also promising candidates for single-photon emitters. Two-dimensional materials combine reduced dimensionality, strong light-matter interactions, and easy integration with nanophotonic structures. Transition metal dichalcogenides like WSe$_2$, MoSe$_2$, and WS$_2$ are well known for hosting strain-localized excitonic quantum emitters\cite{chakraborty2015voltage, palacios2017atomically}. These emitters are usually found at defect or strain-induced sites and can produce single photons at low temperatures. Hexagonal boron nitride is also a strong option for quantum emission at room temperature because of its wide bandgap and ability to tolerate defects\cite{tran2016quantum, kim2018photonic}. In hBN, defect centres show sharp zero-phonon lines and stable light emission. Graphene quantum dots and graphene oxide derivatives have shown promise for tunable fluorescence, but their quantum coherence is still limited \cite{peng2020graphene}.
Figure~\ref{fig:intro_figure} shows Single-photon emitters (SPEs) have been realized across a diverse set of material platforms as explained above, each offering distinct trade-offs in operating temperature, optical coherence, integration readiness, and scalability. Below we summarise the explanation for each set of material shown in Figure~\ref{fig:intro_figure}  provided.

\subsection{0D quantum emitters: quantum dots as near-ideal single-photon sources}

A single InAs quantum dot embedded in a GaAs nanophotonic waveguide, electrically contacted and adiabatically coupled to an on-chip output (adapted from Ref.~\cite{Uppu2020SciAdv}), exemplifies how an atom-like emitter confined in a 0D potential can operate as a bright, nearly deterministic single-photon source when integrated into a photonic environment with a large $\beta$-factor and a stable charge configuration (Fig 2(a)). Early cavity-based devices established that quantum dots can match or even outperform SPDC sources in terms of purity, indistinguishability, and brightness,\cite{Somaschi2016NatPhoton,Ding2016PRL} and Uppu~\textit{et al.} extended this paradigm to planar, scalable waveguide circuits.\cite{Uppu2020SciAdv} From a materials perspective, 0D III--V quantum dots currently set the practical benchmark for on-demand single-photon performance, while major outstanding challenges include wafer-scale spectral uniformity, deterministic positioning, and hybrid integration with heterogeneous material platforms and active photonic circuitry.

\subsection{1D quantum emitters: doped carbon nanotubes in the telecom band}

Within one-dimensional material platforms, dopant-localized excitons in single-walled carbon nanotubes (SWCNTs) coupled to a silicon photonic crystal cavity, as realized in Ref.~\cite{Ishii2018NanoLett}, highlight how CNTs can function as integrated quantum light sources (Fig 2(b)). The SWCNT backbone provides a tunable excitonic system, while $sp^{3}$ functionalization introduces deep, quantum-dot-like traps that yield room-temperature single-photon emission. Earlier studies established solitary dopants in SWCNTs as room-temperature single-photon sources\cite{Ma2015NatNano} and demonstrated telecom-band emission from $sp^{3}$ defects spanning the O, S, C, and L bands.\cite{He2017NatPhotonCNT} In this setting, cavity coupling on a silicon platform achieves Purcell enhancement without degrading antibunching,\cite{Ishii2018NanoLett} advancing ultra-compact, telecom-compatible quantum light sources. From a materials perspective, one-dimensional hosts such as CNTs and semiconductor nanowires offer a natural bridge between nanoscale emitters and established silicon photonics, but still face hurdles in chirality-selective growth, long-term chemical stability, and scalable wafer-level integration relative to 0D and 3D platforms.

\subsection{2D quantum emitters}

Fig 2(c)(taken from Ref.~\cite{Grosso2017NatCommun}) illustrates how point defects in hexagonal boron nitride (hBN) serve as atom-like emitters in an atomically thin host. The figure connects a simple defect model in the hBN lattice to confocal PL maps and $g^{(2)}(\tau)$ traces, demonstrating high-purity single-photon emission at room temperature, and highlights strain as an effective tool to tune the zero-phonon line by several meV\cite{Grosso2017NatCommun}. Since then, significant progress has been made in controlling both the chemical identity and growth of hBN, including carbon-doped thin films that provide highly pure, stable single-photon emission with $g^{(2)}(0)\!\sim\!0.01$ on centimetre scales\cite{Chatterjee2025SciAdv}. From a materials perspective, 2D hosts such as hBN and transition-metal dichalcogenides offer room-temperature operation, mechanical flexibility, and van der Waals integration, making them well-suited for hybrid stacks on silicon, III--V, or ferroelectric photonic platforms. The main challenges are deterministic defect engineering, charge-state control, and achieving spectral stability comparable to the best 0D quantum-dot devices.

\subsection{3D nonlinear hosts: SPDC in engineered lithium niobate waveguides}

A layer-poled thin-film lithium niobate (LN) waveguide, as constructed in Ref.~\cite{Shi2024LPLN}, employs cascaded second-harmonic generation and spontaneous parametric down-conversion (SPDC) to generate photon pairs across multiple telecom bands. In contrast to localized 0D emitters, the active region here is an extended nonlinear optical mode supported by a three-dimensional ferroelectric host, with layer-wise poling and modal engineering used to break symmetry and maximize the effective $\chi^{(2)}$ overlap.\cite{Shi2024LPLN} Related work on periodically poled thin-film LN has yielded high-quality entangled photon pairs and heralded single photons with excellent CAR and $g^{(2)}(0)$ in compact waveguides,\cite{Zhao2020PRL_TFLN} demonstrating that 3D nonlinear media can be developed into on-chip, telecom-ready quantum light sources whose brightness scales with chip length and pump power. More broadly, SPDC-based platforms remain unmatched for producing entangled states and multiplexed photon streams compatible with fiber networks, and are naturally suited to coexist and co-integrate with 0D, 1D, and 2D emitters on heterogeneous quantum photonic chips.

Table1.1 compares representative materials across four key figures of merit: (i) \emph{Operating Temperature}, indicating whether room-temperature operation is feasible or if cryogenic cooling is required; (ii) \emph{Radiative Lifetime} (in ns), connected to the transform-limited linewidth (iii) \emph{Optical Coherence Time} ($T_2$), which quantifies dephasing and sets an upper bound for photon indistinguishability via $T_2 \le 2T_1$.

\begin{table*}[htbp]
\centering
\small
\caption{Key properties of quantum materials for single-photon emission.}
\label{tab:quantum_properties}
\renewcommand{\arraystretch}{1.3}
\setlength{\tabcolsep}{5pt}
\begin{tabular}{|p{4.5cm}|p{3.25cm}|p{2.5cm}|p{3.2cm}|p{3.0cm}|}
\hline
\textbf{Materials} & \textbf{Operating Temp.} & \textbf{Lifetime (ns)} & \textbf{Coherence Time}  \\
\hline
hBN (carbon- or oxygen-related SPEs) \cite{tran2016quantum,bourrellier2016bright,chejanovsky2021single} & Room (300 K) & 1--5 & 0.1--1 $\mu$s (optical)  \\
\hline
MoS$_2$ (localized exciton) \cite{chakraborty2015voltage} & Cryogenic (4 K) & 3--10 & 10--100 ps (optical)  \\
\hline
WSe$_2$ (strain-localized exciton) \cite{Koperski2015} & Cryogenic (4 K) & 2--10 & 10--100 ps (optical)  \\
\hline
WS$_2$ (defect-localized exciton) \cite{palacios2017atomically} & Cryogenic (4 K) & 2--8 & 10--80 ps (optical) \\
\hline
MoSe$_2$ (strain or defect localized) \cite{Schneider2018, BrotonsGisbert2019} & Cryogenic (4 K) & 2--6 & 10--80 ps (optical)  \\
\hline
TMDC heterostructures (e.g., WSe$_2$/MoSe$_2$) \cite{Seyler2019,Alexeev2019} & Cryogenic (4 K) & 100--500 & 100 ps--few ns (optical)  \\
\hline
Graphene quantum dots (SPE behavior) \cite{Zhao2018} & Room (300 K) & 1--10 & Tens of ps--ns \hspace{1mm} (optical) \\
\hline
NV center in diamond \cite{doherty2013nitrogen} & Room (300 K) & $\sim$12 & Up to ms (spin)  \\
\hline
SiV center in diamond \cite{rogers2014multiple,Sipahigil2014} & Cryogenic (4--10 K) & $\sim$1.5 & 10--100 ns  \\
\hline
V$_\text{Si}$ in SiC \cite{castelletto2020silicon,Anderson2019} & Room (300 K) & 6--8 & 100--300 $\mu$s (spin)  \\
\hline
Er$^{3+}$ in Y$_2$SiO$_5$ \cite{Zhong2015,Kindem2020} & Cryogenic (4 K) & $\sim$10 & Up to seconds (spin)  \\
\hline
Silicon Nitride (SiN) \cite{Senichev2021} & Room (300 K) & N/A (Passive) & N/A  \\
\hline
Lithium Niobate (LiNbO$_3$) \cite{Wang2018} & Room (300 K) & N/A (Passive host) & N/A \\
\hline
III--V Quantum Dots (InAs/GaAs, GaN) \cite{Somaschi2016NatPhoton,Ding2016PRL} & Cryogenic (4--10 K) & 0.5--1 & 100 ps--1 ns \\
\hline
Silicon (G-centers, C-centers) \cite{Komza2024} & Cryogenic (4--20 K) & $\sim$6--10 & Up to $\mu$s (spin)  \\
\hline
Aluminum Nitride (AlN) \cite{tran2016quantum} & Room (300 K) & 1--10 & Tens of ps--ns \hspace{1mm}(optical)  \\
\hline
\end{tabular}
\end{table*}

\section{Integration of Single-Photon Emitters with Photonic Devices}
\label{sec:integration}

The integration of a single photon emitter (SPE) with a photonic device in a deterministic manner is an important factor for its use in solid-state quantum technologies. During the coupling of SPEs with photonic devices, it is important to achieve coherent coupling, which requires the emitter to be confined within the photonic mode of the cavity. This confinement can enhance the radiative decay rate by increasing the Purcell factor (the ratio of the emission rate of the emitter in the cavity to its emission rate in free space), achieving a collection efficiency close to unity. The achievable coupling efficiency, defined as $\eta_c = \Gamma_\mathrm{wg}/\Gamma_\mathrm{tot}$, where $\Gamma_\mathrm{wg}$, where $\eta_c$ is the coupling efficiency, $\Gamma_\mathrm{wg}$ is the decay rate into the guided mode and $\Gamma_\mathrm{tot}$ is the total decay rate, is dictated by dipole orientation, photonic mode overlap, and emitter positioning. 

\begin{figure*}[htbp]
    \centering
    \includegraphics[width = 0.98\linewidth]{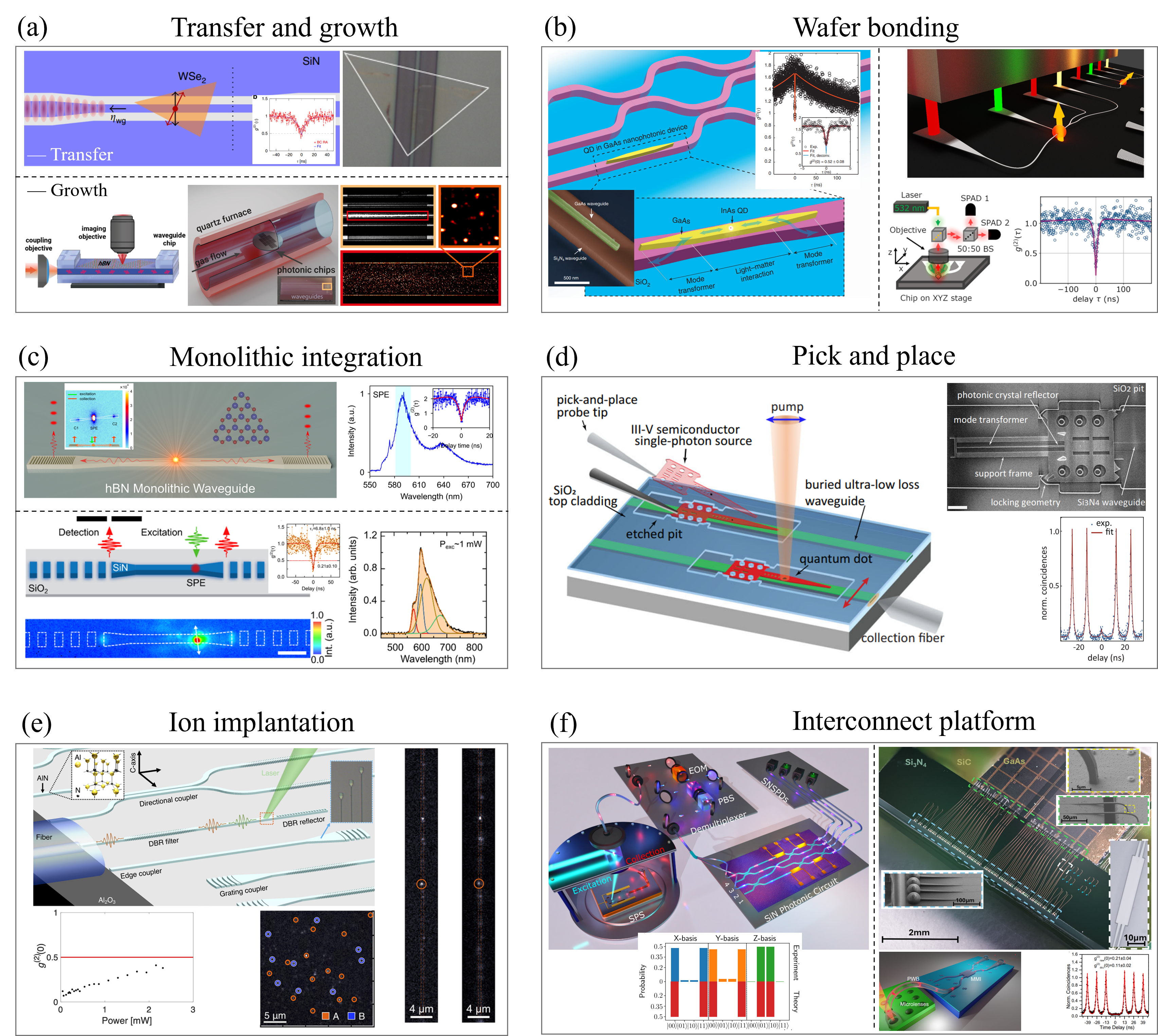}
    \caption{\textbf{Milestones in hybrid and heterogeneous integration of solid-state quantum emitters (QEs) with low-loss QPIC platforms, highlighting routes to scalable, stable, and efficient cavity–emitter interaction systems.} (a) \textbf{Transfer printing} (top) and direct CVD \textbf{growth} (bottom) of 2D materials (WSe$_2$, hBN) on SiN waveguides to realize localized quantum emitters coupled to guided modes; insets: flake–waveguide interface (top) and zoomed optical images of as-grown hBN emitters on the waveguide (bottom). (b) \textbf {Wafer-bonded} bulk emitters on PICs, with a single InAs QD in a GaAs nanowaveguide adiabatically coupled to a SiN chip (left) and NV centers in nanodiamonds coupled to photonic crystal cavities (right) insets: measured $g^{(2)}(0)$ under QD–cavity coupling (left) and cavity-enhanced NV emission (right). (c) \textbf{Monolithic on-chip integration} of hBN, serving as a host of defect-based extrinsic quantum emitters (top), and SiN, acting as a source of intrinsic quantum emitters (bottom), in simple nanophotonic waveguides, enabling generation and routing of single photons on the same platform; insets: PL spectra revealing multiple, spectrally distinct extrinsic and intrinsic emitter species respectively. (d) \textbf{Pick-and-place integration} of heterogeneous quantum microchiplets, with III–V QD sources aligned into prefabricated PIC sockets using micromanipulators; insets: quantum signature of triggered single-photon emission. (e) Deterministic creation of photostable vacancy-based emitters into AlN waveguides via \textbf{He-ion implantation} followed by thermal annealing; inset: confocal image with isolated emitters. (f) \textbf{Photonic interconnects} bridging external quantum emitters and photonic chips: externally generated high-quality quantum emitters interfaced with programmable SiN PICs (left panel, histograms showing qubit measurements) and a III–V QEs-based chip is photonic wire bonded to a SiN chip (right panel, SEM micrograph of the wire-bonded interface). Figure panels are adapted with permission from: (a) ref. \cite{Peyskens2019, Glushkov2021}; (b) ref. \cite{Davanco2017,Schrinner2020}; (c) ref. \cite{Senichev2022, Li2021}; (d) ref. \cite{Chanana2022ULLW}; (e) ref. \cite{Lu2020}; (f) ref. \cite{Wang2023, Pfister2025}.}
    \label{fig:integration}
\end{figure*}

Figure~\ref{fig:integration} explains how single photon emitters can be integrated with various photonic devices. These include transfer and growth, wafer bonding, pick-and-place, doping and ion implantation, parametric nonlinearity-based hybrid integration, monolithic integration, and photonic interconnects. Each method balances coupling precision, material compatibility, and scalability.

\subsection{Transfer and Growth Techniques}
Single photon emitters in two-dimensional (2D) materials can be directly integrated onto photonic chips, providing a high-throughput route to integration. For example, defect states in hexagonal boron nitride (hBN), a type of 2D insulating material, can be grown directly on silicon nitride (Si$_3$N$_4$) photonic chips, while transition metal dichalcogenides (TMDCs), which are semiconducting 2D materials, can be transferred onto these chips Figure~\ref{fig:integration}(a). Single photons generated using this method are in close proximity to the photonic chip, experience reduced dielectric screening (less interference from surrounding materials), minimize lattice mismatch (better alignment of atomic structures), and enable defect-free interfaces with high refractive index contrast to the underlying Si$_3$N$_4$ or aluminum nitride (AlN) photonic layers \cite{Peyskens2019, Glushkov2021}. This methodology offers planar scalability and compatibility with wafer-level fabrication.

\subsection{Wafer Bonding}
Wafer bonding provides  integration of different materials such as III–V semiconductors, diamond, or AlN onto silicon or Si$_3$N$_4$ photonic wafers \cite{Davanco2017,Schrinner2020} Figure~\ref{fig:integration}(b). In this method, a quantum emitter-containing layer is bonded to a photonic chip using adhesive bonding or plasma activation, for substrate removal and photonic circuit patterning\cite{Davanco2017}. The  Purcell factor $F_P$ enhances up to 50 times in optimized cavity–emitter systems, followed by the optical field confinement within the bonded interface\cite{Kim2017}. Bonding also allows alignment of multiple functional layers, such as emitters, modulators, and detectors on the same platform, creating hybrid photonic–electronic quantum chips.

\subsection{Monolithic Integration}
In a monolithic platform, both the photonic circuit and the quantum emitter are integrated within the same material, minimizing fabrication complexities Fig.~\ref{fig:integration}(c). Silicon systems embedding atomic emissive centers distinct atomic-scale defects that can emit single photons within photonic cavities have recently achieved on-chip single-photon emission with high spectral stability \cite{Li2021,Senichev2022}. This configuration eliminates interfacial losses (losses at material boundaries) and guarantees thermal and mechanical robustness. Similarly, diamond photonic circuits incorporating nitrogen-vacancy (NV) or silicon-vacancy (SiV) centers specific types of atomic defects in diamond that can act as quantum emitters have demonstrated coherent spin–photon interfaces \cite{Aharonovich2016}. The optical field enhancement inside such cavities modifies the emitter’s spontaneous emission rate according to Fermi’s Golden Rule, which describes the probability of transitions between quantum states, leading to enhanced radiative efficiency (higher light output per excitation) and reduced lifetime jitter (variability in emission duration)\cite{Aharonovich2016}.

\subsection{Pick-and-Place Deterministic Assembly}
Deterministic placement techniques shown in Fig.~\ref{fig:integration}(d) involve transferring pre-characterized emitters or nanocrystals onto photonic chips by using micro-manipulators or polymer stamps \cite{Chanana2022}. This strategy provides the selection of emitters with narrow linewidths and stable emission, ensuring high device yield. For example, a quantum dot or two-dimensional flakes can be placed in the vicinity of  the cavity to maximize coupling. When coupled to Si$_3$N$_4$ photonic wire structures, coupling efficiencies exceeding 80\% have been reported \cite{Chanana2022}. Such deterministic positioning also supports integration with programmable photonic circuits for reconfigurable quantum networks \cite{Chanana2022}.

\subsection{Doping and Ion Implantation}
It is a technique to generate the defect center directly within the host material in a controllable manner.  Focused ion beams (FIB) or masked implantation can fabricate arrays of emitters with subnanometer precision  \cite{Lu2020} Fig.~\ref{fig:integration}(e). Post-implantation annealing activates the defect states and repairs implantation-induced damage, restoring optical coherence. For example, the NV centers in the diamond or color center in the hBN provide scalable integration with the monolithic photonic circuits as well as CMOS compatibility. The ion implantation energy controls the emitter depth, enabling precise overlap with optical modes and tailoring of the local photonic density of states.

\subsection{Photonic Interconnects}
As the name Interconnect suggests, these architectures link multiple emitters and photonic components into scalable quantum networks. Building on this approach, the large-scale integration of artificial atoms within hybrid photonic circuits has enabled photon-mediated coupling between spatially separated emitters \cite{Wang2023}. Furthermore, low-loss Si$_3$N$_4$ interconnects ($<0.1$ dB/cm) preserve coherence over centimeter-scale distances \cite{Pfister2025}. As a result, these interconnects serve as waveguides, delay lines, and interferometers, enabling quantum interference and entanglement distribution across chip-scale systems.

\section{Conclusion}

The fundamental pillars of quantum photonic technologies are qubits, and single-photon sources are one type of qubit. Over the past decade, researchers have significantly advanced the understanding and engineering of optically active quantum emitters across diverse material classes. In this perspective article, we discuss the application of quantum technologies like quantum sensing, quantum computing, quantum AI, and quantum communication. We discussed the material aspects for single-photon emitters from 0D to 3D and highlighted their key properties. We also examine ways to integrate single-photon emitters with photonic chips for practical applications. Finally, we outline the application of integration for single-photon emitters for quantum technologies.

\section*{Acknowledgement}
We acknowledge funding support from the National Quantum Mission, an initiative of the Department of Science and Technology, Government of India.
\bibliography{ref.bib}

\begin{thebibliography}{147}%
\makeatletter
\providecommand \@ifxundefined [1]{%
 \@ifx{#1\undefined}
}%
\providecommand \@ifnum [1]{%
 \ifnum #1\expandafter \@firstoftwo
 \else \expandafter \@secondoftwo
 \fi
}%
\providecommand \@ifx [1]{%
 \ifx #1\expandafter \@firstoftwo
 \else \expandafter \@secondoftwo
 \fi
}%
\providecommand \natexlab [1]{#1}%
\providecommand \enquote  [1]{``#1''}%
\providecommand \bibnamefont  [1]{#1}%
\providecommand \bibfnamefont [1]{#1}%
\providecommand \citenamefont [1]{#1}%
\providecommand \href@noop [0]{\@secondoftwo}%
\providecommand \href [0]{\begingroup \@sanitize@url \@href}%
\providecommand \@href[1]{\@@startlink{#1}\@@href}%
\providecommand \@@href[1]{\endgroup#1\@@endlink}%
\providecommand \@sanitize@url [0]{\catcode `\\12\catcode `\$12\catcode `\&12\catcode `\#12\catcode `\^12\catcode `\_12\catcode `\%12\relax}%
\providecommand \@@startlink[1]{}%
\providecommand \@@endlink[0]{}%
\providecommand \url  [0]{\begingroup\@sanitize@url \@url }%
\providecommand \@url [1]{\endgroup\@href {#1}{\urlprefix }}%
\providecommand \urlprefix  [0]{URL }%
\providecommand \Eprint [0]{\href }%
\providecommand \doibase [0]{https://doi.org/}%
\providecommand \selectlanguage [0]{\@gobble}%
\providecommand \bibinfo  [0]{\@secondoftwo}%
\providecommand \bibfield  [0]{\@secondoftwo}%
\providecommand \translation [1]{[#1]}%
\providecommand \BibitemOpen [0]{}%
\providecommand \bibitemStop [0]{}%
\providecommand \bibitemNoStop [0]{.\EOS\space}%
\providecommand \EOS [0]{\spacefactor3000\relax}%
\providecommand \BibitemShut  [1]{\csname bibitem#1\endcsname}%
\let\auto@bib@innerbib\@empty
\bibitem [{\citenamefont {Gisin}\ \emph {et~al.}(2002)\citenamefont {Gisin}, \citenamefont {Ribordy}, \citenamefont {Tittel},\ and\ \citenamefont {Zbinden}}]{gisin2002quantum}%
  \BibitemOpen
  \bibfield  {author} {\bibinfo {author} {\bibfnamefont {N.}~\bibnamefont {Gisin}}, \bibinfo {author} {\bibfnamefont {G.}~\bibnamefont {Ribordy}}, \bibinfo {author} {\bibfnamefont {W.}~\bibnamefont {Tittel}},\ and\ \bibinfo {author} {\bibfnamefont {H.}~\bibnamefont {Zbinden}},\ }\bibfield  {title} {\bibinfo {title} {Quantum cryptography},\ }\href@noop {} {\bibfield  {journal} {\bibinfo  {journal} {Reviews of modern physics}\ }\textbf {\bibinfo {volume} {74}},\ \bibinfo {pages} {145} (\bibinfo {year} {2002})}\BibitemShut {NoStop}%
\bibitem [{\citenamefont {O'Brien}\ \emph {et~al.}(2009)\citenamefont {O'Brien}, \citenamefont {Furusawa},\ and\ \citenamefont {Vu{\v{c}}kovi{\'c}}}]{obrien2009photonic}%
  \BibitemOpen
  \bibfield  {author} {\bibinfo {author} {\bibfnamefont {J.~L.}\ \bibnamefont {O'Brien}}, \bibinfo {author} {\bibfnamefont {A.}~\bibnamefont {Furusawa}},\ and\ \bibinfo {author} {\bibfnamefont {J.}~\bibnamefont {Vu{\v{c}}kovi{\'c}}},\ }\bibfield  {title} {\bibinfo {title} {Photonic quantum technologies},\ }\href@noop {} {\bibfield  {journal} {\bibinfo  {journal} {Nature Photonics}\ }\textbf {\bibinfo {volume} {3}},\ \bibinfo {pages} {687} (\bibinfo {year} {2009})}\BibitemShut {NoStop}%
\bibitem [{\citenamefont {Degen}\ \emph {et~al.}(2017{\natexlab{a}})\citenamefont {Degen}, \citenamefont {Reinhard},\ and\ \citenamefont {Cappellaro}}]{degen2017quantum}%
  \BibitemOpen
  \bibfield  {author} {\bibinfo {author} {\bibfnamefont {C.~L.}\ \bibnamefont {Degen}}, \bibinfo {author} {\bibfnamefont {F.}~\bibnamefont {Reinhard}},\ and\ \bibinfo {author} {\bibfnamefont {P.}~\bibnamefont {Cappellaro}},\ }\bibfield  {title} {\bibinfo {title} {Quantum sensing},\ }\href@noop {} {\bibfield  {journal} {\bibinfo  {journal} {Reviews of Modern Physics}\ }\textbf {\bibinfo {volume} {89}},\ \bibinfo {pages} {035002} (\bibinfo {year} {2017}{\natexlab{a}})}\BibitemShut {NoStop}%
\bibitem [{\citenamefont {Bennett}(1992)}]{bennett1992}%
  \BibitemOpen
  \bibfield  {author} {\bibinfo {author} {\bibfnamefont {C.~H.}\ \bibnamefont {Bennett}},\ }\bibfield  {title} {\bibinfo {title} {Experimental quantum cryptography},\ }\href {https://doi.org/10.1007/BF00191318} {\bibfield  {journal} {\bibinfo  {journal} {Journal of Cryptology}\ }\textbf {\bibinfo {volume} {5}},\ \bibinfo {pages} {3} (\bibinfo {year} {1992})}\BibitemShut {NoStop}%
\bibitem [{\citenamefont {Zhang}\ \emph {et~al.}(2025)\citenamefont {Zhang}, \citenamefont {Ding}, \citenamefont {Li}, \citenamefont {Zhang}, \citenamefont {Guo}, \citenamefont {Wang}, \citenamefont {Ning}, \citenamefont {Xu}, \citenamefont {Liu}, \citenamefont {Zhao}, \citenamefont {Zou}, \citenamefont {Wang}, \citenamefont {Cao}, \citenamefont {He}, \citenamefont {Peng}, \citenamefont {Huo}, \citenamefont {Liao}, \citenamefont {Lu}, \citenamefont {Xu},\ and\ \citenamefont {Pan}}]{zhang2025}%
  \BibitemOpen
  \bibfield  {author} {\bibinfo {author} {\bibfnamefont {Y.}~\bibnamefont {Zhang}}, \bibinfo {author} {\bibfnamefont {X.}~\bibnamefont {Ding}}, \bibinfo {author} {\bibfnamefont {Y.}~\bibnamefont {Li}}, \bibinfo {author} {\bibfnamefont {L.}~\bibnamefont {Zhang}}, \bibinfo {author} {\bibfnamefont {Y.-P.}\ \bibnamefont {Guo}}, \bibinfo {author} {\bibfnamefont {G.-Q.}\ \bibnamefont {Wang}}, \bibinfo {author} {\bibfnamefont {Z.}~\bibnamefont {Ning}}, \bibinfo {author} {\bibfnamefont {M.-C.}\ \bibnamefont {Xu}}, \bibinfo {author} {\bibfnamefont {R.-Z.}\ \bibnamefont {Liu}}, \bibinfo {author} {\bibfnamefont {J.-Y.}\ \bibnamefont {Zhao}}, \bibinfo {author} {\bibfnamefont {G.-Y.}\ \bibnamefont {Zou}}, \bibinfo {author} {\bibfnamefont {H.}~\bibnamefont {Wang}}, \bibinfo {author} {\bibfnamefont {Y.}~\bibnamefont {Cao}}, \bibinfo {author} {\bibfnamefont {Y.-M.}\ \bibnamefont {He}}, \bibinfo {author} {\bibfnamefont {C.-Z.}\ \bibnamefont {Peng}}, \bibinfo {author} {\bibfnamefont {Y.-H.}\ \bibnamefont {Huo}}, \bibinfo
  {author} {\bibfnamefont {S.-K.}\ \bibnamefont {Liao}}, \bibinfo {author} {\bibfnamefont {C.-Y.}\ \bibnamefont {Lu}}, \bibinfo {author} {\bibfnamefont {F.}~\bibnamefont {Xu}},\ and\ \bibinfo {author} {\bibfnamefont {J.-W.}\ \bibnamefont {Pan}},\ }\bibfield  {title} {\bibinfo {title} {Experimental single-photon quantum key distribution surpassing the fundamental secret-key-rate limit},\ }\href {https://doi.org/10.1103/PhysRevLett.134.210801} {\bibfield  {journal} {\bibinfo  {journal} {Physical Review Letters}\ }\textbf {\bibinfo {volume} {134}},\ \bibinfo {pages} {210801} (\bibinfo {year} {2025})}\BibitemShut {NoStop}%
\bibitem [{\citenamefont {Laneve}\ \emph {et~al.}(2025)\citenamefont {Laneve}, \citenamefont {Ronco}, \citenamefont {Beccaceci}, \citenamefont {Barigelli}, \citenamefont {Salusti}, \citenamefont {Claro-Rodríguez}, \citenamefont {Pascalis}, \citenamefont {Suprano}, \citenamefont {Chiaudano}, \citenamefont {Sch\"oll}, \citenamefont {Hanschke}, \citenamefont {Krieger}, \citenamefont {Buchinger}, \citenamefont {Silva}, \citenamefont {Neuwirth}, \citenamefont {Stroj}, \citenamefont {H\"ofling}, \citenamefont {Huber}, \citenamefont {Castaneda}, \citenamefont {Carvacho}, \citenamefont {Spagnolo}, \citenamefont {Rota}, \citenamefont {Basset}, \citenamefont {Rastelli}, \citenamefont {Sciarrino}, \citenamefont {J\"ons},\ and\ \citenamefont {Trotta}}]{laneve2025}%
  \BibitemOpen
  \bibfield  {author} {\bibinfo {author} {\bibfnamefont {A.}~\bibnamefont {Laneve}}, \bibinfo {author} {\bibfnamefont {G.}~\bibnamefont {Ronco}}, \bibinfo {author} {\bibfnamefont {M.}~\bibnamefont {Beccaceci}}, \bibinfo {author} {\bibfnamefont {P.}~\bibnamefont {Barigelli}}, \bibinfo {author} {\bibfnamefont {F.}~\bibnamefont {Salusti}}, \bibinfo {author} {\bibfnamefont {N.}~\bibnamefont {Claro-Rodríguez}}, \bibinfo {author} {\bibfnamefont {G.~D.}\ \bibnamefont {Pascalis}}, \bibinfo {author} {\bibfnamefont {A.}~\bibnamefont {Suprano}}, \bibinfo {author} {\bibfnamefont {L.}~\bibnamefont {Chiaudano}}, \bibinfo {author} {\bibfnamefont {E.}~\bibnamefont {Sch\"oll}}, \bibinfo {author} {\bibfnamefont {L.}~\bibnamefont {Hanschke}}, \bibinfo {author} {\bibfnamefont {T.~M.}\ \bibnamefont {Krieger}}, \bibinfo {author} {\bibfnamefont {Q.}~\bibnamefont {Buchinger}}, \bibinfo {author} {\bibfnamefont {S.~F. C.~D.}\ \bibnamefont {Silva}}, \bibinfo {author} {\bibfnamefont {J.}~\bibnamefont {Neuwirth}}, \bibinfo {author}
  {\bibfnamefont {S.}~\bibnamefont {Stroj}}, \bibinfo {author} {\bibfnamefont {S.}~\bibnamefont {H\"ofling}}, \bibinfo {author} {\bibfnamefont {T.}~\bibnamefont {Huber}}, \bibinfo {author} {\bibfnamefont {M.~A.~U.}\ \bibnamefont {Castaneda}}, \bibinfo {author} {\bibfnamefont {G.}~\bibnamefont {Carvacho}}, \bibinfo {author} {\bibfnamefont {N.}~\bibnamefont {Spagnolo}}, \bibinfo {author} {\bibfnamefont {M.~B.}\ \bibnamefont {Rota}}, \bibinfo {author} {\bibfnamefont {F.~B.}\ \bibnamefont {Basset}}, \bibinfo {author} {\bibfnamefont {A.}~\bibnamefont {Rastelli}}, \bibinfo {author} {\bibfnamefont {F.}~\bibnamefont {Sciarrino}}, \bibinfo {author} {\bibfnamefont {K.~D.}\ \bibnamefont {J\"ons}},\ and\ \bibinfo {author} {\bibfnamefont {R.}~\bibnamefont {Trotta}},\ }\bibfield  {title} {\bibinfo {title} {Quantum teleportation with dissimilar quantum dots over a hybrid quantum network},\ }\href {https://doi.org/10.1038/s41467-025-65911-9} {\bibfield  {journal} {\bibinfo  {journal} {Nature Communications}\ }\textbf
  {\bibinfo {volume} {16}},\ \bibinfo {pages} {10028} (\bibinfo {year} {2025})}\BibitemShut {NoStop}%
\bibitem [{\citenamefont {Hou}\ \emph {et~al.}(2025)\citenamefont {Hou}, \citenamefont {Yang}, \citenamefont {Qi}, \citenamefont {Li}, \citenamefont {Li}, \citenamefont {Lin}, \citenamefont {Zheng},\ and\ \citenamefont {Chen}}]{hou2025}%
  \BibitemOpen
  \bibfield  {author} {\bibinfo {author} {\bibfnamefont {Y.}~\bibnamefont {Hou}}, \bibinfo {author} {\bibfnamefont {Y.}~\bibnamefont {Yang}}, \bibinfo {author} {\bibfnamefont {Z.}~\bibnamefont {Qi}}, \bibinfo {author} {\bibfnamefont {H.}~\bibnamefont {Li}}, \bibinfo {author} {\bibfnamefont {Y.}~\bibnamefont {Li}}, \bibinfo {author} {\bibfnamefont {J.}~\bibnamefont {Lin}}, \bibinfo {author} {\bibfnamefont {Y.}~\bibnamefont {Zheng}},\ and\ \bibinfo {author} {\bibfnamefont {X.}~\bibnamefont {Chen}},\ }\bibfield  {title} {\bibinfo {title} {200-km multi-user fully connected quantum entanglement distribution network in noisy environments},\ }\href {https://doi.org/10.1364/OL.569922} {\bibfield  {journal} {\bibinfo  {journal} {Optics Letters}\ }\textbf {\bibinfo {volume} {50}},\ \bibinfo {pages} {6020} (\bibinfo {year} {2025})}\BibitemShut {NoStop}%
\bibitem [{\citenamefont {Xu}\ \emph {et~al.}(2025)\citenamefont {Xu}, \citenamefont {de~Vicente}, \citenamefont {Sun},\ and\ \citenamefont {Yu}}]{xu2025}%
  \BibitemOpen
  \bibfield  {author} {\bibinfo {author} {\bibfnamefont {Z.-P.}\ \bibnamefont {Xu}}, \bibinfo {author} {\bibfnamefont {J.~I.}\ \bibnamefont {de~Vicente}}, \bibinfo {author} {\bibfnamefont {L.-L.}\ \bibnamefont {Sun}},\ and\ \bibinfo {author} {\bibfnamefont {S.}~\bibnamefont {Yu}},\ }\bibfield  {title} {\bibinfo {title} {Quantum network-entanglement measures},\ }\href {https://doi.org/10.22331/q-2025-05-06-1736} {\bibfield  {journal} {\bibinfo  {journal} {Quantum}\ }\textbf {\bibinfo {volume} {9}},\ \bibinfo {pages} {1736} (\bibinfo {year} {2025})}\BibitemShut {NoStop}%
\bibitem [{\citenamefont {Brida}\ \emph {et~al.}(2010)\citenamefont {Brida}, \citenamefont {Genovese},\ and\ \citenamefont {Ruo~Berchera}}]{samantaray2023}%
  \BibitemOpen
  \bibfield  {author} {\bibinfo {author} {\bibfnamefont {G.}~\bibnamefont {Brida}}, \bibinfo {author} {\bibfnamefont {M.}~\bibnamefont {Genovese}},\ and\ \bibinfo {author} {\bibfnamefont {I.}~\bibnamefont {Ruo~Berchera}},\ }\bibfield  {title} {\bibinfo {title} {Experimental realization of sub-shot-noise quantum imaging},\ }\href {https://doi.org/10.1038/nphoton.2010.29} {\bibfield  {journal} {\bibinfo  {journal} {Nature Photonics}\ }\textbf {\bibinfo {volume} {4}},\ \bibinfo {pages} {227–230} (\bibinfo {year} {2010})}\BibitemShut {NoStop}%
\bibitem [{\citenamefont {Farhi}\ \emph {et~al.}(2001)\citenamefont {Farhi}, \citenamefont {Goldstone}, \citenamefont {Gutmann},\ and\ \citenamefont {Sipser}}]{farhi2001}%
  \BibitemOpen
  \bibfield  {author} {\bibinfo {author} {\bibfnamefont {E.}~\bibnamefont {Farhi}}, \bibinfo {author} {\bibfnamefont {J.}~\bibnamefont {Goldstone}}, \bibinfo {author} {\bibfnamefont {S.}~\bibnamefont {Gutmann}},\ and\ \bibinfo {author} {\bibfnamefont {M.}~\bibnamefont {Sipser}},\ }\bibfield  {title} {\bibinfo {title} {Quantum computation by adiabatic evolution},\ }\href@noop {} {\bibfield  {journal} {\bibinfo  {journal} {arXiv preprint quant-ph/0001106}\ } (\bibinfo {year} {2001})}\BibitemShut {NoStop}%
\bibitem [{\citenamefont {Raussendorf}\ and\ \citenamefont {Briegel}(2001)}]{raussendorf2001}%
  \BibitemOpen
  \bibfield  {author} {\bibinfo {author} {\bibfnamefont {R.}~\bibnamefont {Raussendorf}}\ and\ \bibinfo {author} {\bibfnamefont {H.~J.}\ \bibnamefont {Briegel}},\ }\bibfield  {title} {\bibinfo {title} {A one-way quantum computer},\ }\href {https://doi.org/10.1103/PhysRevLett.86.5188} {\bibfield  {journal} {\bibinfo  {journal} {Physical Review Letters}\ }\textbf {\bibinfo {volume} {86}},\ \bibinfo {pages} {5188} (\bibinfo {year} {2001})}\BibitemShut {NoStop}%
\bibitem [{\citenamefont {Kitaev}(2003)}]{kitaev2003}%
  \BibitemOpen
  \bibfield  {author} {\bibinfo {author} {\bibfnamefont {A.~Y.}\ \bibnamefont {Kitaev}},\ }\bibfield  {title} {\bibinfo {title} {Fault-tolerant quantum computation by anyons},\ }\href {https://doi.org/10.1016/S0003-4916(02)00018-0} {\bibfield  {journal} {\bibinfo  {journal} {Annals of Physics}\ }\textbf {\bibinfo {volume} {303}},\ \bibinfo {pages} {2} (\bibinfo {year} {2003})}\BibitemShut {NoStop}%
\bibitem [{\citenamefont {Georgescu}\ \emph {et~al.}(2014)\citenamefont {Georgescu}, \citenamefont {Ashhab},\ and\ \citenamefont {Nori}}]{georgescu2014}%
  \BibitemOpen
  \bibfield  {author} {\bibinfo {author} {\bibfnamefont {I.~M.}\ \bibnamefont {Georgescu}}, \bibinfo {author} {\bibfnamefont {S.}~\bibnamefont {Ashhab}},\ and\ \bibinfo {author} {\bibfnamefont {F.}~\bibnamefont {Nori}},\ }\bibfield  {title} {\bibinfo {title} {Quantum simulation},\ }\href {https://doi.org/10.1103/RevModPhys.86.153} {\bibfield  {journal} {\bibinfo  {journal} {Reviews of Modern Physics}\ }\textbf {\bibinfo {volume} {86}},\ \bibinfo {pages} {153} (\bibinfo {year} {2014})}\BibitemShut {NoStop}%
\bibitem [{\citenamefont {Kjaergaard}\ \emph {et~al.}(2020)\citenamefont {Kjaergaard} \emph {et~al.}}]{kjaergaard2020}%
  \BibitemOpen
  \bibfield  {author} {\bibinfo {author} {\bibfnamefont {M.}~\bibnamefont {Kjaergaard}} \emph {et~al.},\ }\bibfield  {title} {\bibinfo {title} {Superconducting qubits: Current state of play},\ }\href {https://doi.org/10.1146/annurev-conmatphys-031119-050605} {\bibfield  {journal} {\bibinfo  {journal} {Annual Review of Condensed Matter Physics}\ }\textbf {\bibinfo {volume} {11}},\ \bibinfo {pages} {369} (\bibinfo {year} {2020})}\BibitemShut {NoStop}%
\bibitem [{\citenamefont {Saffman}\ \emph {et~al.}(2010)\citenamefont {Saffman}, \citenamefont {Walker},\ and\ \citenamefont {M{\o}lmer}}]{saffman2010}%
  \BibitemOpen
  \bibfield  {author} {\bibinfo {author} {\bibfnamefont {M.}~\bibnamefont {Saffman}}, \bibinfo {author} {\bibfnamefont {T.~G.}\ \bibnamefont {Walker}},\ and\ \bibinfo {author} {\bibfnamefont {K.}~\bibnamefont {M{\o}lmer}},\ }\bibfield  {title} {\bibinfo {title} {Quantum information with rydberg atoms},\ }\href {https://doi.org/10.1103/RevModPhys.82.2313} {\bibfield  {journal} {\bibinfo  {journal} {Reviews of Modern Physics}\ }\textbf {\bibinfo {volume} {82}},\ \bibinfo {pages} {2313} (\bibinfo {year} {2010})}\BibitemShut {NoStop}%
\bibitem [{\citenamefont {Blatt}\ and\ \citenamefont {Wineland}(2008)}]{blatt2008}%
  \BibitemOpen
  \bibfield  {author} {\bibinfo {author} {\bibfnamefont {R.}~\bibnamefont {Blatt}}\ and\ \bibinfo {author} {\bibfnamefont {D.}~\bibnamefont {Wineland}},\ }\bibfield  {title} {\bibinfo {title} {Entangled states of trapped atomic ions},\ }\href {https://doi.org/10.1038/nature07125} {\bibfield  {journal} {\bibinfo  {journal} {Nature}\ }\textbf {\bibinfo {volume} {453}},\ \bibinfo {pages} {1008} (\bibinfo {year} {2008})}\BibitemShut {NoStop}%
\bibitem [{\citenamefont {Moody}\ \emph {et~al.}(2022)\citenamefont {Moody}, \citenamefont {Sorger}, \citenamefont {Blumenthal}, \citenamefont {Juodawlkis}, \citenamefont {Loh}, \citenamefont {Sorace-Agaskar}, \citenamefont {Jones}, \citenamefont {Balram}, \citenamefont {Matthews}, \citenamefont {Laing} \emph {et~al.}}]{moody20222022}%
  \BibitemOpen
  \bibfield  {author} {\bibinfo {author} {\bibfnamefont {G.}~\bibnamefont {Moody}}, \bibinfo {author} {\bibfnamefont {V.~J.}\ \bibnamefont {Sorger}}, \bibinfo {author} {\bibfnamefont {D.~J.}\ \bibnamefont {Blumenthal}}, \bibinfo {author} {\bibfnamefont {P.~W.}\ \bibnamefont {Juodawlkis}}, \bibinfo {author} {\bibfnamefont {W.}~\bibnamefont {Loh}}, \bibinfo {author} {\bibfnamefont {C.}~\bibnamefont {Sorace-Agaskar}}, \bibinfo {author} {\bibfnamefont {A.~E.}\ \bibnamefont {Jones}}, \bibinfo {author} {\bibfnamefont {K.~C.}\ \bibnamefont {Balram}}, \bibinfo {author} {\bibfnamefont {J.~C.}\ \bibnamefont {Matthews}}, \bibinfo {author} {\bibfnamefont {A.}~\bibnamefont {Laing}}, \emph {et~al.},\ }\bibfield  {title} {\bibinfo {title} {2022 roadmap on integrated quantum photonics},\ }\href@noop {} {\bibfield  {journal} {\bibinfo  {journal} {Journal of Physics: Photonics}\ }\textbf {\bibinfo {volume} {4}},\ \bibinfo {pages} {012501} (\bibinfo {year} {2022})}\BibitemShut {NoStop}%
\bibitem [{\citenamefont {Maring}\ \emph {et~al.}(2024)\citenamefont {Maring}, \citenamefont {Fyrillas}, \citenamefont {Pont}, \citenamefont {Ivanov}, \citenamefont {Stepanov}, \citenamefont {Margaria}, \citenamefont {Hease}, \citenamefont {Pishchagin}, \citenamefont {Lemaître}, \citenamefont {Sagnes}, \citenamefont {Au}, \citenamefont {Boissier}, \citenamefont {Bertasi}, \citenamefont {Baert}, \citenamefont {Valdivia}, \citenamefont {Billard}, \citenamefont {Acar}, \citenamefont {Brieussel}, \citenamefont {Mezher}, \citenamefont {Wein}, \citenamefont {Salavrakos}, \citenamefont {Sinnott}, \citenamefont {Fioretto}, \citenamefont {Emeriau}, \citenamefont {Belabas}, \citenamefont {Mansfield}, \citenamefont {Senellart}, \citenamefont {Senellart},\ and\ \citenamefont {Somaschi}}]{Maring2024}%
  \BibitemOpen
  \bibfield  {author} {\bibinfo {author} {\bibfnamefont {N.}~\bibnamefont {Maring}}, \bibinfo {author} {\bibfnamefont {A.}~\bibnamefont {Fyrillas}}, \bibinfo {author} {\bibfnamefont {M.}~\bibnamefont {Pont}}, \bibinfo {author} {\bibfnamefont {E.}~\bibnamefont {Ivanov}}, \bibinfo {author} {\bibfnamefont {P.}~\bibnamefont {Stepanov}}, \bibinfo {author} {\bibfnamefont {N.}~\bibnamefont {Margaria}}, \bibinfo {author} {\bibfnamefont {W.}~\bibnamefont {Hease}}, \bibinfo {author} {\bibfnamefont {A.}~\bibnamefont {Pishchagin}}, \bibinfo {author} {\bibfnamefont {A.}~\bibnamefont {Lemaître}}, \bibinfo {author} {\bibfnamefont {I.}~\bibnamefont {Sagnes}}, \bibinfo {author} {\bibfnamefont {T.~H.}\ \bibnamefont {Au}}, \bibinfo {author} {\bibfnamefont {S.}~\bibnamefont {Boissier}}, \bibinfo {author} {\bibfnamefont {E.}~\bibnamefont {Bertasi}}, \bibinfo {author} {\bibfnamefont {A.}~\bibnamefont {Baert}}, \bibinfo {author} {\bibfnamefont {M.}~\bibnamefont {Valdivia}}, \bibinfo {author} {\bibfnamefont {M.}~\bibnamefont
  {Billard}}, \bibinfo {author} {\bibfnamefont {O.}~\bibnamefont {Acar}}, \bibinfo {author} {\bibfnamefont {A.}~\bibnamefont {Brieussel}}, \bibinfo {author} {\bibfnamefont {R.}~\bibnamefont {Mezher}}, \bibinfo {author} {\bibfnamefont {S.~C.}\ \bibnamefont {Wein}}, \bibinfo {author} {\bibfnamefont {A.}~\bibnamefont {Salavrakos}}, \bibinfo {author} {\bibfnamefont {P.}~\bibnamefont {Sinnott}}, \bibinfo {author} {\bibfnamefont {D.~A.}\ \bibnamefont {Fioretto}}, \bibinfo {author} {\bibfnamefont {P.-E.}\ \bibnamefont {Emeriau}}, \bibinfo {author} {\bibfnamefont {N.}~\bibnamefont {Belabas}}, \bibinfo {author} {\bibfnamefont {S.}~\bibnamefont {Mansfield}}, \bibinfo {author} {\bibfnamefont {P.}~\bibnamefont {Senellart}}, \bibinfo {author} {\bibfnamefont {J.}~\bibnamefont {Senellart}},\ and\ \bibinfo {author} {\bibfnamefont {N.}~\bibnamefont {Somaschi}},\ }\bibfield  {title} {\bibinfo {title} {A versatile single-photon-based quantum computing platform},\ }\href {https://doi.org/10.1038/s41566-024-01403-4} {\bibfield
  {journal} {\bibinfo  {journal} {Nature Photonics}\ }\textbf {\bibinfo {volume} {18}},\ \bibinfo {pages} {603–609} (\bibinfo {year} {2024})}\BibitemShut {NoStop}%
\bibitem [{\citenamefont {Elshaari}\ \emph {et~al.}(2020)\citenamefont {Elshaari}, \citenamefont {Pernice}, \citenamefont {Srinivasan}, \citenamefont {Benson},\ and\ \citenamefont {Zwiller}}]{Elshaari2020}%
  \BibitemOpen
  \bibfield  {author} {\bibinfo {author} {\bibfnamefont {A.~W.}\ \bibnamefont {Elshaari}}, \bibinfo {author} {\bibfnamefont {W.}~\bibnamefont {Pernice}}, \bibinfo {author} {\bibfnamefont {K.}~\bibnamefont {Srinivasan}}, \bibinfo {author} {\bibfnamefont {O.}~\bibnamefont {Benson}},\ and\ \bibinfo {author} {\bibfnamefont {V.}~\bibnamefont {Zwiller}},\ }\bibfield  {title} {\bibinfo {title} {Hybrid integrated quantum photonic circuits},\ }\href {https://doi.org/10.1038/s41566-020-0609-x} {\bibfield  {journal} {\bibinfo  {journal} {Nature Photonics}\ }\textbf {\bibinfo {volume} {14}},\ \bibinfo {pages} {285} (\bibinfo {year} {2020})}\BibitemShut {NoStop}%
\bibitem [{\citenamefont {Aharonovich}\ \emph {et~al.}(2016{\natexlab{a}})\citenamefont {Aharonovich}, \citenamefont {Englund},\ and\ \citenamefont {Toth}}]{Aharonovich2016}%
  \BibitemOpen
  \bibfield  {author} {\bibinfo {author} {\bibfnamefont {I.}~\bibnamefont {Aharonovich}}, \bibinfo {author} {\bibfnamefont {D.}~\bibnamefont {Englund}},\ and\ \bibinfo {author} {\bibfnamefont {M.}~\bibnamefont {Toth}},\ }\bibfield  {title} {\bibinfo {title} {Solid-state single-photon emitters},\ }\href {https://doi.org/10.1038/nphoton.2016.186} {\bibfield  {journal} {\bibinfo  {journal} {Nature Photonics}\ }\textbf {\bibinfo {volume} {10}},\ \bibinfo {pages} {631–641} (\bibinfo {year} {2016}{\natexlab{a}})}\BibitemShut {NoStop}%
\bibitem [{\citenamefont {Kim}\ \emph {et~al.}(2020)\citenamefont {Kim}, \citenamefont {Aghaeimeibodi}, \citenamefont {Carolan}, \citenamefont {Englund},\ and\ \citenamefont {Waks}}]{Kim2020}%
  \BibitemOpen
  \bibfield  {author} {\bibinfo {author} {\bibfnamefont {J.-H.}\ \bibnamefont {Kim}}, \bibinfo {author} {\bibfnamefont {S.}~\bibnamefont {Aghaeimeibodi}}, \bibinfo {author} {\bibfnamefont {J.}~\bibnamefont {Carolan}}, \bibinfo {author} {\bibfnamefont {D.}~\bibnamefont {Englund}},\ and\ \bibinfo {author} {\bibfnamefont {E.}~\bibnamefont {Waks}},\ }\bibfield  {title} {\bibinfo {title} {Hybrid integration methods for on-chip quantum photonics},\ }\href {https://doi.org/10.1364/optica.384118} {\bibfield  {journal} {\bibinfo  {journal} {Optica}\ }\textbf {\bibinfo {volume} {7}},\ \bibinfo {pages} {291} (\bibinfo {year} {2020})}\BibitemShut {NoStop}%
\bibitem [{\citenamefont {Mandal}\ \emph {et~al.}(2024)\citenamefont {Mandal}, \citenamefont {Singh}, \citenamefont {Kumar}, \citenamefont {Shah}, \citenamefont {Vij}, \citenamefont {Majumder}, \citenamefont {Khunte}, \citenamefont {Achanta},\ and\ \citenamefont {Kumar}}]{Mandal2024}%
  \BibitemOpen
  \bibfield  {author} {\bibinfo {author} {\bibfnamefont {K.~K.}\ \bibnamefont {Mandal}}, \bibinfo {author} {\bibfnamefont {A.~K.}\ \bibnamefont {Singh}}, \bibinfo {author} {\bibfnamefont {B.}~\bibnamefont {Kumar}}, \bibinfo {author} {\bibfnamefont {A.~P.}\ \bibnamefont {Shah}}, \bibinfo {author} {\bibfnamefont {R.}~\bibnamefont {Vij}}, \bibinfo {author} {\bibfnamefont {A.}~\bibnamefont {Majumder}}, \bibinfo {author} {\bibfnamefont {J.~J.}\ \bibnamefont {Khunte}}, \bibinfo {author} {\bibfnamefont {V.~G.}\ \bibnamefont {Achanta}},\ and\ \bibinfo {author} {\bibfnamefont {A.}~\bibnamefont {Kumar}},\ }\bibfield  {title} {\bibinfo {title} {Emission engineering in monolithically integrated silicon nitride microring resonators},\ }\href {https://doi.org/10.1021/acsmaterialslett.4c00105} {\bibfield  {journal} {\bibinfo  {journal} {ACS Materials Letters}\ }\textbf {\bibinfo {volume} {6}},\ \bibinfo {pages} {1831–1840} (\bibinfo {year} {2024})}\BibitemShut {NoStop}%
\bibitem [{\citenamefont {Zelaya}\ \emph {et~al.}(2025)\citenamefont {Zelaya}, \citenamefont {Honari-Latifpour}, \citenamefont {Mandal}, \citenamefont {Friedman}, \citenamefont {Madamopoulos},\ and\ \citenamefont {Miri}}]{zelaya2025chip}%
  \BibitemOpen
  \bibfield  {author} {\bibinfo {author} {\bibfnamefont {K.}~\bibnamefont {Zelaya}}, \bibinfo {author} {\bibfnamefont {M.}~\bibnamefont {Honari-Latifpour}}, \bibinfo {author} {\bibfnamefont {K.~K.}\ \bibnamefont {Mandal}}, \bibinfo {author} {\bibfnamefont {J.}~\bibnamefont {Friedman}}, \bibinfo {author} {\bibfnamefont {N.}~\bibnamefont {Madamopoulos}},\ and\ \bibinfo {author} {\bibfnamefont {M.-A.}\ \bibnamefont {Miri}},\ }\bibfield  {title} {\bibinfo {title} {On-chip unitary generation of arbitrary complex spatial photonic states},\ }\href@noop {} {\bibfield  {journal} {\bibinfo  {journal} {Optica}\ }\textbf {\bibinfo {volume} {12}},\ \bibinfo {pages} {1492} (\bibinfo {year} {2025})}\BibitemShut {NoStop}%
\bibitem [{\citenamefont {Kumar~Mandal}\ \emph {et~al.}(2023)\citenamefont {Kumar~Mandal}, \citenamefont {Gupta}, \citenamefont {Kumar}, \citenamefont {Sohoni}, \citenamefont {Venu~Gopal},\ and\ \citenamefont {Kumar}}]{kumar2023photonic}%
  \BibitemOpen
  \bibfield  {author} {\bibinfo {author} {\bibfnamefont {K.}~\bibnamefont {Kumar~Mandal}}, \bibinfo {author} {\bibfnamefont {Y.}~\bibnamefont {Gupta}}, \bibinfo {author} {\bibfnamefont {B.}~\bibnamefont {Kumar}}, \bibinfo {author} {\bibfnamefont {M.}~\bibnamefont {Sohoni}}, \bibinfo {author} {\bibfnamefont {A.}~\bibnamefont {Venu~Gopal}},\ and\ \bibinfo {author} {\bibfnamefont {A.}~\bibnamefont {Kumar}},\ }\bibfield  {title} {\bibinfo {title} {A photonic integrated chip platform for interlayer exciton valley routing},\ }\href@noop {} {\bibfield  {journal} {\bibinfo  {journal} {Journal of applied physics}\ }\textbf {\bibinfo {volume} {133}} (\bibinfo {year} {2023})}\BibitemShut {NoStop}%
\bibitem [{\citenamefont {Singh}\ \emph {et~al.}(2023)\citenamefont {Singh}, \citenamefont {Mandal}, \citenamefont {Gupta}, \citenamefont {Anand~VS}, \citenamefont {Eswaramoorthy}, \citenamefont {Kumar}, \citenamefont {Kala}, \citenamefont {Dixit}, \citenamefont {Achanta},\ and\ \citenamefont {Kumar}}]{singh2023low}%
  \BibitemOpen
  \bibfield  {author} {\bibinfo {author} {\bibfnamefont {A.~K.}\ \bibnamefont {Singh}}, \bibinfo {author} {\bibfnamefont {K.~K.}\ \bibnamefont {Mandal}}, \bibinfo {author} {\bibfnamefont {Y.}~\bibnamefont {Gupta}}, \bibinfo {author} {\bibfnamefont {A.}~\bibnamefont {Anand~VS}}, \bibinfo {author} {\bibfnamefont {L.}~\bibnamefont {Eswaramoorthy}}, \bibinfo {author} {\bibfnamefont {B.}~\bibnamefont {Kumar}}, \bibinfo {author} {\bibfnamefont {A.}~\bibnamefont {Kala}}, \bibinfo {author} {\bibfnamefont {S.}~\bibnamefont {Dixit}}, \bibinfo {author} {\bibfnamefont {V.~G.}\ \bibnamefont {Achanta}},\ and\ \bibinfo {author} {\bibfnamefont {A.}~\bibnamefont {Kumar}},\ }\bibfield  {title} {\bibinfo {title} {Low-cost plasmonic platform for photon-emission engineering of two-dimensional semiconductors},\ }\href@noop {} {\bibfield  {journal} {\bibinfo  {journal} {Physical Review Applied}\ }\textbf {\bibinfo {volume} {19}},\ \bibinfo {pages} {044012} (\bibinfo {year} {2023})}\BibitemShut {NoStop}%
\bibitem [{\citenamefont {Kumar}\ \emph {et~al.}(2023)\citenamefont {Kumar}, \citenamefont {Singh}, \citenamefont {Mandal}, \citenamefont {Sharma}, \citenamefont {Sahoo},\ and\ \citenamefont {Kumar}}]{kumar2023universal}%
  \BibitemOpen
  \bibfield  {author} {\bibinfo {author} {\bibfnamefont {B.}~\bibnamefont {Kumar}}, \bibinfo {author} {\bibfnamefont {A.~K.}\ \bibnamefont {Singh}}, \bibinfo {author} {\bibfnamefont {K.~K.}\ \bibnamefont {Mandal}}, \bibinfo {author} {\bibfnamefont {P.}~\bibnamefont {Sharma}}, \bibinfo {author} {\bibfnamefont {N.~R.}\ \bibnamefont {Sahoo}},\ and\ \bibinfo {author} {\bibfnamefont {A.}~\bibnamefont {Kumar}},\ }\bibfield  {title} {\bibinfo {title} {A universal and stable metasurface for photonic quasi bound state in continuum coupled with two dimensional semiconductors},\ }\href@noop {} {\bibfield  {journal} {\bibinfo  {journal} {Journal of Physics D: Applied Physics}\ }\textbf {\bibinfo {volume} {56}},\ \bibinfo {pages} {425105} (\bibinfo {year} {2023})}\BibitemShut {NoStop}%
\bibitem [{\citenamefont {Singh}\ \emph {et~al.}(2025)\citenamefont {Singh}, \citenamefont {Utkarsh}, \citenamefont {Tieben}, \citenamefont {Mandal}, \citenamefont {Kumar}, \citenamefont {Vij}, \citenamefont {Majumder}, \citenamefont {Shyam}, \citenamefont {Kumar}, \citenamefont {Watanabe} \emph {et~al.}}]{singh2025plasmonic}%
  \BibitemOpen
  \bibfield  {author} {\bibinfo {author} {\bibfnamefont {A.~K.}\ \bibnamefont {Singh}}, \bibinfo {author} {\bibnamefont {Utkarsh}}, \bibinfo {author} {\bibfnamefont {P.}~\bibnamefont {Tieben}}, \bibinfo {author} {\bibfnamefont {K.~K.}\ \bibnamefont {Mandal}}, \bibinfo {author} {\bibfnamefont {B.}~\bibnamefont {Kumar}}, \bibinfo {author} {\bibfnamefont {R.}~\bibnamefont {Vij}}, \bibinfo {author} {\bibfnamefont {A.}~\bibnamefont {Majumder}}, \bibinfo {author} {\bibfnamefont {I.}~\bibnamefont {Shyam}}, \bibinfo {author} {\bibfnamefont {S.}~\bibnamefont {Kumar}}, \bibinfo {author} {\bibfnamefont {K.}~\bibnamefont {Watanabe}}, \emph {et~al.},\ }\bibfield  {title} {\bibinfo {title} {Plasmonic-strain engineering of quantum emitters in hexagonal boron nitride},\ }\href@noop {} {\bibfield  {journal} {\bibinfo  {journal} {Advanced Materials Interfaces}\ ,\ \bibinfo {pages} {2500071}} (\bibinfo {year} {2025})}\BibitemShut {NoStop}%
\bibitem [{\citenamefont {Somaschi}\ \emph {et~al.}(2016{\natexlab{a}})\citenamefont {Somaschi}, \citenamefont {Giesz}, \citenamefont {De~Santis}, \citenamefont {Loredo}, \citenamefont {Almeida}, \citenamefont {Hornecker}, \citenamefont {Portalupi}, \citenamefont {Grange}, \citenamefont {Antón}, \citenamefont {Demory}, \citenamefont {Gómez}, \citenamefont {Sagnes}, \citenamefont {Lanzillotti-Kimura}, \citenamefont {Lemaítre}, \citenamefont {Auffeves}, \citenamefont {White}, \citenamefont {Lanco},\ and\ \citenamefont {Senellart}}]{somaschi2016}%
  \BibitemOpen
  \bibfield  {author} {\bibinfo {author} {\bibfnamefont {N.}~\bibnamefont {Somaschi}}, \bibinfo {author} {\bibfnamefont {V.}~\bibnamefont {Giesz}}, \bibinfo {author} {\bibfnamefont {L.}~\bibnamefont {De~Santis}}, \bibinfo {author} {\bibfnamefont {J.~C.}\ \bibnamefont {Loredo}}, \bibinfo {author} {\bibfnamefont {M.~P.}\ \bibnamefont {Almeida}}, \bibinfo {author} {\bibfnamefont {G.}~\bibnamefont {Hornecker}}, \bibinfo {author} {\bibfnamefont {S.~L.}\ \bibnamefont {Portalupi}}, \bibinfo {author} {\bibfnamefont {T.}~\bibnamefont {Grange}}, \bibinfo {author} {\bibfnamefont {C.}~\bibnamefont {Antón}}, \bibinfo {author} {\bibfnamefont {J.}~\bibnamefont {Demory}}, \bibinfo {author} {\bibfnamefont {C.}~\bibnamefont {Gómez}}, \bibinfo {author} {\bibfnamefont {I.}~\bibnamefont {Sagnes}}, \bibinfo {author} {\bibfnamefont {N.~D.}\ \bibnamefont {Lanzillotti-Kimura}}, \bibinfo {author} {\bibfnamefont {A.}~\bibnamefont {Lemaítre}}, \bibinfo {author} {\bibfnamefont {A.}~\bibnamefont {Auffeves}}, \bibinfo {author}
  {\bibfnamefont {A.~G.}\ \bibnamefont {White}}, \bibinfo {author} {\bibfnamefont {L.}~\bibnamefont {Lanco}},\ and\ \bibinfo {author} {\bibfnamefont {P.}~\bibnamefont {Senellart}},\ }\bibfield  {title} {\bibinfo {title} {Near-optimal single-photon sources in the solid state},\ }\href {https://doi.org/10.1038/nphoton.2016.23} {\bibfield  {journal} {\bibinfo  {journal} {Nature Photonics}\ }\textbf {\bibinfo {volume} {10}},\ \bibinfo {pages} {340–345} (\bibinfo {year} {2016}{\natexlab{a}})}\BibitemShut {NoStop}%
\bibitem [{\citenamefont {He}\ \emph {et~al.}(2013)\citenamefont {He}, \citenamefont {He}, \citenamefont {Wei}, \citenamefont {Wu}, \citenamefont {Atat\"{u}re}, \citenamefont {Schneider}, \citenamefont {H\"{o}fling}, \citenamefont {Kamp}, \citenamefont {Lu},\ and\ \citenamefont {Pan}}]{he2013}%
  \BibitemOpen
  \bibfield  {author} {\bibinfo {author} {\bibfnamefont {Y.-M.}\ \bibnamefont {He}}, \bibinfo {author} {\bibfnamefont {Y.}~\bibnamefont {He}}, \bibinfo {author} {\bibfnamefont {Y.-J.}\ \bibnamefont {Wei}}, \bibinfo {author} {\bibfnamefont {D.}~\bibnamefont {Wu}}, \bibinfo {author} {\bibfnamefont {M.}~\bibnamefont {Atat\"{u}re}}, \bibinfo {author} {\bibfnamefont {C.}~\bibnamefont {Schneider}}, \bibinfo {author} {\bibfnamefont {S.}~\bibnamefont {H\"{o}fling}}, \bibinfo {author} {\bibfnamefont {M.}~\bibnamefont {Kamp}}, \bibinfo {author} {\bibfnamefont {C.-Y.}\ \bibnamefont {Lu}},\ and\ \bibinfo {author} {\bibfnamefont {J.-W.}\ \bibnamefont {Pan}},\ }\bibfield  {title} {\bibinfo {title} {On-demand semiconductor single-photon source with near-unity indistinguishability},\ }\href {https://doi.org/10.1038/nnano.2012.262} {\bibfield  {journal} {\bibinfo  {journal} {Nature Nanotechnology}\ }\textbf {\bibinfo {volume} {8}},\ \bibinfo {pages} {213–217} (\bibinfo {year} {2013})}\BibitemShut {NoStop}%
\bibitem [{\citenamefont {Santori}\ \emph {et~al.}(2002)\citenamefont {Santori}, \citenamefont {Fattal}, \citenamefont {Vučković}, \citenamefont {Solomon},\ and\ \citenamefont {Yamamoto}}]{Santori2002}%
  \BibitemOpen
  \bibfield  {author} {\bibinfo {author} {\bibfnamefont {C.}~\bibnamefont {Santori}}, \bibinfo {author} {\bibfnamefont {D.}~\bibnamefont {Fattal}}, \bibinfo {author} {\bibfnamefont {J.}~\bibnamefont {Vučković}}, \bibinfo {author} {\bibfnamefont {G.~S.}\ \bibnamefont {Solomon}},\ and\ \bibinfo {author} {\bibfnamefont {Y.}~\bibnamefont {Yamamoto}},\ }\bibfield  {title} {\bibinfo {title} {Indistinguishable photons from a single-photon device},\ }\href {https://doi.org/10.1038/nature01086} {\bibfield  {journal} {\bibinfo  {journal} {Nature}\ }\textbf {\bibinfo {volume} {419}},\ \bibinfo {pages} {594–597} (\bibinfo {year} {2002})}\BibitemShut {NoStop}%
\bibitem [{\citenamefont {Doherty}\ \emph {et~al.}(2013)\citenamefont {Doherty}, \citenamefont {Manson}, \citenamefont {Delaney}, \citenamefont {Jelezko}, \citenamefont {Wrachtrup},\ and\ \citenamefont {Hollenberg}}]{doherty2013nitrogen}%
  \BibitemOpen
  \bibfield  {author} {\bibinfo {author} {\bibfnamefont {M.~W.}\ \bibnamefont {Doherty}}, \bibinfo {author} {\bibfnamefont {N.~B.}\ \bibnamefont {Manson}}, \bibinfo {author} {\bibfnamefont {P.}~\bibnamefont {Delaney}}, \bibinfo {author} {\bibfnamefont {F.}~\bibnamefont {Jelezko}}, \bibinfo {author} {\bibfnamefont {J.}~\bibnamefont {Wrachtrup}},\ and\ \bibinfo {author} {\bibfnamefont {L.~C.}\ \bibnamefont {Hollenberg}},\ }\bibfield  {title} {\bibinfo {title} {The nitrogen-vacancy colour centre in diamond},\ }\href {https://doi.org/10.1016/j.physrep.2013.02.001} {\bibfield  {journal} {\bibinfo  {journal} {Physics Reports}\ }\textbf {\bibinfo {volume} {528}},\ \bibinfo {pages} {1–45} (\bibinfo {year} {2013})}\BibitemShut {NoStop}%
\bibitem [{\citenamefont {Rogers}\ \emph {et~al.}(2014)\citenamefont {Rogers}, \citenamefont {Jahnke}, \citenamefont {Teraji}, \citenamefont {Marseglia}, \citenamefont {Muller}, \citenamefont {Naydenov}, \citenamefont {Schauffert}, \citenamefont {Kranz}, \citenamefont {Isoya},\ and\ \citenamefont {Jelezko}}]{rogers2014multiple}%
  \BibitemOpen
  \bibfield  {author} {\bibinfo {author} {\bibfnamefont {L.~J.}\ \bibnamefont {Rogers}}, \bibinfo {author} {\bibfnamefont {K.~D.}\ \bibnamefont {Jahnke}}, \bibinfo {author} {\bibfnamefont {T.}~\bibnamefont {Teraji}}, \bibinfo {author} {\bibfnamefont {L.}~\bibnamefont {Marseglia}}, \bibinfo {author} {\bibfnamefont {C.}~\bibnamefont {Muller}}, \bibinfo {author} {\bibfnamefont {B.}~\bibnamefont {Naydenov}}, \bibinfo {author} {\bibfnamefont {H.}~\bibnamefont {Schauffert}}, \bibinfo {author} {\bibfnamefont {C.}~\bibnamefont {Kranz}}, \bibinfo {author} {\bibfnamefont {J.}~\bibnamefont {Isoya}},\ and\ \bibinfo {author} {\bibfnamefont {F.}~\bibnamefont {Jelezko}},\ }\bibfield  {title} {\bibinfo {title} {Multiple intrinsically identical single-photon emitters in the solid state},\ }\href@noop {} {\bibfield  {journal} {\bibinfo  {journal} {Nature communications}\ }\textbf {\bibinfo {volume} {5}},\ \bibinfo {pages} {1} (\bibinfo {year} {2014})}\BibitemShut {NoStop}%
\bibitem [{\citenamefont {Chakraborty}\ \emph {et~al.}(2015)\citenamefont {Chakraborty}, \citenamefont {Kinnischtzke}, \citenamefont {Goodfellow}, \citenamefont {Beams},\ and\ \citenamefont {Vamivakas}}]{chakraborty2015voltage}%
  \BibitemOpen
  \bibfield  {author} {\bibinfo {author} {\bibfnamefont {C.}~\bibnamefont {Chakraborty}}, \bibinfo {author} {\bibfnamefont {L.}~\bibnamefont {Kinnischtzke}}, \bibinfo {author} {\bibfnamefont {K.~M.}\ \bibnamefont {Goodfellow}}, \bibinfo {author} {\bibfnamefont {R.}~\bibnamefont {Beams}},\ and\ \bibinfo {author} {\bibfnamefont {A.~N.}\ \bibnamefont {Vamivakas}},\ }\bibfield  {title} {\bibinfo {title} {Voltage-controlled quantum light from an atomically thin semiconductor},\ }\href@noop {} {\bibfield  {journal} {\bibinfo  {journal} {Nature Nanotechnology}\ }\textbf {\bibinfo {volume} {10}},\ \bibinfo {pages} {507} (\bibinfo {year} {2015})}\BibitemShut {NoStop}%
\bibitem [{\citenamefont {Tran}\ \emph {et~al.}(2016)\citenamefont {Tran}, \citenamefont {Bray}, \citenamefont {Ford}, \citenamefont {Toth},\ and\ \citenamefont {Aharonovich}}]{tran2016quantum}%
  \BibitemOpen
  \bibfield  {author} {\bibinfo {author} {\bibfnamefont {T.~T.}\ \bibnamefont {Tran}}, \bibinfo {author} {\bibfnamefont {K.}~\bibnamefont {Bray}}, \bibinfo {author} {\bibfnamefont {M.~J.}\ \bibnamefont {Ford}}, \bibinfo {author} {\bibfnamefont {M.}~\bibnamefont {Toth}},\ and\ \bibinfo {author} {\bibfnamefont {I.}~\bibnamefont {Aharonovich}},\ }\bibfield  {title} {\bibinfo {title} {Quantum emission from hexagonal boron nitride monolayers},\ }\href@noop {} {\bibfield  {journal} {\bibinfo  {journal} {Nature Nanotechnology}\ }\textbf {\bibinfo {volume} {11}},\ \bibinfo {pages} {37} (\bibinfo {year} {2016})}\BibitemShut {NoStop}%
\bibitem [{\citenamefont {Jiang}\ \emph {et~al.}(2024{\natexlab{a}})\citenamefont {Jiang}, \citenamefont {Hong}, \citenamefont {Hu}, \citenamefont {Chen}, \citenamefont {Yang}, \citenamefont {Hu}, \citenamefont {Yang}, \citenamefont {Shu}, \citenamefont {Zhao}, \citenamefont {Peng},\ and\ \citenamefont {Du}}]{Jiang2024}%
  \BibitemOpen
  \bibfield  {author} {\bibinfo {author} {\bibfnamefont {M.}~\bibnamefont {Jiang}}, \bibinfo {author} {\bibfnamefont {T.}~\bibnamefont {Hong}}, \bibinfo {author} {\bibfnamefont {D.}~\bibnamefont {Hu}}, \bibinfo {author} {\bibfnamefont {Y.}~\bibnamefont {Chen}}, \bibinfo {author} {\bibfnamefont {F.}~\bibnamefont {Yang}}, \bibinfo {author} {\bibfnamefont {T.}~\bibnamefont {Hu}}, \bibinfo {author} {\bibfnamefont {X.}~\bibnamefont {Yang}}, \bibinfo {author} {\bibfnamefont {J.}~\bibnamefont {Shu}}, \bibinfo {author} {\bibfnamefont {Y.}~\bibnamefont {Zhao}}, \bibinfo {author} {\bibfnamefont {X.}~\bibnamefont {Peng}},\ and\ \bibinfo {author} {\bibfnamefont {J.}~\bibnamefont {Du}},\ }\bibfield  {title} {\bibinfo {title} {Long-baseline quantum sensor network as dark matter haloscope},\ }\href {https://doi.org/10.1038/s41467-024-47566-0} {\bibfield  {journal} {\bibinfo  {journal} {Nature Communications}\ }\textbf {\bibinfo {volume} {15}},\ \bibinfo {pages} {3331} (\bibinfo {year} {2024}{\natexlab{a}})}\BibitemShut
  {NoStop}%
\bibitem [{\citenamefont {Gao}\ \emph {et~al.}(2023{\natexlab{a}})\citenamefont {Gao}, \citenamefont {von Helversen}, \citenamefont {Ant{\'o}n-Solanas}, \citenamefont {Schneider},\ and\ \citenamefont {Heindel}}]{Gao2023}%
  \BibitemOpen
  \bibfield  {author} {\bibinfo {author} {\bibfnamefont {T.}~\bibnamefont {Gao}}, \bibinfo {author} {\bibfnamefont {M.}~\bibnamefont {von Helversen}}, \bibinfo {author} {\bibfnamefont {C.}~\bibnamefont {Ant{\'o}n-Solanas}}, \bibinfo {author} {\bibfnamefont {C.}~\bibnamefont {Schneider}},\ and\ \bibinfo {author} {\bibfnamefont {T.}~\bibnamefont {Heindel}},\ }\bibfield  {title} {\bibinfo {title} {Atomically-thin single-photon sources for quantum communication},\ }\href {https://doi.org/10.1038/s41699-023-00366-4} {\bibfield  {journal} {\bibinfo  {journal} {npj 2D Materials and Applications}\ }\textbf {\bibinfo {volume} {7}},\ \bibinfo {pages} {4} (\bibinfo {year} {2023}{\natexlab{a}})}\BibitemShut {NoStop}%
\bibitem [{\citenamefont {Sakurai}\ \emph {et~al.}(2025{\natexlab{a}})\citenamefont {Sakurai}, \citenamefont {Hayashi}, \citenamefont {Munro},\ and\ \citenamefont {Nemoto}}]{Sakurai:25}%
  \BibitemOpen
  \bibfield  {author} {\bibinfo {author} {\bibfnamefont {A.}~\bibnamefont {Sakurai}}, \bibinfo {author} {\bibfnamefont {A.}~\bibnamefont {Hayashi}}, \bibinfo {author} {\bibfnamefont {W.~J.}\ \bibnamefont {Munro}},\ and\ \bibinfo {author} {\bibfnamefont {K.}~\bibnamefont {Nemoto}},\ }\bibfield  {title} {\bibinfo {title} {Quantum optical reservoir computing powered by boson sampling},\ }\href {https://doi.org/10.1364/OPTICAQ.541432} {\bibfield  {journal} {\bibinfo  {journal} {Optica Quantum}\ }\textbf {\bibinfo {volume} {3}},\ \bibinfo {pages} {238} (\bibinfo {year} {2025}{\natexlab{a}})}\BibitemShut {NoStop}%
\bibitem [{\citenamefont {Chanana}\ \emph {et~al.}(2022{\natexlab{a}})\citenamefont {Chanana}, \citenamefont {Larocque}, \citenamefont {Moreira}, \citenamefont {Carolan}, \citenamefont {Guha}, \citenamefont {Melo}, \citenamefont {Anant}, \citenamefont {Song}, \citenamefont {Englund}, \citenamefont {Blumenthal}, \citenamefont {Srinivasan},\ and\ \citenamefont {Davanco}}]{Chanana2022ULLW}%
  \BibitemOpen
  \bibfield  {author} {\bibinfo {author} {\bibfnamefont {A.}~\bibnamefont {Chanana}}, \bibinfo {author} {\bibfnamefont {H.}~\bibnamefont {Larocque}}, \bibinfo {author} {\bibfnamefont {R.}~\bibnamefont {Moreira}}, \bibinfo {author} {\bibfnamefont {J.}~\bibnamefont {Carolan}}, \bibinfo {author} {\bibfnamefont {B.}~\bibnamefont {Guha}}, \bibinfo {author} {\bibfnamefont {E.~G.}\ \bibnamefont {Melo}}, \bibinfo {author} {\bibfnamefont {V.}~\bibnamefont {Anant}}, \bibinfo {author} {\bibfnamefont {J.~D.}\ \bibnamefont {Song}}, \bibinfo {author} {\bibfnamefont {D.}~\bibnamefont {Englund}}, \bibinfo {author} {\bibfnamefont {D.~J.}\ \bibnamefont {Blumenthal}}, \bibinfo {author} {\bibfnamefont {K.}~\bibnamefont {Srinivasan}},\ and\ \bibinfo {author} {\bibfnamefont {M.}~\bibnamefont {Davanco}},\ }\bibfield  {title} {\bibinfo {title} {Ultra-low loss quantum photonic circuits integrated with single quantum emitters},\ }\href {https://doi.org/10.1038/s41467-022-35332-z} {\bibfield  {journal} {\bibinfo  {journal} {Nature
  Communications}\ }\textbf {\bibinfo {volume} {13}},\ \bibinfo {pages} {7693} (\bibinfo {year} {2022}{\natexlab{a}})}\BibitemShut {NoStop}%
\bibitem [{\citenamefont {Pelucchi}\ \emph {et~al.}(2022)\citenamefont {Pelucchi}, \citenamefont {Fagas}, \citenamefont {Aharonovich}, \citenamefont {Englund}, \citenamefont {Figueroa}, \citenamefont {Gong}, \citenamefont {Hannes}, \citenamefont {Liu}, \citenamefont {Lu}, \citenamefont {Matsuda}, \citenamefont {Pan}, \citenamefont {Schreck}, \citenamefont {Sciarrino}, \citenamefont {Silberhorn}, \citenamefont {Wang},\ and\ \citenamefont {J{\"o}ns}}]{Pelucchi2022PotentialIntegrated}%
  \BibitemOpen
  \bibfield  {author} {\bibinfo {author} {\bibfnamefont {E.}~\bibnamefont {Pelucchi}}, \bibinfo {author} {\bibfnamefont {G.}~\bibnamefont {Fagas}}, \bibinfo {author} {\bibfnamefont {I.}~\bibnamefont {Aharonovich}}, \bibinfo {author} {\bibfnamefont {D.}~\bibnamefont {Englund}}, \bibinfo {author} {\bibfnamefont {E.}~\bibnamefont {Figueroa}}, \bibinfo {author} {\bibfnamefont {Q.}~\bibnamefont {Gong}}, \bibinfo {author} {\bibfnamefont {H.}~\bibnamefont {Hannes}}, \bibinfo {author} {\bibfnamefont {J.}~\bibnamefont {Liu}}, \bibinfo {author} {\bibfnamefont {C.-Y.}\ \bibnamefont {Lu}}, \bibinfo {author} {\bibfnamefont {N.}~\bibnamefont {Matsuda}}, \bibinfo {author} {\bibfnamefont {J.-W.}\ \bibnamefont {Pan}}, \bibinfo {author} {\bibfnamefont {F.}~\bibnamefont {Schreck}}, \bibinfo {author} {\bibfnamefont {F.}~\bibnamefont {Sciarrino}}, \bibinfo {author} {\bibfnamefont {C.}~\bibnamefont {Silberhorn}}, \bibinfo {author} {\bibfnamefont {J.}~\bibnamefont {Wang}},\ and\ \bibinfo {author} {\bibfnamefont {K.~D.}\
  \bibnamefont {J{\"o}ns}},\ }\bibfield  {title} {\bibinfo {title} {The potential and global outlook of integrated photonics for quantum technologies},\ }\href {https://doi.org/10.1038/s42254-021-00398-z} {\bibfield  {journal} {\bibinfo  {journal} {Nature Reviews Physics}\ }\textbf {\bibinfo {volume} {4}},\ \bibinfo {pages} {194} (\bibinfo {year} {2022})}\BibitemShut {NoStop}%
\bibitem [{\citenamefont {Giordani}\ \emph {et~al.}(2023)\citenamefont {Giordani}, \citenamefont {Hoch}, \citenamefont {Carvacho}, \citenamefont {Spagnolo},\ and\ \citenamefont {Sciarrino}}]{Giordani2023IntegratedPhotonicQT}%
  \BibitemOpen
  \bibfield  {author} {\bibinfo {author} {\bibfnamefont {T.}~\bibnamefont {Giordani}}, \bibinfo {author} {\bibfnamefont {F.}~\bibnamefont {Hoch}}, \bibinfo {author} {\bibfnamefont {G.}~\bibnamefont {Carvacho}}, \bibinfo {author} {\bibfnamefont {N.}~\bibnamefont {Spagnolo}},\ and\ \bibinfo {author} {\bibfnamefont {F.}~\bibnamefont {Sciarrino}},\ }\bibfield  {title} {\bibinfo {title} {Integrated photonics in quantum technologies},\ }\href {https://doi.org/10.1007/s40766-023-00040-x} {\bibfield  {journal} {\bibinfo  {journal} {La Rivista del Nuovo Cimento}\ }\textbf {\bibinfo {volume} {46}},\ \bibinfo {pages} {71} (\bibinfo {year} {2023})}\BibitemShut {NoStop}%
\bibitem [{\citenamefont {Ramakrishnan}\ \emph {et~al.}(2023)\citenamefont {Ramakrishnan}, \citenamefont {Ravichandran}, \citenamefont {Mishra}, \citenamefont {Kaushalram}, \citenamefont {Hegde}, \citenamefont {Talabattula},\ and\ \citenamefont {Rohde}}]{Ramakrishnan2023IntegratedPlatformsReview}%
  \BibitemOpen
  \bibfield  {author} {\bibinfo {author} {\bibfnamefont {R.~K.}\ \bibnamefont {Ramakrishnan}}, \bibinfo {author} {\bibfnamefont {A.~B.}\ \bibnamefont {Ravichandran}}, \bibinfo {author} {\bibfnamefont {A.}~\bibnamefont {Mishra}}, \bibinfo {author} {\bibfnamefont {A.}~\bibnamefont {Kaushalram}}, \bibinfo {author} {\bibfnamefont {G.}~\bibnamefont {Hegde}}, \bibinfo {author} {\bibfnamefont {S.}~\bibnamefont {Talabattula}},\ and\ \bibinfo {author} {\bibfnamefont {P.~P.}\ \bibnamefont {Rohde}},\ }\bibfield  {title} {\bibinfo {title} {Integrated photonic platforms for quantum technology: A review},\ }\href {https://doi.org/10.1007/s41683-023-00115-1} {\bibfield  {journal} {\bibinfo  {journal} {ISSS Journal of Micro and Smart Systems}\ ,\ \bibinfo {pages} {1}} (\bibinfo {year} {2023})}\BibitemShut {NoStop}%
\bibitem [{\citenamefont {Degen}\ \emph {et~al.}(2017{\natexlab{b}})\citenamefont {Degen}, \citenamefont {Reinhard},\ and\ \citenamefont {Cappellaro}}]{Degen2017QSensing}%
  \BibitemOpen
  \bibfield  {author} {\bibinfo {author} {\bibfnamefont {C.~L.}\ \bibnamefont {Degen}}, \bibinfo {author} {\bibfnamefont {F.}~\bibnamefont {Reinhard}},\ and\ \bibinfo {author} {\bibfnamefont {P.}~\bibnamefont {Cappellaro}},\ }\bibfield  {title} {\bibinfo {title} {Quantum sensing},\ }\href {https://doi.org/10.1103/RevModPhys.89.035002} {\bibfield  {journal} {\bibinfo  {journal} {Reviews of Modern Physics}\ }\textbf {\bibinfo {volume} {89}},\ \bibinfo {pages} {035002} (\bibinfo {year} {2017}{\natexlab{b}})}\BibitemShut {NoStop}%
\bibitem [{\citenamefont {Kominis}\ \emph {et~al.}(2003)\citenamefont {Kominis}, \citenamefont {Kornack}, \citenamefont {Allred},\ and\ \citenamefont {Romalis}}]{Kominis2003SERF}%
  \BibitemOpen
  \bibfield  {author} {\bibinfo {author} {\bibfnamefont {I.~K.}\ \bibnamefont {Kominis}}, \bibinfo {author} {\bibfnamefont {T.~W.}\ \bibnamefont {Kornack}}, \bibinfo {author} {\bibfnamefont {J.~C.}\ \bibnamefont {Allred}},\ and\ \bibinfo {author} {\bibfnamefont {M.~V.}\ \bibnamefont {Romalis}},\ }\bibfield  {title} {\bibinfo {title} {A subfemtotesla multichannel atomic magnetometer},\ }\href {https://doi.org/10.1038/nature01484} {\bibfield  {journal} {\bibinfo  {journal} {Nature}\ }\textbf {\bibinfo {volume} {422}},\ \bibinfo {pages} {596} (\bibinfo {year} {2003})}\BibitemShut {NoStop}%
\bibitem [{\citenamefont {Sheng}\ \emph {et~al.}(2013)\citenamefont {Sheng}, \citenamefont {Li}, \citenamefont {Dural},\ and\ \citenamefont {Romalis}}]{Sheng2013Multipass}%
  \BibitemOpen
  \bibfield  {author} {\bibinfo {author} {\bibfnamefont {D.}~\bibnamefont {Sheng}}, \bibinfo {author} {\bibfnamefont {S.}~\bibnamefont {Li}}, \bibinfo {author} {\bibfnamefont {N.}~\bibnamefont {Dural}},\ and\ \bibinfo {author} {\bibfnamefont {M.~V.}\ \bibnamefont {Romalis}},\ }\bibfield  {title} {\bibinfo {title} {Subfemtotesla scalar atomic magnetometry using multipass cells},\ }\href {https://doi.org/10.1103/PhysRevLett.110.160802} {\bibfield  {journal} {\bibinfo  {journal} {Physical Review Letters}\ }\textbf {\bibinfo {volume} {110}},\ \bibinfo {pages} {160802} (\bibinfo {year} {2013})}\BibitemShut {NoStop}%
\bibitem [{\citenamefont {Jiang}\ \emph {et~al.}(2024{\natexlab{b}})\citenamefont {Jiang}, \citenamefont {Hong}, \citenamefont {Hu}, \citenamefont {Chen}, \citenamefont {Yang}, \citenamefont {Hu}, \citenamefont {Yang}, \citenamefont {Shu}, \citenamefont {Zhao}, \citenamefont {Peng},\ and\ \citenamefont {Du}}]{Jiang2024DPDM}%
  \BibitemOpen
  \bibfield  {author} {\bibinfo {author} {\bibfnamefont {M.}~\bibnamefont {Jiang}}, \bibinfo {author} {\bibfnamefont {T.}~\bibnamefont {Hong}}, \bibinfo {author} {\bibfnamefont {D.}~\bibnamefont {Hu}}, \bibinfo {author} {\bibfnamefont {Y.}~\bibnamefont {Chen}}, \bibinfo {author} {\bibfnamefont {F.}~\bibnamefont {Yang}}, \bibinfo {author} {\bibfnamefont {T.}~\bibnamefont {Hu}}, \bibinfo {author} {\bibfnamefont {X.}~\bibnamefont {Yang}}, \bibinfo {author} {\bibfnamefont {J.}~\bibnamefont {Shu}}, \bibinfo {author} {\bibfnamefont {Y.}~\bibnamefont {Zhao}}, \bibinfo {author} {\bibfnamefont {X.}~\bibnamefont {Peng}},\ and\ \bibinfo {author} {\bibfnamefont {J.}~\bibnamefont {Du}},\ }\bibfield  {title} {\bibinfo {title} {Long-baseline quantum sensor network as dark matter haloscope},\ }\href {https://doi.org/10.1038/s41467-024-47566-0} {\bibfield  {journal} {\bibinfo  {journal} {Nature Communications}\ }\textbf {\bibinfo {volume} {15}},\ \bibinfo {pages} {3331} (\bibinfo {year} {2024}{\natexlab{b}})}\BibitemShut
  {NoStop}%
\bibitem [{\citenamefont {Chaudhuri}\ \emph {et~al.}(2015)\citenamefont {Chaudhuri}, \citenamefont {Graham}, \citenamefont {Irwin}, \citenamefont {Mardon}, \citenamefont {Rajendran},\ and\ \citenamefont {Zhao}}]{Chaudhuri2015Radio}%
  \BibitemOpen
  \bibfield  {author} {\bibinfo {author} {\bibfnamefont {S.}~\bibnamefont {Chaudhuri}}, \bibinfo {author} {\bibfnamefont {P.~W.}\ \bibnamefont {Graham}}, \bibinfo {author} {\bibfnamefont {K.}~\bibnamefont {Irwin}}, \bibinfo {author} {\bibfnamefont {J.}~\bibnamefont {Mardon}}, \bibinfo {author} {\bibfnamefont {S.}~\bibnamefont {Rajendran}},\ and\ \bibinfo {author} {\bibfnamefont {Y.}~\bibnamefont {Zhao}},\ }\bibfield  {title} {\bibinfo {title} {Radio for hidden-photon dark matter detection},\ }\href {https://doi.org/10.1103/PhysRevD.92.075012} {\bibfield  {journal} {\bibinfo  {journal} {Physical Review D}\ }\textbf {\bibinfo {volume} {92}},\ \bibinfo {pages} {075012} (\bibinfo {year} {2015})}\BibitemShut {NoStop}%
\bibitem [{\citenamefont {Caputo}\ \emph {et~al.}(2021)\citenamefont {Caputo}, \citenamefont {Millar}, \citenamefont {O'Hare},\ and\ \citenamefont {Vitagliano}}]{Caputo2021DarkPhotonHandbook}%
  \BibitemOpen
  \bibfield  {author} {\bibinfo {author} {\bibfnamefont {A.}~\bibnamefont {Caputo}}, \bibinfo {author} {\bibfnamefont {A.~J.}\ \bibnamefont {Millar}}, \bibinfo {author} {\bibfnamefont {C.~A.~J.}\ \bibnamefont {O'Hare}},\ and\ \bibinfo {author} {\bibfnamefont {E.}~\bibnamefont {Vitagliano}},\ }\bibfield  {title} {\bibinfo {title} {Dark photon limits: a handbook},\ }\href {https://doi.org/10.1103/PhysRevD.104.095029} {\bibfield  {journal} {\bibinfo  {journal} {Physical Review D}\ }\textbf {\bibinfo {volume} {104}},\ \bibinfo {pages} {095029} (\bibinfo {year} {2021})}\BibitemShut {NoStop}%
\bibitem [{\citenamefont {Boto}\ \emph {et~al.}(2018)\citenamefont {Boto}, \citenamefont {Holmes}, \citenamefont {Leggett}, \citenamefont {Roberts}, \citenamefont {Shah}, \citenamefont {Meyer}, \citenamefont {Mu{\~{n}}oz}, \citenamefont {Mullinger}, \citenamefont {Tierney}, \citenamefont {Bestmann}, \citenamefont {Barnes}, \citenamefont {Bowtell},\ and\ \citenamefont {Brookes}}]{Boto2018WearableMEG}%
  \BibitemOpen
  \bibfield  {author} {\bibinfo {author} {\bibfnamefont {E.}~\bibnamefont {Boto}}, \bibinfo {author} {\bibfnamefont {N.}~\bibnamefont {Holmes}}, \bibinfo {author} {\bibfnamefont {J.}~\bibnamefont {Leggett}}, \bibinfo {author} {\bibfnamefont {G.}~\bibnamefont {Roberts}}, \bibinfo {author} {\bibfnamefont {V.}~\bibnamefont {Shah}}, \bibinfo {author} {\bibfnamefont {S.~S.}\ \bibnamefont {Meyer}}, \bibinfo {author} {\bibfnamefont {L.~D.}\ \bibnamefont {Mu{\~{n}}oz}}, \bibinfo {author} {\bibfnamefont {K.~J.}\ \bibnamefont {Mullinger}}, \bibinfo {author} {\bibfnamefont {T.~M.}\ \bibnamefont {Tierney}}, \bibinfo {author} {\bibfnamefont {S.}~\bibnamefont {Bestmann}}, \bibinfo {author} {\bibfnamefont {G.~R.}\ \bibnamefont {Barnes}}, \bibinfo {author} {\bibfnamefont {R.}~\bibnamefont {Bowtell}},\ and\ \bibinfo {author} {\bibfnamefont {M.~J.}\ \bibnamefont {Brookes}},\ }\bibfield  {title} {\bibinfo {title} {Moving magnetoencephalography towards real-world applications with a wearable system},\ }\href
  {https://doi.org/10.1038/nature26147} {\bibfield  {journal} {\bibinfo  {journal} {Nature}\ }\textbf {\bibinfo {volume} {555}},\ \bibinfo {pages} {657} (\bibinfo {year} {2018})}\BibitemShut {NoStop}%
\bibitem [{\citenamefont {Aslam}\ \emph {et~al.}(2023)\citenamefont {Aslam}, \citenamefont {Zhou}, \citenamefont {Urbach}, \citenamefont {Turner}, \citenamefont {Walsworth}, \citenamefont {Lukin},\ and\ \citenamefont {Park}}]{Aslam2023BioQSens}%
  \BibitemOpen
  \bibfield  {author} {\bibinfo {author} {\bibfnamefont {N.}~\bibnamefont {Aslam}}, \bibinfo {author} {\bibfnamefont {H.}~\bibnamefont {Zhou}}, \bibinfo {author} {\bibfnamefont {E.~K.}\ \bibnamefont {Urbach}}, \bibinfo {author} {\bibfnamefont {M.~J.}\ \bibnamefont {Turner}}, \bibinfo {author} {\bibfnamefont {R.~L.}\ \bibnamefont {Walsworth}}, \bibinfo {author} {\bibfnamefont {M.~D.}\ \bibnamefont {Lukin}},\ and\ \bibinfo {author} {\bibfnamefont {H.}~\bibnamefont {Park}},\ }\bibfield  {title} {\bibinfo {title} {Quantum sensors for biomedical applications},\ }\href {https://doi.org/10.1038/s42254-022-00566-4} {\bibfield  {journal} {\bibinfo  {journal} {Nature Reviews Physics}\ }\textbf {\bibinfo {volume} {5}},\ \bibinfo {pages} {157} (\bibinfo {year} {2023})}\BibitemShut {NoStop}%
\bibitem [{\citenamefont {Kimble}(2008)}]{Kimble2008QI}%
  \BibitemOpen
  \bibfield  {author} {\bibinfo {author} {\bibfnamefont {H.~J.}\ \bibnamefont {Kimble}},\ }\bibfield  {title} {\bibinfo {title} {The quantum internet},\ }\href {https://doi.org/10.1038/nature07127} {\bibfield  {journal} {\bibinfo  {journal} {Nature}\ }\textbf {\bibinfo {volume} {453}},\ \bibinfo {pages} {1023} (\bibinfo {year} {2008})}\BibitemShut {NoStop}%
\bibitem [{\citenamefont {Wang}(2005)}]{Wang2005Decoy}%
  \BibitemOpen
  \bibfield  {author} {\bibinfo {author} {\bibfnamefont {X.-B.}\ \bibnamefont {Wang}},\ }\bibfield  {title} {\bibinfo {title} {Beating the photon-number-splitting attack in practical quantum cryptography},\ }\href {https://doi.org/10.1103/PhysRevLett.94.230503} {\bibfield  {journal} {\bibinfo  {journal} {Physical Review Letters}\ }\textbf {\bibinfo {volume} {94}},\ \bibinfo {pages} {230503} (\bibinfo {year} {2005})}\BibitemShut {NoStop}%
\bibitem [{\citenamefont {Ma}\ \emph {et~al.}(2005)\citenamefont {Ma}, \citenamefont {Qi}, \citenamefont {Zhao},\ and\ \citenamefont {Lo}}]{Ma2005Decoy}%
  \BibitemOpen
  \bibfield  {author} {\bibinfo {author} {\bibfnamefont {X.}~\bibnamefont {Ma}}, \bibinfo {author} {\bibfnamefont {B.}~\bibnamefont {Qi}}, \bibinfo {author} {\bibfnamefont {Y.}~\bibnamefont {Zhao}},\ and\ \bibinfo {author} {\bibfnamefont {H.-K.}\ \bibnamefont {Lo}},\ }\bibfield  {title} {\bibinfo {title} {Practical decoy state for quantum key distribution},\ }\href {https://doi.org/10.1103/PhysRevA.72.012326} {\bibfield  {journal} {\bibinfo  {journal} {Physical Review A}\ }\textbf {\bibinfo {volume} {72}},\ \bibinfo {pages} {012326} (\bibinfo {year} {2005})}\BibitemShut {NoStop}%
\bibitem [{\citenamefont {Aharonovich}\ \emph {et~al.}(2016{\natexlab{b}})\citenamefont {Aharonovich}, \citenamefont {Englund},\ and\ \citenamefont {Toth}}]{Aharonovich2016SPE}%
  \BibitemOpen
  \bibfield  {author} {\bibinfo {author} {\bibfnamefont {I.}~\bibnamefont {Aharonovich}}, \bibinfo {author} {\bibfnamefont {D.}~\bibnamefont {Englund}},\ and\ \bibinfo {author} {\bibfnamefont {M.}~\bibnamefont {Toth}},\ }\bibfield  {title} {\bibinfo {title} {Solid-state single-photon emitters},\ }\href {https://doi.org/10.1038/nphoton.2016.186} {\bibfield  {journal} {\bibinfo  {journal} {Nature Photonics}\ }\textbf {\bibinfo {volume} {10}},\ \bibinfo {pages} {631} (\bibinfo {year} {2016}{\natexlab{b}})}\BibitemShut {NoStop}%
\bibitem [{\citenamefont {Mak}\ and\ \citenamefont {Shan}(2016)}]{Mak2016TMDCReview}%
  \BibitemOpen
  \bibfield  {author} {\bibinfo {author} {\bibfnamefont {K.~F.}\ \bibnamefont {Mak}}\ and\ \bibinfo {author} {\bibfnamefont {J.}~\bibnamefont {Shan}},\ }\bibfield  {title} {\bibinfo {title} {Photonics and optoelectronics of 2d semiconductor transition metal dichalcogenides},\ }\href {https://doi.org/10.1038/nphoton.2015.282} {\bibfield  {journal} {\bibinfo  {journal} {Nature Photonics}\ }\textbf {\bibinfo {volume} {10}},\ \bibinfo {pages} {216} (\bibinfo {year} {2016})}\BibitemShut {NoStop}%
\bibitem [{\citenamefont {Turunen}\ \emph {et~al.}(2022)\citenamefont {Turunen}, \citenamefont {Brotons-Gisbert}, \citenamefont {Dai}, \citenamefont {Wang}, \citenamefont {Scerri}, \citenamefont {Bonato}, \citenamefont {J{\"o}ns}, \citenamefont {Sun},\ and\ \citenamefont {Gerardot}}]{Turunen2022Review}%
  \BibitemOpen
  \bibfield  {author} {\bibinfo {author} {\bibfnamefont {M.}~\bibnamefont {Turunen}}, \bibinfo {author} {\bibfnamefont {M.}~\bibnamefont {Brotons-Gisbert}}, \bibinfo {author} {\bibfnamefont {Y.}~\bibnamefont {Dai}}, \bibinfo {author} {\bibfnamefont {Y.}~\bibnamefont {Wang}}, \bibinfo {author} {\bibfnamefont {E.}~\bibnamefont {Scerri}}, \bibinfo {author} {\bibfnamefont {C.}~\bibnamefont {Bonato}}, \bibinfo {author} {\bibfnamefont {K.~D.}\ \bibnamefont {J{\"o}ns}}, \bibinfo {author} {\bibfnamefont {Z.}~\bibnamefont {Sun}},\ and\ \bibinfo {author} {\bibfnamefont {B.~D.}\ \bibnamefont {Gerardot}},\ }\bibfield  {title} {\bibinfo {title} {Quantum photonics with layered 2d materials},\ }\href {https://doi.org/10.1038/s42254-021-00410-9} {\bibfield  {journal} {\bibinfo  {journal} {Nature Reviews Physics}\ }\textbf {\bibinfo {volume} {4}},\ \bibinfo {pages} {219} (\bibinfo {year} {2022})}\BibitemShut {NoStop}%
\bibitem [{\citenamefont {Gao}\ \emph {et~al.}(2023{\natexlab{b}})\citenamefont {Gao}, \citenamefont {von Helversen}, \citenamefont {Ant{\'o}n-Solanas}, \citenamefont {Schneider},\ and\ \citenamefont {Heindel}}]{Gao2023TMDCQKD}%
  \BibitemOpen
  \bibfield  {author} {\bibinfo {author} {\bibfnamefont {T.}~\bibnamefont {Gao}}, \bibinfo {author} {\bibfnamefont {M.}~\bibnamefont {von Helversen}}, \bibinfo {author} {\bibfnamefont {C.}~\bibnamefont {Ant{\'o}n-Solanas}}, \bibinfo {author} {\bibfnamefont {C.}~\bibnamefont {Schneider}},\ and\ \bibinfo {author} {\bibfnamefont {T.}~\bibnamefont {Heindel}},\ }\bibfield  {title} {\bibinfo {title} {Atomically-thin single-photon sources for quantum communication},\ }\href {https://doi.org/10.1038/s41699-023-00366-4} {\bibfield  {journal} {\bibinfo  {journal} {npj 2D Materials and Applications}\ }\textbf {\bibinfo {volume} {7}},\ \bibinfo {pages} {4} (\bibinfo {year} {2023}{\natexlab{b}})}\BibitemShut {NoStop}%
\bibitem [{\citenamefont {Ursin}\ \emph {et~al.}(2007)\citenamefont {Ursin}, \citenamefont {Tiefenbacher}, \citenamefont {Schmitt-Manderbach}, \citenamefont {Weier}, \citenamefont {Scheidl}, \citenamefont {Lindenthal}, \citenamefont {Blauensteiner}, \citenamefont {Jennewein}, \citenamefont {Perdigues}, \citenamefont {Trojek}, \citenamefont {{\"O}mer}, \citenamefont {F{\"u}rst}, \citenamefont {Meyenburg}, \citenamefont {Rarity}, \citenamefont {Sodnik}, \citenamefont {Barbieri}, \citenamefont {Weinfurter},\ and\ \citenamefont {Zeilinger}}]{Ursin2007FSO}%
  \BibitemOpen
  \bibfield  {author} {\bibinfo {author} {\bibfnamefont {R.}~\bibnamefont {Ursin}}, \bibinfo {author} {\bibfnamefont {F.}~\bibnamefont {Tiefenbacher}}, \bibinfo {author} {\bibfnamefont {T.}~\bibnamefont {Schmitt-Manderbach}}, \bibinfo {author} {\bibfnamefont {H.}~\bibnamefont {Weier}}, \bibinfo {author} {\bibfnamefont {T.}~\bibnamefont {Scheidl}}, \bibinfo {author} {\bibfnamefont {M.}~\bibnamefont {Lindenthal}}, \bibinfo {author} {\bibfnamefont {B.}~\bibnamefont {Blauensteiner}}, \bibinfo {author} {\bibfnamefont {T.}~\bibnamefont {Jennewein}}, \bibinfo {author} {\bibfnamefont {J.}~\bibnamefont {Perdigues}}, \bibinfo {author} {\bibfnamefont {P.}~\bibnamefont {Trojek}}, \bibinfo {author} {\bibfnamefont {B.}~\bibnamefont {{\"O}mer}}, \bibinfo {author} {\bibfnamefont {M.}~\bibnamefont {F{\"u}rst}}, \bibinfo {author} {\bibfnamefont {M.}~\bibnamefont {Meyenburg}}, \bibinfo {author} {\bibfnamefont {J.}~\bibnamefont {Rarity}}, \bibinfo {author} {\bibfnamefont {Z.}~\bibnamefont {Sodnik}}, \bibinfo {author}
  {\bibfnamefont {C.}~\bibnamefont {Barbieri}}, \bibinfo {author} {\bibfnamefont {H.}~\bibnamefont {Weinfurter}},\ and\ \bibinfo {author} {\bibfnamefont {A.}~\bibnamefont {Zeilinger}},\ }\bibfield  {title} {\bibinfo {title} {Entanglement-based quantum communication over 144 km},\ }\href {https://doi.org/10.1038/nphys629} {\bibfield  {journal} {\bibinfo  {journal} {Nature Physics}\ }\textbf {\bibinfo {volume} {3}},\ \bibinfo {pages} {481} (\bibinfo {year} {2007})}\BibitemShut {NoStop}%
\bibitem [{\citenamefont {Liao}\ \emph {et~al.}(2017)\citenamefont {Liao}, \citenamefont {Cai}, \citenamefont {Liu}, \citenamefont {Zhang}, \citenamefont {Li}, \citenamefont {Ren}, \citenamefont {Yin}, \citenamefont {Shen}, \citenamefont {Cao}, \citenamefont {Li}, \citenamefont {Li}, \citenamefont {Chen}, \citenamefont {Sun}, \citenamefont {Jia}, \citenamefont {Wu}, \citenamefont {Jiang}, \citenamefont {Wang}, \citenamefont {Huang}, \citenamefont {Wang}, \citenamefont {Zhou}, \citenamefont {Deng}, \citenamefont {Xi}, \citenamefont {Ma}, \citenamefont {Hu}, \citenamefont {Zhang}, \citenamefont {Chen}, \citenamefont {Liu}, \citenamefont {Wang}, \citenamefont {Zhu}, \citenamefont {Lu}, \citenamefont {Shu}, \citenamefont {Peng}, \citenamefont {Wang},\ and\ \citenamefont {Pan}}]{Liao2017Satellite}%
  \BibitemOpen
  \bibfield  {author} {\bibinfo {author} {\bibfnamefont {S.-K.}\ \bibnamefont {Liao}}, \bibinfo {author} {\bibfnamefont {W.-Q.}\ \bibnamefont {Cai}}, \bibinfo {author} {\bibfnamefont {W.-Y.}\ \bibnamefont {Liu}}, \bibinfo {author} {\bibfnamefont {L.}~\bibnamefont {Zhang}}, \bibinfo {author} {\bibfnamefont {Y.}~\bibnamefont {Li}}, \bibinfo {author} {\bibfnamefont {J.-G.}\ \bibnamefont {Ren}}, \bibinfo {author} {\bibfnamefont {J.}~\bibnamefont {Yin}}, \bibinfo {author} {\bibfnamefont {Q.}~\bibnamefont {Shen}}, \bibinfo {author} {\bibfnamefont {Y.}~\bibnamefont {Cao}}, \bibinfo {author} {\bibfnamefont {Z.-P.}\ \bibnamefont {Li}}, \bibinfo {author} {\bibfnamefont {F.-Z.}\ \bibnamefont {Li}}, \bibinfo {author} {\bibfnamefont {X.-W.}\ \bibnamefont {Chen}}, \bibinfo {author} {\bibfnamefont {L.-H.}\ \bibnamefont {Sun}}, \bibinfo {author} {\bibfnamefont {J.-J.}\ \bibnamefont {Jia}}, \bibinfo {author} {\bibfnamefont {J.-C.}\ \bibnamefont {Wu}}, \bibinfo {author} {\bibfnamefont {X.-J.}\ \bibnamefont {Jiang}}, \bibinfo
  {author} {\bibfnamefont {J.-F.}\ \bibnamefont {Wang}}, \bibinfo {author} {\bibfnamefont {Y.-M.}\ \bibnamefont {Huang}}, \bibinfo {author} {\bibfnamefont {Q.}~\bibnamefont {Wang}}, \bibinfo {author} {\bibfnamefont {Y.-L.}\ \bibnamefont {Zhou}}, \bibinfo {author} {\bibfnamefont {L.}~\bibnamefont {Deng}}, \bibinfo {author} {\bibfnamefont {T.}~\bibnamefont {Xi}}, \bibinfo {author} {\bibfnamefont {L.}~\bibnamefont {Ma}}, \bibinfo {author} {\bibfnamefont {T.}~\bibnamefont {Hu}}, \bibinfo {author} {\bibfnamefont {Q.}~\bibnamefont {Zhang}}, \bibinfo {author} {\bibfnamefont {Y.-A.}\ \bibnamefont {Chen}}, \bibinfo {author} {\bibfnamefont {N.-L.}\ \bibnamefont {Liu}}, \bibinfo {author} {\bibfnamefont {X.-B.}\ \bibnamefont {Wang}}, \bibinfo {author} {\bibfnamefont {Z.-C.}\ \bibnamefont {Zhu}}, \bibinfo {author} {\bibfnamefont {C.-Y.}\ \bibnamefont {Lu}}, \bibinfo {author} {\bibfnamefont {R.}~\bibnamefont {Shu}}, \bibinfo {author} {\bibfnamefont {C.-Z.}\ \bibnamefont {Peng}}, \bibinfo {author} {\bibfnamefont {J.-Y.}\
  \bibnamefont {Wang}},\ and\ \bibinfo {author} {\bibfnamefont {J.-W.}\ \bibnamefont {Pan}},\ }\bibfield  {title} {\bibinfo {title} {Satellite-to-ground quantum key distribution},\ }\href {https://doi.org/10.1038/nature23655} {\bibfield  {journal} {\bibinfo  {journal} {Nature}\ }\textbf {\bibinfo {volume} {549}},\ \bibinfo {pages} {43} (\bibinfo {year} {2017})}\BibitemShut {NoStop}%
\bibitem [{\citenamefont {Gottesman}\ \emph {et~al.}(2004)\citenamefont {Gottesman}, \citenamefont {Lo}, \citenamefont {Lutkenhaus},\ and\ \citenamefont {Preskill}}]{Gottesman2004GLLP}%
  \BibitemOpen
  \bibfield  {author} {\bibinfo {author} {\bibfnamefont {D.}~\bibnamefont {Gottesman}}, \bibinfo {author} {\bibfnamefont {H.-K.}\ \bibnamefont {Lo}}, \bibinfo {author} {\bibfnamefont {N.}~\bibnamefont {Lutkenhaus}},\ and\ \bibinfo {author} {\bibfnamefont {J.}~\bibnamefont {Preskill}},\ }\bibfield  {title} {\bibinfo {title} {Security of quantum key distribution with imperfect devices},\ }in\ \href {https://doi.org/10.1109/ISIT.2004.1365172} {\emph {\bibinfo {booktitle} {International Symposium onInformation Theory, 2004. ISIT 2004. Proceedings.}}}\ (\bibinfo {year} {2004})\BibitemShut {NoStop}%
\bibitem [{\citenamefont {Waks}\ \emph {et~al.}(2002{\natexlab{a}})\citenamefont {Waks}, \citenamefont {Santori},\ and\ \citenamefont {Yamamoto}}]{Waks2002SubPoisson}%
  \BibitemOpen
  \bibfield  {author} {\bibinfo {author} {\bibfnamefont {E.}~\bibnamefont {Waks}}, \bibinfo {author} {\bibfnamefont {C.}~\bibnamefont {Santori}},\ and\ \bibinfo {author} {\bibfnamefont {Y.}~\bibnamefont {Yamamoto}},\ }\bibfield  {title} {\bibinfo {title} {Security aspects of quantum key distribution with sub-poisson light},\ }\href {https://doi.org/10.1103/PhysRevA.66.042315} {\bibfield  {journal} {\bibinfo  {journal} {Physical Review A}\ }\textbf {\bibinfo {volume} {66}},\ \bibinfo {pages} {042315} (\bibinfo {year} {2002}{\natexlab{a}})}\BibitemShut {NoStop}%
\bibitem [{\citenamefont {Takeoka}\ \emph {et~al.}(2014)\citenamefont {Takeoka}, \citenamefont {Guha},\ and\ \citenamefont {Wilde}}]{Takeoka2014RateLoss}%
  \BibitemOpen
  \bibfield  {author} {\bibinfo {author} {\bibfnamefont {M.}~\bibnamefont {Takeoka}}, \bibinfo {author} {\bibfnamefont {S.}~\bibnamefont {Guha}},\ and\ \bibinfo {author} {\bibfnamefont {M.~M.}\ \bibnamefont {Wilde}},\ }\bibfield  {title} {\bibinfo {title} {Fundamental rate-loss tradeoff for optical quantum key distribution},\ }\href {https://doi.org/10.1038/ncomms6235} {\bibfield  {journal} {\bibinfo  {journal} {Nature Communications}\ }\textbf {\bibinfo {volume} {5}},\ \bibinfo {pages} {5235} (\bibinfo {year} {2014})}\BibitemShut {NoStop}%
\bibitem [{\citenamefont {Pirandola}\ \emph {et~al.}(2017)\citenamefont {Pirandola}, \citenamefont {Laurenza}, \citenamefont {Ottaviani},\ and\ \citenamefont {Banchi}}]{Pirandola2017Repeaterless}%
  \BibitemOpen
  \bibfield  {author} {\bibinfo {author} {\bibfnamefont {S.}~\bibnamefont {Pirandola}}, \bibinfo {author} {\bibfnamefont {R.}~\bibnamefont {Laurenza}}, \bibinfo {author} {\bibfnamefont {C.}~\bibnamefont {Ottaviani}},\ and\ \bibinfo {author} {\bibfnamefont {L.}~\bibnamefont {Banchi}},\ }\bibfield  {title} {\bibinfo {title} {Fundamental limits of repeaterless quantum communications},\ }\href {https://doi.org/10.1038/ncomms15043} {\bibfield  {journal} {\bibinfo  {journal} {Nature Communications}\ }\textbf {\bibinfo {volume} {8}},\ \bibinfo {pages} {15043} (\bibinfo {year} {2017})}\BibitemShut {NoStop}%
\bibitem [{\citenamefont {Kupko}\ \emph {et~al.}(2020)\citenamefont {Kupko}, \citenamefont {von Helversen}, \citenamefont {Rickert}, \citenamefont {Schulze}, \citenamefont {Strittmatter}, \citenamefont {Gschrey}, \citenamefont {Rodt}, \citenamefont {Reitzenstein},\ and\ \citenamefont {Heindel}}]{Kupko2020Tools}%
  \BibitemOpen
  \bibfield  {author} {\bibinfo {author} {\bibfnamefont {T.}~\bibnamefont {Kupko}}, \bibinfo {author} {\bibfnamefont {M.}~\bibnamefont {von Helversen}}, \bibinfo {author} {\bibfnamefont {L.}~\bibnamefont {Rickert}}, \bibinfo {author} {\bibfnamefont {J.-H.}\ \bibnamefont {Schulze}}, \bibinfo {author} {\bibfnamefont {A.}~\bibnamefont {Strittmatter}}, \bibinfo {author} {\bibfnamefont {M.}~\bibnamefont {Gschrey}}, \bibinfo {author} {\bibfnamefont {S.}~\bibnamefont {Rodt}}, \bibinfo {author} {\bibfnamefont {S.}~\bibnamefont {Reitzenstein}},\ and\ \bibinfo {author} {\bibfnamefont {T.}~\bibnamefont {Heindel}},\ }\bibfield  {title} {\bibinfo {title} {Tools for the performance optimization of single-photon quantum key distribution},\ }\href {https://doi.org/10.1038/s41534-020-0253-8} {\bibfield  {journal} {\bibinfo  {journal} {npj Quantum Information}\ }\textbf {\bibinfo {volume} {6}},\ \bibinfo {pages} {29} (\bibinfo {year} {2020})}\BibitemShut {NoStop}%
\bibitem [{\citenamefont {Vajner}\ \emph {et~al.}(2022)\citenamefont {Vajner}, \citenamefont {Rickert}, \citenamefont {Gao}, \citenamefont {Kaymazlar},\ and\ \citenamefont {Heindel}}]{Vajner2022QDReview}%
  \BibitemOpen
  \bibfield  {author} {\bibinfo {author} {\bibfnamefont {D.~A.}\ \bibnamefont {Vajner}}, \bibinfo {author} {\bibfnamefont {L.}~\bibnamefont {Rickert}}, \bibinfo {author} {\bibfnamefont {T.}~\bibnamefont {Gao}}, \bibinfo {author} {\bibfnamefont {K.}~\bibnamefont {Kaymazlar}},\ and\ \bibinfo {author} {\bibfnamefont {T.}~\bibnamefont {Heindel}},\ }\bibfield  {title} {\bibinfo {title} {Quantum communication using semiconductor quantum dots},\ }\href {https://doi.org/10.1002/qute.202100116} {\bibfield  {journal} {\bibinfo  {journal} {Advanced Quantum Technologies}\ }\textbf {\bibinfo {volume} {5}},\ \bibinfo {pages} {2100116} (\bibinfo {year} {2022})}\BibitemShut {NoStop}%
\bibitem [{\citenamefont {Waks}\ \emph {et~al.}(2002{\natexlab{b}})\citenamefont {Waks}, \citenamefont {Inoue}, \citenamefont {Santori}, \citenamefont {Fattal}, \citenamefont {Vuckovic}, \citenamefont {Solomon},\ and\ \citenamefont {Yamamoto}}]{Waks2002Turnstile}%
  \BibitemOpen
  \bibfield  {author} {\bibinfo {author} {\bibfnamefont {E.}~\bibnamefont {Waks}}, \bibinfo {author} {\bibfnamefont {K.}~\bibnamefont {Inoue}}, \bibinfo {author} {\bibfnamefont {C.}~\bibnamefont {Santori}}, \bibinfo {author} {\bibfnamefont {D.}~\bibnamefont {Fattal}}, \bibinfo {author} {\bibfnamefont {J.}~\bibnamefont {Vuckovic}}, \bibinfo {author} {\bibfnamefont {G.~S.}\ \bibnamefont {Solomon}},\ and\ \bibinfo {author} {\bibfnamefont {Y.}~\bibnamefont {Yamamoto}},\ }\bibfield  {title} {\bibinfo {title} {Quantum cryptography with a photon turnstile},\ }\href {https://doi.org/10.1038/nature01240} {\bibfield  {journal} {\bibinfo  {journal} {Nature}\ }\textbf {\bibinfo {volume} {420}},\ \bibinfo {pages} {762} (\bibinfo {year} {2002}{\natexlab{b}})}\BibitemShut {NoStop}%
\bibitem [{\citenamefont {Leifgen}\ \emph {et~al.}(2014)\citenamefont {Leifgen}, \citenamefont {Schröder}, \citenamefont {Gädeke}, \citenamefont {Riemann}, \citenamefont {Métillon}, \citenamefont {Neu}, \citenamefont {Hepp}, \citenamefont {Arend}, \citenamefont {Becher}, \citenamefont {Lauritsen},\ and\ \citenamefont {Benson}}]{Leifgen2014NVQKD}%
  \BibitemOpen
  \bibfield  {author} {\bibinfo {author} {\bibfnamefont {M.}~\bibnamefont {Leifgen}}, \bibinfo {author} {\bibfnamefont {T.}~\bibnamefont {Schröder}}, \bibinfo {author} {\bibfnamefont {F.}~\bibnamefont {Gädeke}}, \bibinfo {author} {\bibfnamefont {R.}~\bibnamefont {Riemann}}, \bibinfo {author} {\bibfnamefont {V.}~\bibnamefont {Métillon}}, \bibinfo {author} {\bibfnamefont {E.}~\bibnamefont {Neu}}, \bibinfo {author} {\bibfnamefont {C.}~\bibnamefont {Hepp}}, \bibinfo {author} {\bibfnamefont {C.}~\bibnamefont {Arend}}, \bibinfo {author} {\bibfnamefont {C.}~\bibnamefont {Becher}}, \bibinfo {author} {\bibfnamefont {K.}~\bibnamefont {Lauritsen}},\ and\ \bibinfo {author} {\bibfnamefont {O.}~\bibnamefont {Benson}},\ }\bibfield  {title} {\bibinfo {title} {Evaluation of nitrogen- and silicon-vacancy defect centres as single photon sources in quantum key distribution},\ }\href {https://doi.org/10.1088/1367-2630/16/2/023021} {\bibfield  {journal} {\bibinfo  {journal} {New Journal of Physics}\ }\textbf {\bibinfo {volume}
  {16}},\ \bibinfo {pages} {023021} (\bibinfo {year} {2014})}\BibitemShut {NoStop}%
\bibitem [{\citenamefont {Takemoto}\ \emph {et~al.}(2015)\citenamefont {Takemoto}, \citenamefont {Nambu}, \citenamefont {Miyazawa}, \citenamefont {Sakuma}, \citenamefont {Yamamoto}, \citenamefont {Yorozu},\ and\ \citenamefont {Arakawa}}]{Takemoto2015QD120km}%
  \BibitemOpen
  \bibfield  {author} {\bibinfo {author} {\bibfnamefont {K.}~\bibnamefont {Takemoto}}, \bibinfo {author} {\bibfnamefont {Y.}~\bibnamefont {Nambu}}, \bibinfo {author} {\bibfnamefont {T.}~\bibnamefont {Miyazawa}}, \bibinfo {author} {\bibfnamefont {Y.}~\bibnamefont {Sakuma}}, \bibinfo {author} {\bibfnamefont {T.}~\bibnamefont {Yamamoto}}, \bibinfo {author} {\bibfnamefont {S.}~\bibnamefont {Yorozu}},\ and\ \bibinfo {author} {\bibfnamefont {Y.}~\bibnamefont {Arakawa}},\ }\bibfield  {title} {\bibinfo {title} {Quantum key distribution over 120 km using ultrahigh purity single-photon source and superconducting single-photon detectors},\ }\href {https://doi.org/10.1038/srep14383} {\bibfield  {journal} {\bibinfo  {journal} {Scientific Reports}\ }\textbf {\bibinfo {volume} {5}},\ \bibinfo {pages} {14383} (\bibinfo {year} {2015})}\BibitemShut {NoStop}%
\bibitem [{\citenamefont {Luo}\ \emph {et~al.}(2018)\citenamefont {Luo}, \citenamefont {Shepard}, \citenamefont {Ardelean}, \citenamefont {Rhodes}, \citenamefont {Kim}, \citenamefont {Barmak}, \citenamefont {Hone},\ and\ \citenamefont {Strauf}}]{Luo2018Cavity}%
  \BibitemOpen
  \bibfield  {author} {\bibinfo {author} {\bibfnamefont {Y.}~\bibnamefont {Luo}}, \bibinfo {author} {\bibfnamefont {G.~D.}\ \bibnamefont {Shepard}}, \bibinfo {author} {\bibfnamefont {J.~V.}\ \bibnamefont {Ardelean}}, \bibinfo {author} {\bibfnamefont {D.~A.}\ \bibnamefont {Rhodes}}, \bibinfo {author} {\bibfnamefont {B.}~\bibnamefont {Kim}}, \bibinfo {author} {\bibfnamefont {K.}~\bibnamefont {Barmak}}, \bibinfo {author} {\bibfnamefont {J.~C.}\ \bibnamefont {Hone}},\ and\ \bibinfo {author} {\bibfnamefont {S.}~\bibnamefont {Strauf}},\ }\bibfield  {title} {\bibinfo {title} {Deterministic coupling of site-controlled quantum emitters in monolayer wse$_2$ to plasmonic nanocavities},\ }\href {https://doi.org/10.1038/s41565-018-0250-5} {\bibfield  {journal} {\bibinfo  {journal} {Nature Nanotechnology}\ }\textbf {\bibinfo {volume} {13}},\ \bibinfo {pages} {1137} (\bibinfo {year} {2018})}\BibitemShut {NoStop}%
\bibitem [{\citenamefont {Iff}\ \emph {et~al.}(2021)\citenamefont {Iff}, \citenamefont {Buchinger}, \citenamefont {Moczała-Dusanowska}, \citenamefont {Kamp}, \citenamefont {Betzold}, \citenamefont {Davanco}, \citenamefont {Srinivasan}, \citenamefont {Tongay}, \citenamefont {Antón-Solanas}, \citenamefont {H{\"o}fling},\ and\ \citenamefont {Schneider}}]{Iff2021Purcell}%
  \BibitemOpen
  \bibfield  {author} {\bibinfo {author} {\bibfnamefont {O.}~\bibnamefont {Iff}}, \bibinfo {author} {\bibfnamefont {Q.}~\bibnamefont {Buchinger}}, \bibinfo {author} {\bibfnamefont {M.}~\bibnamefont {Moczała-Dusanowska}}, \bibinfo {author} {\bibfnamefont {M.}~\bibnamefont {Kamp}}, \bibinfo {author} {\bibfnamefont {S.}~\bibnamefont {Betzold}}, \bibinfo {author} {\bibfnamefont {M.}~\bibnamefont {Davanco}}, \bibinfo {author} {\bibfnamefont {K.}~\bibnamefont {Srinivasan}}, \bibinfo {author} {\bibfnamefont {S.}~\bibnamefont {Tongay}}, \bibinfo {author} {\bibfnamefont {C.}~\bibnamefont {Antón-Solanas}}, \bibinfo {author} {\bibfnamefont {S.}~\bibnamefont {H{\"o}fling}},\ and\ \bibinfo {author} {\bibfnamefont {C.}~\bibnamefont {Schneider}},\ }\bibfield  {title} {\bibinfo {title} {Purcell-enhanced single photon source based on a deterministically placed wse$_2$ monolayer quantum dot in a circular bragg grating cavity},\ }\href {https://doi.org/10.1021/acs.nanolett.1c00525} {\bibfield  {journal} {\bibinfo  {journal}
  {Nano Letters}\ }\textbf {\bibinfo {volume} {21}},\ \bibinfo {pages} {4715} (\bibinfo {year} {2021})}\BibitemShut {NoStop}%
\bibitem [{\citenamefont {Sparrow}\ \emph {et~al.}(2018)\citenamefont {Sparrow}, \citenamefont {Mart{\'i}n-L{\'o}pez}, \citenamefont {Maraviglia}, \citenamefont {Neville}, \citenamefont {Harrold}, \citenamefont {Carolan}, \citenamefont {Joglekar}, \citenamefont {Hashimoto}, \citenamefont {Matsuda}, \citenamefont {O'Brien}, \citenamefont {Tew},\ and\ \citenamefont {Laing}}]{Sparrow2018}%
  \BibitemOpen
  \bibfield  {author} {\bibinfo {author} {\bibfnamefont {C.}~\bibnamefont {Sparrow}}, \bibinfo {author} {\bibfnamefont {E.}~\bibnamefont {Mart{\'i}n-L{\'o}pez}}, \bibinfo {author} {\bibfnamefont {N.}~\bibnamefont {Maraviglia}}, \bibinfo {author} {\bibfnamefont {A.}~\bibnamefont {Neville}}, \bibinfo {author} {\bibfnamefont {C.}~\bibnamefont {Harrold}}, \bibinfo {author} {\bibfnamefont {J.}~\bibnamefont {Carolan}}, \bibinfo {author} {\bibfnamefont {Y.~N.}\ \bibnamefont {Joglekar}}, \bibinfo {author} {\bibfnamefont {T.}~\bibnamefont {Hashimoto}}, \bibinfo {author} {\bibfnamefont {N.}~\bibnamefont {Matsuda}}, \bibinfo {author} {\bibfnamefont {J.~L.}\ \bibnamefont {O'Brien}}, \bibinfo {author} {\bibfnamefont {D.~P.}\ \bibnamefont {Tew}},\ and\ \bibinfo {author} {\bibfnamefont {A.}~\bibnamefont {Laing}},\ }\bibfield  {title} {\bibinfo {title} {Simulating the vibrational quantum dynamics of molecules using photonics},\ }\href {https://doi.org/10.1038/s41586-018-0152-9} {\bibfield  {journal} {\bibinfo  {journal}
  {Nature}\ }\textbf {\bibinfo {volume} {557}},\ \bibinfo {pages} {660} (\bibinfo {year} {2018})}\BibitemShut {NoStop}%
\bibitem [{\citenamefont {Steinbrecher}\ \emph {et~al.}(2019)\citenamefont {Steinbrecher}, \citenamefont {Olson}, \citenamefont {Englund},\ and\ \citenamefont {Carolan}}]{Steinbrecher2019}%
  \BibitemOpen
  \bibfield  {author} {\bibinfo {author} {\bibfnamefont {G.~R.}\ \bibnamefont {Steinbrecher}}, \bibinfo {author} {\bibfnamefont {J.~P.}\ \bibnamefont {Olson}}, \bibinfo {author} {\bibfnamefont {D.}~\bibnamefont {Englund}},\ and\ \bibinfo {author} {\bibfnamefont {J.}~\bibnamefont {Carolan}},\ }\bibfield  {title} {\bibinfo {title} {Quantum optical neural networks},\ }\href {https://doi.org/10.1038/s41534-019-0174-7} {\bibfield  {journal} {\bibinfo  {journal} {npj Quantum Information}\ }\textbf {\bibinfo {volume} {5}},\ \bibinfo {pages} {60} (\bibinfo {year} {2019})}\BibitemShut {NoStop}%
\bibitem [{\citenamefont {Choi}\ \emph {et~al.}(2019)\citenamefont {Choi}, \citenamefont {Pant}, \citenamefont {Guha},\ and\ \citenamefont {Englund}}]{Choi2019}%
  \BibitemOpen
  \bibfield  {author} {\bibinfo {author} {\bibfnamefont {H.}~\bibnamefont {Choi}}, \bibinfo {author} {\bibfnamefont {M.}~\bibnamefont {Pant}}, \bibinfo {author} {\bibfnamefont {S.}~\bibnamefont {Guha}},\ and\ \bibinfo {author} {\bibfnamefont {D.}~\bibnamefont {Englund}},\ }\bibfield  {title} {\bibinfo {title} {Percolation-based architecture for cluster state creation using photon-mediated entanglement between atomic memories},\ }\href {https://doi.org/10.1038/s41534-019-0215-2} {\bibfield  {journal} {\bibinfo  {journal} {npj Quantum Information}\ }\textbf {\bibinfo {volume} {5}},\ \bibinfo {pages} {104} (\bibinfo {year} {2019})}\BibitemShut {NoStop}%
\bibitem [{\citenamefont {Rudolph}(2017)}]{Rudolph2017}%
  \BibitemOpen
  \bibfield  {author} {\bibinfo {author} {\bibfnamefont {T.}~\bibnamefont {Rudolph}},\ }\bibfield  {title} {\bibinfo {title} {Why i am optimistic about the silicon-photonic route to quantum computing},\ }\href {https://doi.org/10.1063/1.4976737} {\bibfield  {journal} {\bibinfo  {journal} {APL Photonics}\ }\textbf {\bibinfo {volume} {2}},\ \bibinfo {pages} {030901} (\bibinfo {year} {2017})}\BibitemShut {NoStop}%
\bibitem [{\citenamefont {Wang}\ \emph {et~al.}(2019)\citenamefont {Wang}, \citenamefont {Qin}, \citenamefont {Ding}, \citenamefont {Chen}, \citenamefont {Chen}, \citenamefont {You}, \citenamefont {He}, \citenamefont {Jiang}, \citenamefont {You}, \citenamefont {Wang}, \citenamefont {Schneider}, \citenamefont {Renema}, \citenamefont {H{\"o}fling}, \citenamefont {Lu},\ and\ \citenamefont {Pan}}]{Wang2019Boson}%
  \BibitemOpen
  \bibfield  {author} {\bibinfo {author} {\bibfnamefont {H.}~\bibnamefont {Wang}}, \bibinfo {author} {\bibfnamefont {J.}~\bibnamefont {Qin}}, \bibinfo {author} {\bibfnamefont {X.}~\bibnamefont {Ding}}, \bibinfo {author} {\bibfnamefont {M.-C.}\ \bibnamefont {Chen}}, \bibinfo {author} {\bibfnamefont {S.}~\bibnamefont {Chen}}, \bibinfo {author} {\bibfnamefont {X.}~\bibnamefont {You}}, \bibinfo {author} {\bibfnamefont {Y.-M.}\ \bibnamefont {He}}, \bibinfo {author} {\bibfnamefont {X.}~\bibnamefont {Jiang}}, \bibinfo {author} {\bibfnamefont {L.}~\bibnamefont {You}}, \bibinfo {author} {\bibfnamefont {Z.}~\bibnamefont {Wang}}, \bibinfo {author} {\bibfnamefont {C.}~\bibnamefont {Schneider}}, \bibinfo {author} {\bibfnamefont {J.~J.}\ \bibnamefont {Renema}}, \bibinfo {author} {\bibfnamefont {S.}~\bibnamefont {H{\"o}fling}}, \bibinfo {author} {\bibfnamefont {C.-Y.}\ \bibnamefont {Lu}},\ and\ \bibinfo {author} {\bibfnamefont {J.-W.}\ \bibnamefont {Pan}},\ }\bibfield  {title} {\bibinfo {title} {Boson sampling with 20 input
  photons and a 60-mode interferometer in a $10^{14}$-dimensional hilbert space},\ }\href {https://doi.org/10.1103/PhysRevLett.123.250503} {\bibfield  {journal} {\bibinfo  {journal} {Physical Review Letters}\ }\textbf {\bibinfo {volume} {123}},\ \bibinfo {pages} {250503} (\bibinfo {year} {2019})}\BibitemShut {NoStop}%
\bibitem [{\citenamefont {Brod}\ \emph {et~al.}(2019)\citenamefont {Brod}, \citenamefont {Galv{\~a}o}, \citenamefont {Crespi}, \citenamefont {Osellame}, \citenamefont {Spagnolo},\ and\ \citenamefont {Sciarrino}}]{Brod2019}%
  \BibitemOpen
  \bibfield  {author} {\bibinfo {author} {\bibfnamefont {D.~J.}\ \bibnamefont {Brod}}, \bibinfo {author} {\bibfnamefont {E.~F.}\ \bibnamefont {Galv{\~a}o}}, \bibinfo {author} {\bibfnamefont {A.}~\bibnamefont {Crespi}}, \bibinfo {author} {\bibfnamefont {R.}~\bibnamefont {Osellame}}, \bibinfo {author} {\bibfnamefont {N.}~\bibnamefont {Spagnolo}},\ and\ \bibinfo {author} {\bibfnamefont {F.}~\bibnamefont {Sciarrino}},\ }\bibfield  {title} {\bibinfo {title} {Photonic implementation of boson sampling: a review},\ }\href {https://doi.org/10.1117/1.AP.1.3.034001} {\bibfield  {journal} {\bibinfo  {journal} {Advanced Photonics}\ }\textbf {\bibinfo {volume} {1}},\ \bibinfo {pages} {034001} (\bibinfo {year} {2019})}\BibitemShut {NoStop}%
\bibitem [{\citenamefont {Deshpande}\ \emph {et~al.}(2022)\citenamefont {Deshpande}, \citenamefont {Mehta}, \citenamefont {Vincent}, \citenamefont {Quesada}, \citenamefont {Hinsche}, \citenamefont {Ioannou}, \citenamefont {Madsen}, \citenamefont {Lavoie}, \citenamefont {Qi}, \citenamefont {Eisert}, \citenamefont {Hangleiter}, \citenamefont {Fefferman},\ and\ \citenamefont {Dhand}}]{Deshpande2022}%
  \BibitemOpen
  \bibfield  {author} {\bibinfo {author} {\bibfnamefont {A.}~\bibnamefont {Deshpande}}, \bibinfo {author} {\bibfnamefont {A.}~\bibnamefont {Mehta}}, \bibinfo {author} {\bibfnamefont {T.}~\bibnamefont {Vincent}}, \bibinfo {author} {\bibfnamefont {N.}~\bibnamefont {Quesada}}, \bibinfo {author} {\bibfnamefont {M.}~\bibnamefont {Hinsche}}, \bibinfo {author} {\bibfnamefont {M.}~\bibnamefont {Ioannou}}, \bibinfo {author} {\bibfnamefont {L.}~\bibnamefont {Madsen}}, \bibinfo {author} {\bibfnamefont {J.}~\bibnamefont {Lavoie}}, \bibinfo {author} {\bibfnamefont {H.}~\bibnamefont {Qi}}, \bibinfo {author} {\bibfnamefont {J.}~\bibnamefont {Eisert}}, \bibinfo {author} {\bibfnamefont {D.}~\bibnamefont {Hangleiter}}, \bibinfo {author} {\bibfnamefont {B.}~\bibnamefont {Fefferman}},\ and\ \bibinfo {author} {\bibfnamefont {I.}~\bibnamefont {Dhand}},\ }\bibfield  {title} {\bibinfo {title} {Quantum computational advantage via high-dimensional gaussian boson sampling},\ }\href {https://doi.org/10.1126/sciadv.abi7894} {\bibfield
  {journal} {\bibinfo  {journal} {Science Advances}\ }\textbf {\bibinfo {volume} {8}},\ \bibinfo {pages} {eabi7894} (\bibinfo {year} {2022})},\ \Eprint {https://arxiv.org/abs/https://www.science.org/doi/pdf/10.1126/sciadv.abi7894} {https://www.science.org/doi/pdf/10.1126/sciadv.abi7894} \BibitemShut {NoStop}%
\bibitem [{\citenamefont {Liu}\ \emph {et~al.}(2021)\citenamefont {Liu}, \citenamefont {Jin}, \citenamefont {Cheng}, \citenamefont {Puckett}, \citenamefont {Behunin}, \citenamefont {Nelson}, \citenamefont {Rakich},\ and\ \citenamefont {Blumenthal}}]{Liu2021}%
  \BibitemOpen
  \bibfield  {author} {\bibinfo {author} {\bibfnamefont {K.}~\bibnamefont {Liu}}, \bibinfo {author} {\bibfnamefont {N.}~\bibnamefont {Jin}}, \bibinfo {author} {\bibfnamefont {H.}~\bibnamefont {Cheng}}, \bibinfo {author} {\bibfnamefont {M.~W.}\ \bibnamefont {Puckett}}, \bibinfo {author} {\bibfnamefont {R.~O.}\ \bibnamefont {Behunin}}, \bibinfo {author} {\bibfnamefont {K.~D.}\ \bibnamefont {Nelson}}, \bibinfo {author} {\bibfnamefont {P.~T.}\ \bibnamefont {Rakich}},\ and\ \bibinfo {author} {\bibfnamefont {D.~J.}\ \bibnamefont {Blumenthal}},\ }\bibfield  {title} {\bibinfo {title} {720 million quality factor integrated all-waveguide photonic resonator},\ }in\ \href {https://doi.org/10.1109/DRC52342.2021.9467132} {\emph {\bibinfo {booktitle} {2021 Device Research Conference (DRC)}}}\ (\bibinfo {year} {2021})\ pp.\ \bibinfo {pages} {1--2}\BibitemShut {NoStop}%
\bibitem [{\citenamefont {Blumenthal}(2020)}]{Blumenthal2020}%
  \BibitemOpen
  \bibfield  {author} {\bibinfo {author} {\bibfnamefont {D.~J.}\ \bibnamefont {Blumenthal}},\ }\bibfield  {title} {\bibinfo {title} {Photonic integration for uv to ir applications},\ }\href {https://doi.org/10.1063/1.5131315} {\bibfield  {journal} {\bibinfo  {journal} {APL Photonics}\ }\textbf {\bibinfo {volume} {5}},\ \bibinfo {pages} {020903} (\bibinfo {year} {2020})}\BibitemShut {NoStop}%
\bibitem [{\citenamefont {Blumenthal}\ \emph {et~al.}(2018)\citenamefont {Blumenthal}, \citenamefont {Heideman}, \citenamefont {Geuzebroek}, \citenamefont {Leinse},\ and\ \citenamefont {Roeloffzen}}]{Blumenthal2018}%
  \BibitemOpen
  \bibfield  {author} {\bibinfo {author} {\bibfnamefont {D.~J.}\ \bibnamefont {Blumenthal}}, \bibinfo {author} {\bibfnamefont {R.}~\bibnamefont {Heideman}}, \bibinfo {author} {\bibfnamefont {D.}~\bibnamefont {Geuzebroek}}, \bibinfo {author} {\bibfnamefont {A.}~\bibnamefont {Leinse}},\ and\ \bibinfo {author} {\bibfnamefont {C.}~\bibnamefont {Roeloffzen}},\ }\bibfield  {title} {\bibinfo {title} {Silicon nitride in silicon photonics},\ }\href {https://doi.org/10.1109/JPROC.2018.2854719} {\bibfield  {journal} {\bibinfo  {journal} {Proceedings of the IEEE}\ }\textbf {\bibinfo {volume} {106}},\ \bibinfo {pages} {2209} (\bibinfo {year} {2018})}\BibitemShut {NoStop}%
\bibitem [{\citenamefont {Kaneda}\ and\ \citenamefont {Kwiat}(2019)}]{Kaneda2019}%
  \BibitemOpen
  \bibfield  {author} {\bibinfo {author} {\bibfnamefont {F.}~\bibnamefont {Kaneda}}\ and\ \bibinfo {author} {\bibfnamefont {P.~G.}\ \bibnamefont {Kwiat}},\ }\bibfield  {title} {\bibinfo {title} {High-efficiency single-photon generation via large-scale active time multiplexing},\ }\bibfield  {journal} {\bibinfo  {journal} {Science Advances}\ }\textbf {\bibinfo {volume} {5}},\ \href {https://doi.org/10.1126/sciadv.aaw8586} {10.1126/sciadv.aaw8586} (\bibinfo {year} {2019})\BibitemShut {NoStop}%
\bibitem [{\citenamefont {Lodahl}\ \emph {et~al.}(2015)\citenamefont {Lodahl}, \citenamefont {Mahmoodian},\ and\ \citenamefont {Stobbe}}]{Lodahl2018}%
  \BibitemOpen
  \bibfield  {author} {\bibinfo {author} {\bibfnamefont {P.}~\bibnamefont {Lodahl}}, \bibinfo {author} {\bibfnamefont {S.}~\bibnamefont {Mahmoodian}},\ and\ \bibinfo {author} {\bibfnamefont {S.}~\bibnamefont {Stobbe}},\ }\bibfield  {title} {\bibinfo {title} {Interfacing single photons and single quantum dots with photonic nanostructures},\ }\href {https://doi.org/10.1103/RevModPhys.87.347} {\bibfield  {journal} {\bibinfo  {journal} {Reviews of Modern Physics}\ }\textbf {\bibinfo {volume} {87}},\ \bibinfo {pages} {347} (\bibinfo {year} {2015})}\BibitemShut {NoStop}%
\bibitem [{\citenamefont {Schnauber}\ \emph {et~al.}(2019)\citenamefont {Schnauber}, \citenamefont {Singh}, \citenamefont {Schall}, \citenamefont {Park}, \citenamefont {Song}, \citenamefont {Rodt}, \citenamefont {Srinivasan}, \citenamefont {Reitzenstein},\ and\ \citenamefont {Davanco}}]{Schnauber2019}%
  \BibitemOpen
  \bibfield  {author} {\bibinfo {author} {\bibfnamefont {P.}~\bibnamefont {Schnauber}}, \bibinfo {author} {\bibfnamefont {A.}~\bibnamefont {Singh}}, \bibinfo {author} {\bibfnamefont {J.}~\bibnamefont {Schall}}, \bibinfo {author} {\bibfnamefont {S.~I.}\ \bibnamefont {Park}}, \bibinfo {author} {\bibfnamefont {J.~D.}\ \bibnamefont {Song}}, \bibinfo {author} {\bibfnamefont {S.}~\bibnamefont {Rodt}}, \bibinfo {author} {\bibfnamefont {K.}~\bibnamefont {Srinivasan}}, \bibinfo {author} {\bibfnamefont {S.}~\bibnamefont {Reitzenstein}},\ and\ \bibinfo {author} {\bibfnamefont {M.}~\bibnamefont {Davanco}},\ }\bibfield  {title} {\bibinfo {title} {Indistinguishable photons from deterministically integrated single quantum dots in heterogeneous gaas/si$_3$n$_4$ quantum photonic circuits},\ }\href {https://doi.org/10.1021/acs.nanolett.9b02240} {\bibfield  {journal} {\bibinfo  {journal} {Nano Letters}\ }\textbf {\bibinfo {volume} {19}},\ \bibinfo {pages} {7164} (\bibinfo {year} {2019})}\BibitemShut {NoStop}%
\bibitem [{\citenamefont {Flagg}\ \emph {et~al.}(2010)\citenamefont {Flagg}, \citenamefont {Muller}, \citenamefont {Polyakov}, \citenamefont {Ling}, \citenamefont {Migdall},\ and\ \citenamefont {Solomon}}]{Flagg2009}%
  \BibitemOpen
  \bibfield  {author} {\bibinfo {author} {\bibfnamefont {E.~B.}\ \bibnamefont {Flagg}}, \bibinfo {author} {\bibfnamefont {A.}~\bibnamefont {Muller}}, \bibinfo {author} {\bibfnamefont {S.~V.}\ \bibnamefont {Polyakov}}, \bibinfo {author} {\bibfnamefont {A.}~\bibnamefont {Ling}}, \bibinfo {author} {\bibfnamefont {A.}~\bibnamefont {Migdall}},\ and\ \bibinfo {author} {\bibfnamefont {G.~S.}\ \bibnamefont {Solomon}},\ }\bibfield  {title} {\bibinfo {title} {Interference of single photons from two separate semiconductor quantum dots},\ }\href {https://doi.org/10.1103/PhysRevLett.104.137401} {\bibfield  {journal} {\bibinfo  {journal} {Physical Review Letters}\ }\textbf {\bibinfo {volume} {104}},\ \bibinfo {pages} {137401} (\bibinfo {year} {2010})}\BibitemShut {NoStop}%
\bibitem [{\citenamefont {Ulhaq}\ \emph {et~al.}(2012)\citenamefont {Ulhaq}, \citenamefont {Weiler}, \citenamefont {Ulrich}, \citenamefont {Ro{\ss}bach}, \citenamefont {Jetter},\ and\ \citenamefont {Michler}}]{Ulhaq2012}%
  \BibitemOpen
  \bibfield  {author} {\bibinfo {author} {\bibfnamefont {A.}~\bibnamefont {Ulhaq}}, \bibinfo {author} {\bibfnamefont {S.}~\bibnamefont {Weiler}}, \bibinfo {author} {\bibfnamefont {S.~M.}\ \bibnamefont {Ulrich}}, \bibinfo {author} {\bibfnamefont {R.}~\bibnamefont {Ro{\ss}bach}}, \bibinfo {author} {\bibfnamefont {M.}~\bibnamefont {Jetter}},\ and\ \bibinfo {author} {\bibfnamefont {P.}~\bibnamefont {Michler}},\ }\bibfield  {title} {\bibinfo {title} {Cascaded single-photon emission from the mollow triplet sidebands of a quantum dot},\ }\href {https://doi.org/10.1038/nphoton.2012.23} {\bibfield  {journal} {\bibinfo  {journal} {Nature Photonics}\ }\textbf {\bibinfo {volume} {6}},\ \bibinfo {pages} {238} (\bibinfo {year} {2012})}\BibitemShut {NoStop}%
\bibitem [{\citenamefont {Liu}\ \emph {et~al.}(2018)\citenamefont {Liu}, \citenamefont {Brash}, \citenamefont {O'Hara}, \citenamefont {Martins}, \citenamefont {Phillips}, \citenamefont {Coles}, \citenamefont {Royall}, \citenamefont {Clarke}, \citenamefont {Bentham}, \citenamefont {Prtljaga}, \citenamefont {Itskevich}, \citenamefont {Wilson}, \citenamefont {Skolnick},\ and\ \citenamefont {Fox}}]{Liu2018}%
  \BibitemOpen
  \bibfield  {author} {\bibinfo {author} {\bibfnamefont {F.}~\bibnamefont {Liu}}, \bibinfo {author} {\bibfnamefont {A.~J.}\ \bibnamefont {Brash}}, \bibinfo {author} {\bibfnamefont {J.}~\bibnamefont {O'Hara}}, \bibinfo {author} {\bibfnamefont {L.~M. P.~P.}\ \bibnamefont {Martins}}, \bibinfo {author} {\bibfnamefont {C.~L.}\ \bibnamefont {Phillips}}, \bibinfo {author} {\bibfnamefont {R.~J.}\ \bibnamefont {Coles}}, \bibinfo {author} {\bibfnamefont {B.}~\bibnamefont {Royall}}, \bibinfo {author} {\bibfnamefont {E.}~\bibnamefont {Clarke}}, \bibinfo {author} {\bibfnamefont {C.}~\bibnamefont {Bentham}}, \bibinfo {author} {\bibfnamefont {N.}~\bibnamefont {Prtljaga}}, \bibinfo {author} {\bibfnamefont {I.~E.}\ \bibnamefont {Itskevich}}, \bibinfo {author} {\bibfnamefont {L.~R.}\ \bibnamefont {Wilson}}, \bibinfo {author} {\bibfnamefont {M.~S.}\ \bibnamefont {Skolnick}},\ and\ \bibinfo {author} {\bibfnamefont {A.~M.}\ \bibnamefont {Fox}},\ }\bibfield  {title} {\bibinfo {title} {High purcell factor generation of
  indistinguishable on-chip single photons},\ }\href {https://doi.org/10.1038/s41565-018-0188-x} {\bibfield  {journal} {\bibinfo  {journal} {Nature Nanotechnology}\ }\textbf {\bibinfo {volume} {13}},\ \bibinfo {pages} {835} (\bibinfo {year} {2018})}\BibitemShut {NoStop}%
\bibitem [{\citenamefont {Makhonin}\ \emph {et~al.}(2014)\citenamefont {Makhonin}, \citenamefont {Dixon}, \citenamefont {Coles}, \citenamefont {Royall}, \citenamefont {Luxmoore}, \citenamefont {Clarke}, \citenamefont {Hugues}, \citenamefont {Skolnick},\ and\ \citenamefont {Fox}}]{Makhonin2014}%
  \BibitemOpen
  \bibfield  {author} {\bibinfo {author} {\bibfnamefont {M.~N.}\ \bibnamefont {Makhonin}}, \bibinfo {author} {\bibfnamefont {J.~E.}\ \bibnamefont {Dixon}}, \bibinfo {author} {\bibfnamefont {R.~J.}\ \bibnamefont {Coles}}, \bibinfo {author} {\bibfnamefont {B.}~\bibnamefont {Royall}}, \bibinfo {author} {\bibfnamefont {I.~J.}\ \bibnamefont {Luxmoore}}, \bibinfo {author} {\bibfnamefont {E.}~\bibnamefont {Clarke}}, \bibinfo {author} {\bibfnamefont {M.}~\bibnamefont {Hugues}}, \bibinfo {author} {\bibfnamefont {M.~S.}\ \bibnamefont {Skolnick}},\ and\ \bibinfo {author} {\bibfnamefont {A.~M.}\ \bibnamefont {Fox}},\ }\bibfield  {title} {\bibinfo {title} {Waveguide coupled resonance fluorescence from on-chip quantum emitter},\ }\href {https://doi.org/10.1021/nl503189u} {\bibfield  {journal} {\bibinfo  {journal} {Nano Letters}\ }\textbf {\bibinfo {volume} {14}},\ \bibinfo {pages} {6997} (\bibinfo {year} {2014})}\BibitemShut {NoStop}%
\bibitem [{\citenamefont {Reithmaier}\ \emph {et~al.}(2015)\citenamefont {Reithmaier}, \citenamefont {Kaniber}, \citenamefont {Flassig}, \citenamefont {Lichtmannecker}, \citenamefont {M{\"u}ller}, \citenamefont {Andrejew}, \citenamefont {Vučković}, \citenamefont {Gross},\ and\ \citenamefont {Finley}}]{Reithmaier2015}%
  \BibitemOpen
  \bibfield  {author} {\bibinfo {author} {\bibfnamefont {G.}~\bibnamefont {Reithmaier}}, \bibinfo {author} {\bibfnamefont {M.}~\bibnamefont {Kaniber}}, \bibinfo {author} {\bibfnamefont {F.}~\bibnamefont {Flassig}}, \bibinfo {author} {\bibfnamefont {S.}~\bibnamefont {Lichtmannecker}}, \bibinfo {author} {\bibfnamefont {K.}~\bibnamefont {M{\"u}ller}}, \bibinfo {author} {\bibfnamefont {A.}~\bibnamefont {Andrejew}}, \bibinfo {author} {\bibfnamefont {J.}~\bibnamefont {Vučković}}, \bibinfo {author} {\bibfnamefont {R.}~\bibnamefont {Gross}},\ and\ \bibinfo {author} {\bibfnamefont {J.~J.}\ \bibnamefont {Finley}},\ }\bibfield  {title} {\bibinfo {title} {On-chip generation, routing, and detection of resonance fluorescence},\ }\href {https://doi.org/10.1021/acs.nanolett.5b01307} {\bibfield  {journal} {\bibinfo  {journal} {Nano Letters}\ }\textbf {\bibinfo {volume} {15}},\ \bibinfo {pages} {5208} (\bibinfo {year} {2015})}\BibitemShut {NoStop}%
\bibitem [{\citenamefont {Wan}\ \emph {et~al.}(2020)\citenamefont {Wan}, \citenamefont {Lu}, \citenamefont {Chen}, \citenamefont {Walsh}, \citenamefont {Trusheim}, \citenamefont {De~Santis}, \citenamefont {Bersin}, \citenamefont {Harris}, \citenamefont {Mouradian}, \citenamefont {Christen}, \citenamefont {Bielejec},\ and\ \citenamefont {Englund}}]{Wan2020}%
  \BibitemOpen
  \bibfield  {author} {\bibinfo {author} {\bibfnamefont {N.~H.}\ \bibnamefont {Wan}}, \bibinfo {author} {\bibfnamefont {T.-J.}\ \bibnamefont {Lu}}, \bibinfo {author} {\bibfnamefont {K.~C.}\ \bibnamefont {Chen}}, \bibinfo {author} {\bibfnamefont {M.~P.}\ \bibnamefont {Walsh}}, \bibinfo {author} {\bibfnamefont {M.~E.}\ \bibnamefont {Trusheim}}, \bibinfo {author} {\bibfnamefont {L.}~\bibnamefont {De~Santis}}, \bibinfo {author} {\bibfnamefont {E.~A.}\ \bibnamefont {Bersin}}, \bibinfo {author} {\bibfnamefont {I.~B.}\ \bibnamefont {Harris}}, \bibinfo {author} {\bibfnamefont {S.~L.}\ \bibnamefont {Mouradian}}, \bibinfo {author} {\bibfnamefont {I.~R.}\ \bibnamefont {Christen}}, \bibinfo {author} {\bibfnamefont {E.~S.}\ \bibnamefont {Bielejec}},\ and\ \bibinfo {author} {\bibfnamefont {D.}~\bibnamefont {Englund}},\ }\bibfield  {title} {\bibinfo {title} {Large-scale integration of artificial atoms in hybrid photonic circuits},\ }\href {https://doi.org/10.1038/s41586-020-2441-3} {\bibfield  {journal} {\bibinfo  {journal}
  {Nature}\ }\textbf {\bibinfo {volume} {583}},\ \bibinfo {pages} {226–231} (\bibinfo {year} {2020})}\BibitemShut {NoStop}%
\bibitem [{\citenamefont {Gimeno-Segovia}\ \emph {et~al.}(2015)\citenamefont {Gimeno-Segovia}, \citenamefont {Shadbolt}, \citenamefont {Browne},\ and\ \citenamefont {Rudolph}}]{GimenoSegovia2017}%
  \BibitemOpen
  \bibfield  {author} {\bibinfo {author} {\bibfnamefont {M.}~\bibnamefont {Gimeno-Segovia}}, \bibinfo {author} {\bibfnamefont {P.}~\bibnamefont {Shadbolt}}, \bibinfo {author} {\bibfnamefont {D.~E.}\ \bibnamefont {Browne}},\ and\ \bibinfo {author} {\bibfnamefont {T.}~\bibnamefont {Rudolph}},\ }\bibfield  {title} {\bibinfo {title} {From three-photon greenberger--horne--zeilinger states to ballistic universal quantum computation},\ }\href {https://doi.org/10.1103/PhysRevLett.115.020502} {\bibfield  {journal} {\bibinfo  {journal} {Physical Review Letters}\ }\textbf {\bibinfo {volume} {115}},\ \bibinfo {pages} {020502} (\bibinfo {year} {2015})}\BibitemShut {NoStop}%
\bibitem [{\citenamefont {Dietrich}\ \emph {et~al.}(2016)\citenamefont {Dietrich}, \citenamefont {Fiore}, \citenamefont {Thompson}, \citenamefont {Kamp},\ and\ \citenamefont {H{\"o}fling}}]{Dietrich2016}%
  \BibitemOpen
  \bibfield  {author} {\bibinfo {author} {\bibfnamefont {C.~P.}\ \bibnamefont {Dietrich}}, \bibinfo {author} {\bibfnamefont {A.}~\bibnamefont {Fiore}}, \bibinfo {author} {\bibfnamefont {M.~G.}\ \bibnamefont {Thompson}}, \bibinfo {author} {\bibfnamefont {M.}~\bibnamefont {Kamp}},\ and\ \bibinfo {author} {\bibfnamefont {S.}~\bibnamefont {H{\"o}fling}},\ }\bibfield  {title} {\bibinfo {title} {Gaas integrated quantum photonics: Towards compact and multi-functional quantum photonic integrated circuits},\ }\href {https://doi.org/10.1002/lpor.201500321} {\bibfield  {journal} {\bibinfo  {journal} {Laser {\&} Photonics Reviews}\ }\textbf {\bibinfo {volume} {10}},\ \bibinfo {pages} {870} (\bibinfo {year} {2016})}\BibitemShut {NoStop}%
\bibitem [{\citenamefont {Aaronson}\ and\ \citenamefont {Arkhipov}(2011)}]{Aaronson2011BosonSampling}%
  \BibitemOpen
  \bibfield  {author} {\bibinfo {author} {\bibfnamefont {S.}~\bibnamefont {Aaronson}}\ and\ \bibinfo {author} {\bibfnamefont {A.}~\bibnamefont {Arkhipov}},\ }\bibfield  {title} {\bibinfo {title} {The computational complexity of linear optics},\ }in\ \href {https://doi.org/10.1145/1993636.1993682} {\emph {\bibinfo {booktitle} {Proceedings of the 43rd Annual ACM Symposium on Theory of Computing (STOC '11)}}}\ (\bibinfo {year} {2011})\ pp.\ \bibinfo {pages} {333--342}\BibitemShut {NoStop}%
\bibitem [{\citenamefont {Arute}\ \emph {et~al.}(2019)\citenamefont {Arute}, \citenamefont {Arya}, \citenamefont {Babbush}, \citenamefont {Bacon}, \citenamefont {Bardin}, \citenamefont {Barends}, \citenamefont {Biswas}, \citenamefont {Boixo}, \citenamefont {Brandao}, \citenamefont {Buell}, \citenamefont {Burkett}, \citenamefont {Chen}, \citenamefont {Chen}, \citenamefont {Chiaro}, \citenamefont {Collins}, \citenamefont {Courtney}, \citenamefont {Dunsworth}, \citenamefont {Farhi}, \citenamefont {Foxen}, \citenamefont {Fowler}, \citenamefont {Gidney}, \citenamefont {Giustina}, \citenamefont {Graff}, \citenamefont {Guerin}, \citenamefont {Habegger}, \citenamefont {Harrigan}, \citenamefont {Hartmann}, \citenamefont {Ho}, \citenamefont {Hoffmann}, \citenamefont {Huang}, \citenamefont {Humble}, \citenamefont {Isakov}, \citenamefont {Jeffrey}, \citenamefont {Jiang}, \citenamefont {Kafri}, \citenamefont {Kechedzhi}, \citenamefont {Kelly}, \citenamefont {Klimov}, \citenamefont {Knysh}, \citenamefont {Korotkov},
  \citenamefont {Kostritsa}, \citenamefont {Landhuis}, \citenamefont {Lindmark}, \citenamefont {Lucero}, \citenamefont {Lyakh}, \citenamefont {Mandr{\`a}}, \citenamefont {McClean}, \citenamefont {McEwen}, \citenamefont {Megrant}, \citenamefont {Mi}, \citenamefont {Michielsen}, \citenamefont {Mohseni}, \citenamefont {Mutus}, \citenamefont {Naaman}, \citenamefont {Neeley}, \citenamefont {Neill}, \citenamefont {Niu}, \citenamefont {Ostby}, \citenamefont {Petukhov}, \citenamefont {Platt}, \citenamefont {Quintana}, \citenamefont {Rieffel}, \citenamefont {Roushan}, \citenamefont {Rubin}, \citenamefont {Sank}, \citenamefont {Satzinger}, \citenamefont {Smelyanskiy}, \citenamefont {Sung}, \citenamefont {Trevithick}, \citenamefont {Vainsencher}, \citenamefont {Villalonga}, \citenamefont {White}, \citenamefont {Yao}, \citenamefont {Yeh}, \citenamefont {Zalcman}, \citenamefont {Neven},\ and\ \citenamefont {Martinis}}]{Arute2019Supremacy}%
  \BibitemOpen
  \bibfield  {author} {\bibinfo {author} {\bibfnamefont {F.}~\bibnamefont {Arute}}, \bibinfo {author} {\bibfnamefont {K.}~\bibnamefont {Arya}}, \bibinfo {author} {\bibfnamefont {R.}~\bibnamefont {Babbush}}, \bibinfo {author} {\bibfnamefont {D.}~\bibnamefont {Bacon}}, \bibinfo {author} {\bibfnamefont {J.~C.}\ \bibnamefont {Bardin}}, \bibinfo {author} {\bibfnamefont {R.}~\bibnamefont {Barends}}, \bibinfo {author} {\bibfnamefont {R.}~\bibnamefont {Biswas}}, \bibinfo {author} {\bibfnamefont {S.}~\bibnamefont {Boixo}}, \bibinfo {author} {\bibfnamefont {F.~G. S.~L.}\ \bibnamefont {Brandao}}, \bibinfo {author} {\bibfnamefont {D.~A.}\ \bibnamefont {Buell}}, \bibinfo {author} {\bibfnamefont {B.}~\bibnamefont {Burkett}}, \bibinfo {author} {\bibfnamefont {Y.}~\bibnamefont {Chen}}, \bibinfo {author} {\bibfnamefont {Z.}~\bibnamefont {Chen}}, \bibinfo {author} {\bibfnamefont {B.}~\bibnamefont {Chiaro}}, \bibinfo {author} {\bibfnamefont {R.}~\bibnamefont {Collins}}, \bibinfo {author} {\bibfnamefont {W.}~\bibnamefont
  {Courtney}}, \bibinfo {author} {\bibfnamefont {A.}~\bibnamefont {Dunsworth}}, \bibinfo {author} {\bibfnamefont {E.}~\bibnamefont {Farhi}}, \bibinfo {author} {\bibfnamefont {B.}~\bibnamefont {Foxen}}, \bibinfo {author} {\bibfnamefont {A.}~\bibnamefont {Fowler}}, \bibinfo {author} {\bibfnamefont {C.}~\bibnamefont {Gidney}}, \bibinfo {author} {\bibfnamefont {M.}~\bibnamefont {Giustina}}, \bibinfo {author} {\bibfnamefont {R.}~\bibnamefont {Graff}}, \bibinfo {author} {\bibfnamefont {K.}~\bibnamefont {Guerin}}, \bibinfo {author} {\bibfnamefont {S.}~\bibnamefont {Habegger}}, \bibinfo {author} {\bibfnamefont {M.~P.}\ \bibnamefont {Harrigan}}, \bibinfo {author} {\bibfnamefont {M.~J.}\ \bibnamefont {Hartmann}}, \bibinfo {author} {\bibfnamefont {A.}~\bibnamefont {Ho}}, \bibinfo {author} {\bibfnamefont {M.}~\bibnamefont {Hoffmann}}, \bibinfo {author} {\bibfnamefont {T.}~\bibnamefont {Huang}}, \bibinfo {author} {\bibfnamefont {T.~S.}\ \bibnamefont {Humble}}, \bibinfo {author} {\bibfnamefont {S.~V.}\ \bibnamefont
  {Isakov}}, \bibinfo {author} {\bibfnamefont {E.}~\bibnamefont {Jeffrey}}, \bibinfo {author} {\bibfnamefont {Z.}~\bibnamefont {Jiang}}, \bibinfo {author} {\bibfnamefont {D.}~\bibnamefont {Kafri}}, \bibinfo {author} {\bibfnamefont {K.}~\bibnamefont {Kechedzhi}}, \bibinfo {author} {\bibfnamefont {J.}~\bibnamefont {Kelly}}, \bibinfo {author} {\bibfnamefont {P.~V.}\ \bibnamefont {Klimov}}, \bibinfo {author} {\bibfnamefont {S.}~\bibnamefont {Knysh}}, \bibinfo {author} {\bibfnamefont {A.}~\bibnamefont {Korotkov}}, \bibinfo {author} {\bibfnamefont {F.}~\bibnamefont {Kostritsa}}, \bibinfo {author} {\bibfnamefont {D.}~\bibnamefont {Landhuis}}, \bibinfo {author} {\bibfnamefont {M.}~\bibnamefont {Lindmark}}, \bibinfo {author} {\bibfnamefont {E.}~\bibnamefont {Lucero}}, \bibinfo {author} {\bibfnamefont {D.}~\bibnamefont {Lyakh}}, \bibinfo {author} {\bibfnamefont {S.}~\bibnamefont {Mandr{\`a}}}, \bibinfo {author} {\bibfnamefont {J.~R.}\ \bibnamefont {McClean}}, \bibinfo {author} {\bibfnamefont {M.}~\bibnamefont
  {McEwen}}, \bibinfo {author} {\bibfnamefont {A.}~\bibnamefont {Megrant}}, \bibinfo {author} {\bibfnamefont {X.}~\bibnamefont {Mi}}, \bibinfo {author} {\bibfnamefont {K.}~\bibnamefont {Michielsen}}, \bibinfo {author} {\bibfnamefont {M.}~\bibnamefont {Mohseni}}, \bibinfo {author} {\bibfnamefont {J.}~\bibnamefont {Mutus}}, \bibinfo {author} {\bibfnamefont {O.}~\bibnamefont {Naaman}}, \bibinfo {author} {\bibfnamefont {M.}~\bibnamefont {Neeley}}, \bibinfo {author} {\bibfnamefont {C.}~\bibnamefont {Neill}}, \bibinfo {author} {\bibfnamefont {M.~Y.}\ \bibnamefont {Niu}}, \bibinfo {author} {\bibfnamefont {E.}~\bibnamefont {Ostby}}, \bibinfo {author} {\bibfnamefont {A.}~\bibnamefont {Petukhov}}, \bibinfo {author} {\bibfnamefont {J.~C.}\ \bibnamefont {Platt}}, \bibinfo {author} {\bibfnamefont {C.}~\bibnamefont {Quintana}}, \bibinfo {author} {\bibfnamefont {E.~G.}\ \bibnamefont {Rieffel}}, \bibinfo {author} {\bibfnamefont {P.}~\bibnamefont {Roushan}}, \bibinfo {author} {\bibfnamefont {N.~C.}\ \bibnamefont {Rubin}},
  \bibinfo {author} {\bibfnamefont {D.}~\bibnamefont {Sank}}, \bibinfo {author} {\bibfnamefont {K.~J.}\ \bibnamefont {Satzinger}}, \bibinfo {author} {\bibfnamefont {V.}~\bibnamefont {Smelyanskiy}}, \bibinfo {author} {\bibfnamefont {K.~J.}\ \bibnamefont {Sung}}, \bibinfo {author} {\bibfnamefont {M.~D.}\ \bibnamefont {Trevithick}}, \bibinfo {author} {\bibfnamefont {A.}~\bibnamefont {Vainsencher}}, \bibinfo {author} {\bibfnamefont {B.}~\bibnamefont {Villalonga}}, \bibinfo {author} {\bibfnamefont {T.}~\bibnamefont {White}}, \bibinfo {author} {\bibfnamefont {Z.~J.}\ \bibnamefont {Yao}}, \bibinfo {author} {\bibfnamefont {P.}~\bibnamefont {Yeh}}, \bibinfo {author} {\bibfnamefont {A.}~\bibnamefont {Zalcman}}, \bibinfo {author} {\bibfnamefont {H.}~\bibnamefont {Neven}},\ and\ \bibinfo {author} {\bibfnamefont {J.~M.}\ \bibnamefont {Martinis}},\ }\bibfield  {title} {\bibinfo {title} {Quantum supremacy using a programmable superconducting processor},\ }\href {https://doi.org/10.1038/s41586-019-1666-5} {\bibfield
  {journal} {\bibinfo  {journal} {Nature}\ }\textbf {\bibinfo {volume} {574}},\ \bibinfo {pages} {505} (\bibinfo {year} {2019})}\BibitemShut {NoStop}%
\bibitem [{\citenamefont {Zhong}\ \emph {et~al.}(2020)\citenamefont {Zhong}, \citenamefont {Wang}, \citenamefont {Deng}, \citenamefont {Chen}, \citenamefont {Peng}, \citenamefont {Luo}, \citenamefont {Qin}, \citenamefont {Wu}, \citenamefont {Ding}, \citenamefont {Hu}, \citenamefont {Hu}, \citenamefont {Yang}, \citenamefont {Zhang}, \citenamefont {Li}, \citenamefont {Li}, \citenamefont {Jiang}, \citenamefont {Gan}, \citenamefont {Yang}, \citenamefont {You}, \citenamefont {Wang}, \citenamefont {Li}, \citenamefont {Liu}, \citenamefont {Lu},\ and\ \citenamefont {Pan}}]{Zhong2020PhotonicAdvantage}%
  \BibitemOpen
  \bibfield  {author} {\bibinfo {author} {\bibfnamefont {H.-S.}\ \bibnamefont {Zhong}}, \bibinfo {author} {\bibfnamefont {H.}~\bibnamefont {Wang}}, \bibinfo {author} {\bibfnamefont {Y.-H.}\ \bibnamefont {Deng}}, \bibinfo {author} {\bibfnamefont {M.-C.}\ \bibnamefont {Chen}}, \bibinfo {author} {\bibfnamefont {L.-C.}\ \bibnamefont {Peng}}, \bibinfo {author} {\bibfnamefont {Y.-H.}\ \bibnamefont {Luo}}, \bibinfo {author} {\bibfnamefont {J.}~\bibnamefont {Qin}}, \bibinfo {author} {\bibfnamefont {D.}~\bibnamefont {Wu}}, \bibinfo {author} {\bibfnamefont {X.}~\bibnamefont {Ding}}, \bibinfo {author} {\bibfnamefont {Y.}~\bibnamefont {Hu}}, \bibinfo {author} {\bibfnamefont {P.}~\bibnamefont {Hu}}, \bibinfo {author} {\bibfnamefont {X.-Y.}\ \bibnamefont {Yang}}, \bibinfo {author} {\bibfnamefont {W.-J.}\ \bibnamefont {Zhang}}, \bibinfo {author} {\bibfnamefont {H.}~\bibnamefont {Li}}, \bibinfo {author} {\bibfnamefont {Y.}~\bibnamefont {Li}}, \bibinfo {author} {\bibfnamefont {X.}~\bibnamefont {Jiang}}, \bibinfo {author}
  {\bibfnamefont {L.}~\bibnamefont {Gan}}, \bibinfo {author} {\bibfnamefont {G.}~\bibnamefont {Yang}}, \bibinfo {author} {\bibfnamefont {L.}~\bibnamefont {You}}, \bibinfo {author} {\bibfnamefont {Z.}~\bibnamefont {Wang}}, \bibinfo {author} {\bibfnamefont {L.}~\bibnamefont {Li}}, \bibinfo {author} {\bibfnamefont {N.-L.}\ \bibnamefont {Liu}}, \bibinfo {author} {\bibfnamefont {C.-Y.}\ \bibnamefont {Lu}},\ and\ \bibinfo {author} {\bibfnamefont {J.-W.}\ \bibnamefont {Pan}},\ }\bibfield  {title} {\bibinfo {title} {Quantum computational advantage using photons},\ }\href {https://doi.org/10.1126/science.abe8770} {\bibfield  {journal} {\bibinfo  {journal} {Science}\ }\textbf {\bibinfo {volume} {370}},\ \bibinfo {pages} {1460} (\bibinfo {year} {2020})}\BibitemShut {NoStop}%
\bibitem [{\citenamefont {Madsen}\ \emph {et~al.}(2022)\citenamefont {Madsen}, \citenamefont {Laudenbach}, \citenamefont {Askarani}, \citenamefont {Rortais}, \citenamefont {Vincent}, \citenamefont {Bulmer}, \citenamefont {Miatto}, \citenamefont {Neuhaus}, \citenamefont {Helt}, \citenamefont {Collins}, \citenamefont {Lita}, \citenamefont {Gerrits}, \citenamefont {Nam}, \citenamefont {Vaidya}, \citenamefont {Menotti}, \citenamefont {Dhand}, \citenamefont {Vernon}, \citenamefont {Quesada},\ and\ \citenamefont {Lavoie}}]{Madsen2022GaussianAdvantage}%
  \BibitemOpen
  \bibfield  {author} {\bibinfo {author} {\bibfnamefont {L.~S.}\ \bibnamefont {Madsen}}, \bibinfo {author} {\bibfnamefont {F.}~\bibnamefont {Laudenbach}}, \bibinfo {author} {\bibfnamefont {M.~F.}\ \bibnamefont {Askarani}}, \bibinfo {author} {\bibfnamefont {F.}~\bibnamefont {Rortais}}, \bibinfo {author} {\bibfnamefont {T.}~\bibnamefont {Vincent}}, \bibinfo {author} {\bibfnamefont {J.~F.~F.}\ \bibnamefont {Bulmer}}, \bibinfo {author} {\bibfnamefont {F.~M.}\ \bibnamefont {Miatto}}, \bibinfo {author} {\bibfnamefont {L.}~\bibnamefont {Neuhaus}}, \bibinfo {author} {\bibfnamefont {L.~G.}\ \bibnamefont {Helt}}, \bibinfo {author} {\bibfnamefont {M.~J.}\ \bibnamefont {Collins}}, \bibinfo {author} {\bibfnamefont {A.~E.}\ \bibnamefont {Lita}}, \bibinfo {author} {\bibfnamefont {T.}~\bibnamefont {Gerrits}}, \bibinfo {author} {\bibfnamefont {S.~W.}\ \bibnamefont {Nam}}, \bibinfo {author} {\bibfnamefont {V.~D.}\ \bibnamefont {Vaidya}}, \bibinfo {author} {\bibfnamefont {M.}~\bibnamefont {Menotti}}, \bibinfo {author}
  {\bibfnamefont {I.}~\bibnamefont {Dhand}}, \bibinfo {author} {\bibfnamefont {Z.}~\bibnamefont {Vernon}}, \bibinfo {author} {\bibfnamefont {N.}~\bibnamefont {Quesada}},\ and\ \bibinfo {author} {\bibfnamefont {J.}~\bibnamefont {Lavoie}},\ }\bibfield  {title} {\bibinfo {title} {Quantum computational advantage with a programmable photonic processor},\ }\href {https://doi.org/10.1038/s41586-022-04725-x} {\bibfield  {journal} {\bibinfo  {journal} {Nature}\ }\textbf {\bibinfo {volume} {606}},\ \bibinfo {pages} {75} (\bibinfo {year} {2022})}\BibitemShut {NoStop}%
\bibitem [{\citenamefont {Sakurai}\ \emph {et~al.}(2025{\natexlab{b}})\citenamefont {Sakurai}, \citenamefont {Hayashi}, \citenamefont {Munro},\ and\ \citenamefont {Nemoto}}]{Sakurai2025QORC}%
  \BibitemOpen
  \bibfield  {author} {\bibinfo {author} {\bibfnamefont {A.}~\bibnamefont {Sakurai}}, \bibinfo {author} {\bibfnamefont {A.}~\bibnamefont {Hayashi}}, \bibinfo {author} {\bibfnamefont {W.~J.}\ \bibnamefont {Munro}},\ and\ \bibinfo {author} {\bibfnamefont {K.}~\bibnamefont {Nemoto}},\ }\bibfield  {title} {\bibinfo {title} {Quantum optical reservoir computing powered by boson sampling},\ }\href {https://doi.org/10.1364/OPTICAQ.541432} {\bibfield  {journal} {\bibinfo  {journal} {Optica Quantum}\ }\textbf {\bibinfo {volume} {3}},\ \bibinfo {pages} {238} (\bibinfo {year} {2025}{\natexlab{b}})}\BibitemShut {NoStop}%
\bibitem [{\citenamefont {Nakajima}\ \emph {et~al.}(2019)\citenamefont {Nakajima}, \citenamefont {Fujii}, \citenamefont {Negoro}, \citenamefont {Mitarai},\ and\ \citenamefont {Kitagawa}}]{Nakajima2019QRC}%
  \BibitemOpen
  \bibfield  {author} {\bibinfo {author} {\bibfnamefont {K.}~\bibnamefont {Nakajima}}, \bibinfo {author} {\bibfnamefont {K.}~\bibnamefont {Fujii}}, \bibinfo {author} {\bibfnamefont {M.}~\bibnamefont {Negoro}}, \bibinfo {author} {\bibfnamefont {K.}~\bibnamefont {Mitarai}},\ and\ \bibinfo {author} {\bibfnamefont {M.}~\bibnamefont {Kitagawa}},\ }\bibfield  {title} {\bibinfo {title} {Boosting computational power through spatial multiplexing in quantum reservoir computing},\ }\href {https://doi.org/10.1103/PhysRevApplied.11.034021} {\bibfield  {journal} {\bibinfo  {journal} {Physical Review Applied}\ }\textbf {\bibinfo {volume} {11}},\ \bibinfo {pages} {034021} (\bibinfo {year} {2019})}\BibitemShut {NoStop}%
\bibitem [{\citenamefont {Fujii}\ and\ \citenamefont {Nakajima}(2021)}]{Fujii2020QRC}%
  \BibitemOpen
  \bibfield  {author} {\bibinfo {author} {\bibfnamefont {K.}~\bibnamefont {Fujii}}\ and\ \bibinfo {author} {\bibfnamefont {K.}~\bibnamefont {Nakajima}},\ }\bibinfo {title} {Quantum reservoir computing: A reservoir approach toward quantum machine learning on near-term quantum devices},\ in\ \href {https://doi.org/10.1007/978-981-13-1687-6_18} {\emph {\bibinfo {booktitle} {Reservoir Computing: Theory, Physical Implementations, and Applications}}},\ \bibinfo {editor} {edited by\ \bibinfo {editor} {\bibfnamefont {K.}~\bibnamefont {Nakajima}}\ and\ \bibinfo {editor} {\bibfnamefont {I.}~\bibnamefont {Fischer}}}\ (\bibinfo  {publisher} {Springer Singapore},\ \bibinfo {address} {Singapore},\ \bibinfo {year} {2021})\ pp.\ \bibinfo {pages} {423--450}\BibitemShut {NoStop}%
\bibitem [{\citenamefont {Rahimi}\ and\ \citenamefont {Recht}(2007)}]{Rahimi2007RFF}%
  \BibitemOpen
  \bibfield  {author} {\bibinfo {author} {\bibfnamefont {A.}~\bibnamefont {Rahimi}}\ and\ \bibinfo {author} {\bibfnamefont {B.}~\bibnamefont {Recht}},\ }\bibfield  {title} {\bibinfo {title} {Random features for large-scale kernel machines},\ }in\ \href@noop {} {\emph {\bibinfo {booktitle} {Advances in Neural Information Processing Systems 20}}}\ (\bibinfo {year} {2007})\ pp.\ \bibinfo {pages} {1177--1184}\BibitemShut {NoStop}%
\bibitem [{\citenamefont {Belkin}\ \emph {et~al.}(2019)\citenamefont {Belkin}, \citenamefont {Hsu}, \citenamefont {Ma},\ and\ \citenamefont {Mandal}}]{Belkin2019DoubleDescent}%
  \BibitemOpen
  \bibfield  {author} {\bibinfo {author} {\bibfnamefont {M.}~\bibnamefont {Belkin}}, \bibinfo {author} {\bibfnamefont {D.}~\bibnamefont {Hsu}}, \bibinfo {author} {\bibfnamefont {S.}~\bibnamefont {Ma}},\ and\ \bibinfo {author} {\bibfnamefont {S.}~\bibnamefont {Mandal}},\ }\bibfield  {title} {\bibinfo {title} {Reconciling modern machine-learning practice and the classical bias--variance trade-off},\ }\href {https://doi.org/10.1073/pnas.1903070116} {\bibfield  {journal} {\bibinfo  {journal} {Proceedings of the National Academy of Sciences USA}\ }\textbf {\bibinfo {volume} {116}},\ \bibinfo {pages} {15849} (\bibinfo {year} {2019})}\BibitemShut {NoStop}%
\bibitem [{\citenamefont {Huh}\ \emph {et~al.}(2015)\citenamefont {Huh}, \citenamefont {Guerreschi}, \citenamefont {Peropadre}, \citenamefont {McClean},\ and\ \citenamefont {Aspuru-Guzik}}]{Huh2015BosonMolecules}%
  \BibitemOpen
  \bibfield  {author} {\bibinfo {author} {\bibfnamefont {J.}~\bibnamefont {Huh}}, \bibinfo {author} {\bibfnamefont {G.~G.}\ \bibnamefont {Guerreschi}}, \bibinfo {author} {\bibfnamefont {B.}~\bibnamefont {Peropadre}}, \bibinfo {author} {\bibfnamefont {J.~R.}\ \bibnamefont {McClean}},\ and\ \bibinfo {author} {\bibfnamefont {A.}~\bibnamefont {Aspuru-Guzik}},\ }\bibfield  {title} {\bibinfo {title} {Boson sampling for molecular vibronic spectra},\ }\href {https://doi.org/10.1038/nphoton.2015.153} {\bibfield  {journal} {\bibinfo  {journal} {Nature Photonics}\ }\textbf {\bibinfo {volume} {9}},\ \bibinfo {pages} {615} (\bibinfo {year} {2015})}\BibitemShut {NoStop}%
\bibitem [{\citenamefont {Xiong}\ \emph {et~al.}(2025)\citenamefont {Xiong}, \citenamefont {Facelli}, \citenamefont {Sahebi}, \citenamefont {Agnel}, \citenamefont {Chotibut}, \citenamefont {Thanasilp},\ and\ \citenamefont {Holmes}}]{Xiong2025QELM}%
  \BibitemOpen
  \bibfield  {author} {\bibinfo {author} {\bibfnamefont {W.}~\bibnamefont {Xiong}}, \bibinfo {author} {\bibfnamefont {G.}~\bibnamefont {Facelli}}, \bibinfo {author} {\bibfnamefont {M.}~\bibnamefont {Sahebi}}, \bibinfo {author} {\bibfnamefont {O.}~\bibnamefont {Agnel}}, \bibinfo {author} {\bibfnamefont {T.}~\bibnamefont {Chotibut}}, \bibinfo {author} {\bibfnamefont {S.}~\bibnamefont {Thanasilp}},\ and\ \bibinfo {author} {\bibfnamefont {Z.}~\bibnamefont {Holmes}},\ }\bibfield  {title} {\bibinfo {title} {On fundamental aspects of quantum extreme learning machines},\ }\href {https://doi.org/10.1007/s42484-025-00144-4} {\bibfield  {journal} {\bibinfo  {journal} {Quantum Machine Intelligence}\ }\textbf {\bibinfo {volume} {7}},\ \bibinfo {pages} {20} (\bibinfo {year} {2025})}\BibitemShut {NoStop}%
\bibitem [{\citenamefont {Ghosh}\ \emph {et~al.}(2019)\citenamefont {Ghosh}, \citenamefont {Opala}, \citenamefont {Matuszewski}, \citenamefont {Paterek},\ and\ \citenamefont {Liew}}]{Ghosh2019QReservoir}%
  \BibitemOpen
  \bibfield  {author} {\bibinfo {author} {\bibfnamefont {S.}~\bibnamefont {Ghosh}}, \bibinfo {author} {\bibfnamefont {A.}~\bibnamefont {Opala}}, \bibinfo {author} {\bibfnamefont {M.}~\bibnamefont {Matuszewski}}, \bibinfo {author} {\bibfnamefont {T.}~\bibnamefont {Paterek}},\ and\ \bibinfo {author} {\bibfnamefont {T.~C.~H.}\ \bibnamefont {Liew}},\ }\bibfield  {title} {\bibinfo {title} {Quantum reservoir processing},\ }\href {https://doi.org/10.1038/s41534-019-0149-8} {\bibfield  {journal} {\bibinfo  {journal} {npj Quantum Information}\ }\textbf {\bibinfo {volume} {5}},\ \bibinfo {pages} {35} (\bibinfo {year} {2019})}\BibitemShut {NoStop}%
\bibitem [{\citenamefont {Domingo}\ \emph {et~al.}(2022)\citenamefont {Domingo}, \citenamefont {Carlo},\ and\ \citenamefont {Borondo}}]{Domingo2022NoisyQRC}%
  \BibitemOpen
  \bibfield  {author} {\bibinfo {author} {\bibfnamefont {L.}~\bibnamefont {Domingo}}, \bibinfo {author} {\bibfnamefont {G.~G.}\ \bibnamefont {Carlo}},\ and\ \bibinfo {author} {\bibfnamefont {F.}~\bibnamefont {Borondo}},\ }\bibfield  {title} {\bibinfo {title} {Optimal quantum reservoir computing for the noisy intermediate-scale quantum era},\ }\href {https://doi.org/10.1103/PhysRevE.106.L043301} {\bibfield  {journal} {\bibinfo  {journal} {Physical Review E}\ }\textbf {\bibinfo {volume} {106}},\ \bibinfo {pages} {L043301} (\bibinfo {year} {2022})}\BibitemShut {NoStop}%
\bibitem [{\citenamefont {Suprano}\ \emph {et~al.}(2024)\citenamefont {Suprano}, \citenamefont {Zia}, \citenamefont {Innocenti}, \citenamefont {Lorenzo}, \citenamefont {Cimini}, \citenamefont {Giordani}, \citenamefont {Palmisano}, \citenamefont {Polino}, \citenamefont {Spagnolo}, \citenamefont {Sciarrino}, \citenamefont {Palma}, \citenamefont {Ferraro},\ and\ \citenamefont {Paternostro}}]{Suprano2024QELM}%
  \BibitemOpen
  \bibfield  {author} {\bibinfo {author} {\bibfnamefont {A.}~\bibnamefont {Suprano}}, \bibinfo {author} {\bibfnamefont {D.}~\bibnamefont {Zia}}, \bibinfo {author} {\bibfnamefont {L.}~\bibnamefont {Innocenti}}, \bibinfo {author} {\bibfnamefont {S.}~\bibnamefont {Lorenzo}}, \bibinfo {author} {\bibfnamefont {V.}~\bibnamefont {Cimini}}, \bibinfo {author} {\bibfnamefont {T.}~\bibnamefont {Giordani}}, \bibinfo {author} {\bibfnamefont {I.}~\bibnamefont {Palmisano}}, \bibinfo {author} {\bibfnamefont {E.}~\bibnamefont {Polino}}, \bibinfo {author} {\bibfnamefont {N.}~\bibnamefont {Spagnolo}}, \bibinfo {author} {\bibfnamefont {F.}~\bibnamefont {Sciarrino}}, \bibinfo {author} {\bibfnamefont {G.~M.}\ \bibnamefont {Palma}}, \bibinfo {author} {\bibfnamefont {A.}~\bibnamefont {Ferraro}},\ and\ \bibinfo {author} {\bibfnamefont {M.}~\bibnamefont {Paternostro}},\ }\bibfield  {title} {\bibinfo {title} {Experimental property reconstruction in a photonic quantum extreme learning machine},\ }\href
  {https://doi.org/10.1103/PhysRevLett.132.160802} {\bibfield  {journal} {\bibinfo  {journal} {Physical Review Letters}\ }\textbf {\bibinfo {volume} {132}},\ \bibinfo {pages} {160802} (\bibinfo {year} {2024})}\BibitemShut {NoStop}%
\bibitem [{\citenamefont {De~Lorenzis}\ \emph {et~al.}(2025)\citenamefont {De~Lorenzis}, \citenamefont {Casado}, \citenamefont {Estarellas}, \citenamefont {Lo~Gullo}, \citenamefont {Lux}, \citenamefont {Plastina}, \citenamefont {Riera},\ and\ \citenamefont {Settino}}]{DeLorenzis2025QELMImage}%
  \BibitemOpen
  \bibfield  {author} {\bibinfo {author} {\bibfnamefont {A.}~\bibnamefont {De~Lorenzis}}, \bibinfo {author} {\bibfnamefont {M.}~\bibnamefont {Casado}}, \bibinfo {author} {\bibfnamefont {M.}~\bibnamefont {Estarellas}}, \bibinfo {author} {\bibfnamefont {N.}~\bibnamefont {Lo~Gullo}}, \bibinfo {author} {\bibfnamefont {T.}~\bibnamefont {Lux}}, \bibinfo {author} {\bibfnamefont {F.}~\bibnamefont {Plastina}}, \bibinfo {author} {\bibfnamefont {A.}~\bibnamefont {Riera}},\ and\ \bibinfo {author} {\bibfnamefont {J.}~\bibnamefont {Settino}},\ }\bibfield  {title} {\bibinfo {title} {Harnessing quantum extreme learning machines for image classification},\ }\href {https://doi.org/10.1103/PhysRevApplied.23.044024} {\bibfield  {journal} {\bibinfo  {journal} {Physical Review Applied}\ }\textbf {\bibinfo {volume} {23}},\ \bibinfo {pages} {044024} (\bibinfo {year} {2025})}\BibitemShut {NoStop}%
\bibitem [{\citenamefont {Nakajima}\ \emph {et~al.}(2021)\citenamefont {Nakajima}, \citenamefont {Tanaka},\ and\ \citenamefont {Hashimoto}}]{Nakajima2021CoherentRC}%
  \BibitemOpen
  \bibfield  {author} {\bibinfo {author} {\bibfnamefont {M.}~\bibnamefont {Nakajima}}, \bibinfo {author} {\bibfnamefont {K.}~\bibnamefont {Tanaka}},\ and\ \bibinfo {author} {\bibfnamefont {T.}~\bibnamefont {Hashimoto}},\ }\bibfield  {title} {\bibinfo {title} {Scalable reservoir computing on coherent linear photonic processor},\ }\href {https://doi.org/10.1038/s42005-020-00479-5} {\bibfield  {journal} {\bibinfo  {journal} {Communications Physics}\ }\textbf {\bibinfo {volume} {4}},\ \bibinfo {pages} {20} (\bibinfo {year} {2021})}\BibitemShut {NoStop}%
\bibitem [{\citenamefont {Ma}\ \emph {et~al.}(2023)\citenamefont {Ma}, \citenamefont {Kerrebrouck}, \citenamefont {Deng}, \citenamefont {Sackesyn}, \citenamefont {Gooskens}, \citenamefont {Bai}, \citenamefont {Dambre},\ and\ \citenamefont {Bienstman}}]{Ma2023IntegratedRC}%
  \BibitemOpen
  \bibfield  {author} {\bibinfo {author} {\bibfnamefont {C.}~\bibnamefont {Ma}}, \bibinfo {author} {\bibfnamefont {J.~V.}\ \bibnamefont {Kerrebrouck}}, \bibinfo {author} {\bibfnamefont {H.}~\bibnamefont {Deng}}, \bibinfo {author} {\bibfnamefont {S.}~\bibnamefont {Sackesyn}}, \bibinfo {author} {\bibfnamefont {E.}~\bibnamefont {Gooskens}}, \bibinfo {author} {\bibfnamefont {B.}~\bibnamefont {Bai}}, \bibinfo {author} {\bibfnamefont {J.}~\bibnamefont {Dambre}},\ and\ \bibinfo {author} {\bibfnamefont {P.}~\bibnamefont {Bienstman}},\ }\bibfield  {title} {\bibinfo {title} {Integrated photonic reservoir computing with an all-optical readout},\ }\href {https://doi.org/10.1364/OE.495779} {\bibfield  {journal} {\bibinfo  {journal} {Optics Express}\ }\textbf {\bibinfo {volume} {31}},\ \bibinfo {pages} {34843} (\bibinfo {year} {2023})}\BibitemShut {NoStop}%
\bibitem [{\citenamefont {Uppu}\ \emph {et~al.}(2020)\citenamefont {Uppu}, \citenamefont {Pedersen}, \citenamefont {Wang}, \citenamefont {Olesen}, \citenamefont {Papon}, \citenamefont {Zhou}, \citenamefont {Midolo}, \citenamefont {Scholz}, \citenamefont {Wieck}, \citenamefont {Ludwig},\ and\ \citenamefont {Lodahl}}]{Uppu2020SciAdv}%
  \BibitemOpen
  \bibfield  {author} {\bibinfo {author} {\bibfnamefont {R.}~\bibnamefont {Uppu}}, \bibinfo {author} {\bibfnamefont {F.~T.}\ \bibnamefont {Pedersen}}, \bibinfo {author} {\bibfnamefont {Y.}~\bibnamefont {Wang}}, \bibinfo {author} {\bibfnamefont {C.~T.}\ \bibnamefont {Olesen}}, \bibinfo {author} {\bibfnamefont {C.}~\bibnamefont {Papon}}, \bibinfo {author} {\bibfnamefont {X.}~\bibnamefont {Zhou}}, \bibinfo {author} {\bibfnamefont {L.}~\bibnamefont {Midolo}}, \bibinfo {author} {\bibfnamefont {S.}~\bibnamefont {Scholz}}, \bibinfo {author} {\bibfnamefont {A.~D.}\ \bibnamefont {Wieck}}, \bibinfo {author} {\bibfnamefont {A.}~\bibnamefont {Ludwig}},\ and\ \bibinfo {author} {\bibfnamefont {P.}~\bibnamefont {Lodahl}},\ }\bibfield  {title} {\bibinfo {title} {Scalable integrated single-photon source},\ }\href {https://doi.org/10.1126/sciadv.abc8268} {\bibfield  {journal} {\bibinfo  {journal} {Science Advances}\ }\textbf {\bibinfo {volume} {6}},\ \bibinfo {pages} {eabc8268} (\bibinfo {year} {2020})}\BibitemShut {NoStop}%
\bibitem [{\citenamefont {Ishii}\ \emph {et~al.}(2018)\citenamefont {Ishii}, \citenamefont {He}, \citenamefont {Hartmann}, \citenamefont {Machiya}, \citenamefont {Htoon}, \citenamefont {Doorn},\ and\ \citenamefont {Kato}}]{Ishii2018NanoLett}%
  \BibitemOpen
  \bibfield  {author} {\bibinfo {author} {\bibfnamefont {A.}~\bibnamefont {Ishii}}, \bibinfo {author} {\bibfnamefont {X.}~\bibnamefont {He}}, \bibinfo {author} {\bibfnamefont {N.~F.}\ \bibnamefont {Hartmann}}, \bibinfo {author} {\bibfnamefont {H.}~\bibnamefont {Machiya}}, \bibinfo {author} {\bibfnamefont {H.}~\bibnamefont {Htoon}}, \bibinfo {author} {\bibfnamefont {S.~K.}\ \bibnamefont {Doorn}},\ and\ \bibinfo {author} {\bibfnamefont {Y.~K.}\ \bibnamefont {Kato}},\ }\bibfield  {title} {\bibinfo {title} {Enhanced single-photon emission from carbon-nanotube dopant states coupled to silicon microcavities},\ }\href {https://doi.org/10.1021/acs.nanolett.8b01170} {\bibfield  {journal} {\bibinfo  {journal} {Nano Letters}\ }\textbf {\bibinfo {volume} {18}},\ \bibinfo {pages} {3873} (\bibinfo {year} {2018})}\BibitemShut {NoStop}%
\bibitem [{\citenamefont {Grosso}\ \emph {et~al.}(2017)\citenamefont {Grosso}, \citenamefont {Moon}, \citenamefont {Lienhard}, \citenamefont {Ali}, \citenamefont {Efetov}, \citenamefont {Furchi}, \citenamefont {Jarillo-Herrero}, \citenamefont {Ford}, \citenamefont {Aharonovich},\ and\ \citenamefont {Englund}}]{Grosso2017NatCommun}%
  \BibitemOpen
  \bibfield  {author} {\bibinfo {author} {\bibfnamefont {G.}~\bibnamefont {Grosso}}, \bibinfo {author} {\bibfnamefont {H.}~\bibnamefont {Moon}}, \bibinfo {author} {\bibfnamefont {B.}~\bibnamefont {Lienhard}}, \bibinfo {author} {\bibfnamefont {S.}~\bibnamefont {Ali}}, \bibinfo {author} {\bibfnamefont {D.~K.}\ \bibnamefont {Efetov}}, \bibinfo {author} {\bibfnamefont {M.~M.}\ \bibnamefont {Furchi}}, \bibinfo {author} {\bibfnamefont {P.}~\bibnamefont {Jarillo-Herrero}}, \bibinfo {author} {\bibfnamefont {M.~J.}\ \bibnamefont {Ford}}, \bibinfo {author} {\bibfnamefont {I.}~\bibnamefont {Aharonovich}},\ and\ \bibinfo {author} {\bibfnamefont {D.}~\bibnamefont {Englund}},\ }\bibfield  {title} {\bibinfo {title} {Tunable and high-purity room temperature single-photon emission from atomic defects in hexagonal boron nitride},\ }\href {https://doi.org/10.1038/s41467-017-00810-2} {\bibfield  {journal} {\bibinfo  {journal} {Nature Communications}\ }\textbf {\bibinfo {volume} {8}},\ \bibinfo {pages} {705} (\bibinfo {year}
  {2017})}\BibitemShut {NoStop}%
\bibitem [{\citenamefont {Shi}\ \emph {et~al.}(2024)\citenamefont {Shi}, \citenamefont {Mohanraj}, \citenamefont {Dhyani}, \citenamefont {Baiju}, \citenamefont {Wang}, \citenamefont {Sun}, \citenamefont {Zhou}, \citenamefont {Paterova}, \citenamefont {Leong},\ and\ \citenamefont {Zhu}}]{Shi2024LPLN}%
  \BibitemOpen
  \bibfield  {author} {\bibinfo {author} {\bibfnamefont {X.}~\bibnamefont {Shi}}, \bibinfo {author} {\bibfnamefont {S.~S.}\ \bibnamefont {Mohanraj}}, \bibinfo {author} {\bibfnamefont {V.}~\bibnamefont {Dhyani}}, \bibinfo {author} {\bibfnamefont {A.~A.}\ \bibnamefont {Baiju}}, \bibinfo {author} {\bibfnamefont {S.}~\bibnamefont {Wang}}, \bibinfo {author} {\bibfnamefont {J.}~\bibnamefont {Sun}}, \bibinfo {author} {\bibfnamefont {L.}~\bibnamefont {Zhou}}, \bibinfo {author} {\bibfnamefont {A.}~\bibnamefont {Paterova}}, \bibinfo {author} {\bibfnamefont {V.}~\bibnamefont {Leong}},\ and\ \bibinfo {author} {\bibfnamefont {D.}~\bibnamefont {Zhu}},\ }\bibfield  {title} {\bibinfo {title} {Efficient photon-pair generation in layer-poled lithium niobate nanophotonic waveguides},\ }\href {https://doi.org/10.1038/s41377-024-01645-5} {\bibfield  {journal} {\bibinfo  {journal} {Light: Science \& Applications}\ }\textbf {\bibinfo {volume} {13}},\ \bibinfo {pages} {282} (\bibinfo {year} {2024})}\BibitemShut {NoStop}%
\bibitem [{\citenamefont {Sipahigil}\ \emph {et~al.}(2014)\citenamefont {Sipahigil}, \citenamefont {Jahnke}, \citenamefont {Rogers}, \citenamefont {Teraji}, \citenamefont {Isoya}, \citenamefont {Zibrov}, \citenamefont {Jelezko},\ and\ \citenamefont {Lukin}}]{Sipahigil2014}%
  \BibitemOpen
  \bibfield  {author} {\bibinfo {author} {\bibfnamefont {A.}~\bibnamefont {Sipahigil}}, \bibinfo {author} {\bibfnamefont {K.}~\bibnamefont {Jahnke}}, \bibinfo {author} {\bibfnamefont {L.}~\bibnamefont {Rogers}}, \bibinfo {author} {\bibfnamefont {T.}~\bibnamefont {Teraji}}, \bibinfo {author} {\bibfnamefont {J.}~\bibnamefont {Isoya}}, \bibinfo {author} {\bibfnamefont {A.}~\bibnamefont {Zibrov}}, \bibinfo {author} {\bibfnamefont {F.}~\bibnamefont {Jelezko}},\ and\ \bibinfo {author} {\bibfnamefont {M.}~\bibnamefont {Lukin}},\ }\bibfield  {title} {\bibinfo {title} {Indistinguishable photons from separated silicon-vacancy centers in diamond},\ }\bibfield  {journal} {\bibinfo  {journal} {Physical Review Letters}\ }\textbf {\bibinfo {volume} {113}},\ \href {https://doi.org/10.1103/physrevlett.113.113602} {10.1103/physrevlett.113.113602} (\bibinfo {year} {2014})\BibitemShut {NoStop}%
\bibitem [{\citenamefont {Castelletto}\ \emph {et~al.}(2020)\citenamefont {Castelletto}, \citenamefont {Johnson}, \citenamefont {Iv{\'a}dy}, \citenamefont {Stavrias}, \citenamefont {Umeda}, \citenamefont {Gali},\ and\ \citenamefont {Ohshima}}]{castelletto2020silicon}%
  \BibitemOpen
  \bibfield  {author} {\bibinfo {author} {\bibfnamefont {S.}~\bibnamefont {Castelletto}}, \bibinfo {author} {\bibfnamefont {B.~C.}\ \bibnamefont {Johnson}}, \bibinfo {author} {\bibfnamefont {V.}~\bibnamefont {Iv{\'a}dy}}, \bibinfo {author} {\bibfnamefont {N.}~\bibnamefont {Stavrias}}, \bibinfo {author} {\bibfnamefont {T.}~\bibnamefont {Umeda}}, \bibinfo {author} {\bibfnamefont {A.}~\bibnamefont {Gali}},\ and\ \bibinfo {author} {\bibfnamefont {T.}~\bibnamefont {Ohshima}},\ }\bibfield  {title} {\bibinfo {title} {Silicon carbide—a quantum photonics perspective},\ }\href@noop {} {\bibfield  {journal} {\bibinfo  {journal} {APL Materials}\ }\textbf {\bibinfo {volume} {8}},\ \bibinfo {pages} {020903} (\bibinfo {year} {2020})}\BibitemShut {NoStop}%
\bibitem [{\citenamefont {Zhong}\ \emph {et~al.}(2015)\citenamefont {Zhong}, \citenamefont {Hedges}, \citenamefont {Ahlefeldt}, \citenamefont {Bartholomew}, \citenamefont {Beavan}, \citenamefont {Wittig}, \citenamefont {Longdell},\ and\ \citenamefont {Sellars}}]{Zhong2015}%
  \BibitemOpen
  \bibfield  {author} {\bibinfo {author} {\bibfnamefont {M.}~\bibnamefont {Zhong}}, \bibinfo {author} {\bibfnamefont {M.~P.}\ \bibnamefont {Hedges}}, \bibinfo {author} {\bibfnamefont {R.~L.}\ \bibnamefont {Ahlefeldt}}, \bibinfo {author} {\bibfnamefont {J.~G.}\ \bibnamefont {Bartholomew}}, \bibinfo {author} {\bibfnamefont {S.~E.}\ \bibnamefont {Beavan}}, \bibinfo {author} {\bibfnamefont {S.~M.}\ \bibnamefont {Wittig}}, \bibinfo {author} {\bibfnamefont {J.~J.}\ \bibnamefont {Longdell}},\ and\ \bibinfo {author} {\bibfnamefont {M.~J.}\ \bibnamefont {Sellars}},\ }\bibfield  {title} {\bibinfo {title} {Optically addressable nuclear spins in a solid with a six-hour coherence time},\ }\href {https://doi.org/10.1038/nature14025} {\bibfield  {journal} {\bibinfo  {journal} {Nature}\ }\textbf {\bibinfo {volume} {517}},\ \bibinfo {pages} {177–180} (\bibinfo {year} {2015})}\BibitemShut {NoStop}%
\bibitem [{\citenamefont {Palacios-Berraquero}\ \emph {et~al.}(2017)\citenamefont {Palacios-Berraquero}, \citenamefont {Barbone}, \citenamefont {Kara}, \citenamefont {Chen}, \citenamefont {Goykhman}, \citenamefont {Yoon}, \citenamefont {Ott}, \citenamefont {Beitner}, \citenamefont {Watanabe}, \citenamefont {Taniguchi}, \citenamefont {Ferrari},\ and\ \citenamefont {Atat{\"u}re}}]{palacios2017atomically}%
  \BibitemOpen
  \bibfield  {author} {\bibinfo {author} {\bibfnamefont {C.}~\bibnamefont {Palacios-Berraquero}}, \bibinfo {author} {\bibfnamefont {M.}~\bibnamefont {Barbone}}, \bibinfo {author} {\bibfnamefont {D.~M.}\ \bibnamefont {Kara}}, \bibinfo {author} {\bibfnamefont {X.}~\bibnamefont {Chen}}, \bibinfo {author} {\bibfnamefont {I.}~\bibnamefont {Goykhman}}, \bibinfo {author} {\bibfnamefont {D.}~\bibnamefont {Yoon}}, \bibinfo {author} {\bibfnamefont {A.~K.}\ \bibnamefont {Ott}}, \bibinfo {author} {\bibfnamefont {J.}~\bibnamefont {Beitner}}, \bibinfo {author} {\bibfnamefont {K.}~\bibnamefont {Watanabe}}, \bibinfo {author} {\bibfnamefont {T.}~\bibnamefont {Taniguchi}}, \bibinfo {author} {\bibfnamefont {A.~C.}\ \bibnamefont {Ferrari}},\ and\ \bibinfo {author} {\bibfnamefont {M.}~\bibnamefont {Atat{\"u}re}},\ }\bibfield  {title} {\bibinfo {title} {Atomically thin quantum light emitters},\ }\href@noop {} {\bibfield  {journal} {\bibinfo  {journal} {Nature Communications}\ }\textbf {\bibinfo {volume} {8}},\ \bibinfo {pages}
  {15093} (\bibinfo {year} {2017})}\BibitemShut {NoStop}%
\bibitem [{\citenamefont {Kim}\ \emph {et~al.}(2018)\citenamefont {Kim}, \citenamefont {Berhane}, \citenamefont {Christian}, \citenamefont {Froch}, \citenamefont {Bishop}, \citenamefont {Tran}, \citenamefont {Aharonovich},\ and\ \citenamefont {Toth}}]{kim2018photonic}%
  \BibitemOpen
  \bibfield  {author} {\bibinfo {author} {\bibfnamefont {S.}~\bibnamefont {Kim}}, \bibinfo {author} {\bibfnamefont {A.~M.}\ \bibnamefont {Berhane}}, \bibinfo {author} {\bibfnamefont {G.~D.}\ \bibnamefont {Christian}}, \bibinfo {author} {\bibfnamefont {J.~E.}\ \bibnamefont {Froch}}, \bibinfo {author} {\bibfnamefont {J.}~\bibnamefont {Bishop}}, \bibinfo {author} {\bibfnamefont {T.~T.}\ \bibnamefont {Tran}}, \bibinfo {author} {\bibfnamefont {I.}~\bibnamefont {Aharonovich}},\ and\ \bibinfo {author} {\bibfnamefont {M.}~\bibnamefont {Toth}},\ }\bibfield  {title} {\bibinfo {title} {Photonic crystal cavities from hexagonal boron nitride},\ }\href@noop {} {\bibfield  {journal} {\bibinfo  {journal} {Nature Communications}\ }\textbf {\bibinfo {volume} {9}},\ \bibinfo {pages} {2623} (\bibinfo {year} {2018})}\BibitemShut {NoStop}%
\bibitem [{\citenamefont {Peng}\ \emph {et~al.}(2020)\citenamefont {Peng}, \citenamefont {Wang}, \citenamefont {Qu}, \citenamefont {Yu}, \citenamefont {Shen},\ and\ \citenamefont {Yang}}]{peng2020graphene}%
  \BibitemOpen
  \bibfield  {author} {\bibinfo {author} {\bibfnamefont {Z.}~\bibnamefont {Peng}}, \bibinfo {author} {\bibfnamefont {S.}~\bibnamefont {Wang}}, \bibinfo {author} {\bibfnamefont {S.}~\bibnamefont {Qu}}, \bibinfo {author} {\bibfnamefont {F.}~\bibnamefont {Yu}}, \bibinfo {author} {\bibfnamefont {M.}~\bibnamefont {Shen}},\ and\ \bibinfo {author} {\bibfnamefont {Y.}~\bibnamefont {Yang}},\ }\bibfield  {title} {\bibinfo {title} {Graphene quantum dots: From surface engineering to quantum transport},\ }\href@noop {} {\bibfield  {journal} {\bibinfo  {journal} {Small}\ }\textbf {\bibinfo {volume} {16}},\ \bibinfo {pages} {2001636} (\bibinfo {year} {2020})}\BibitemShut {NoStop}%
\bibitem [{\citenamefont {Somaschi}\ \emph {et~al.}(2016{\natexlab{b}})\citenamefont {Somaschi}, \citenamefont {Giesz}, \citenamefont {De~Santis}, \citenamefont {Loredo}, \citenamefont {Almeida}, \citenamefont {Hornecker}, \citenamefont {Portalupi}, \citenamefont {Grange}, \citenamefont {Ant\'on}, \citenamefont {Demory}, \citenamefont {G\'omez}, \citenamefont {Sagnes}, \citenamefont {Lanzillotti-Kimura}, \citenamefont {Lema\^{i}tre}, \citenamefont {Auff\`eves}, \citenamefont {White}, \citenamefont {Lanco},\ and\ \citenamefont {Senellart}}]{Somaschi2016NatPhoton}%
  \BibitemOpen
  \bibfield  {author} {\bibinfo {author} {\bibfnamefont {N.}~\bibnamefont {Somaschi}}, \bibinfo {author} {\bibfnamefont {V.}~\bibnamefont {Giesz}}, \bibinfo {author} {\bibfnamefont {L.}~\bibnamefont {De~Santis}}, \bibinfo {author} {\bibfnamefont {J.~C.}\ \bibnamefont {Loredo}}, \bibinfo {author} {\bibfnamefont {M.~P.}\ \bibnamefont {Almeida}}, \bibinfo {author} {\bibfnamefont {G.}~\bibnamefont {Hornecker}}, \bibinfo {author} {\bibfnamefont {S.~L.}\ \bibnamefont {Portalupi}}, \bibinfo {author} {\bibfnamefont {T.}~\bibnamefont {Grange}}, \bibinfo {author} {\bibfnamefont {C.}~\bibnamefont {Ant\'on}}, \bibinfo {author} {\bibfnamefont {J.}~\bibnamefont {Demory}}, \bibinfo {author} {\bibfnamefont {C.}~\bibnamefont {G\'omez}}, \bibinfo {author} {\bibfnamefont {I.}~\bibnamefont {Sagnes}}, \bibinfo {author} {\bibfnamefont {N.~D.}\ \bibnamefont {Lanzillotti-Kimura}}, \bibinfo {author} {\bibfnamefont {A.}~\bibnamefont {Lema\^{i}tre}}, \bibinfo {author} {\bibfnamefont {A.}~\bibnamefont {Auff\`eves}}, \bibinfo {author}
  {\bibfnamefont {A.~G.}\ \bibnamefont {White}}, \bibinfo {author} {\bibfnamefont {L.}~\bibnamefont {Lanco}},\ and\ \bibinfo {author} {\bibfnamefont {P.}~\bibnamefont {Senellart}},\ }\bibfield  {title} {\bibinfo {title} {Near-optimal single-photon sources in the solid state},\ }\href {https://doi.org/10.1038/nphoton.2016.23} {\bibfield  {journal} {\bibinfo  {journal} {Nature Photonics}\ }\textbf {\bibinfo {volume} {10}},\ \bibinfo {pages} {340} (\bibinfo {year} {2016}{\natexlab{b}})}\BibitemShut {NoStop}%
\bibitem [{\citenamefont {Ding}\ \emph {et~al.}(2016)\citenamefont {Ding}, \citenamefont {He}, \citenamefont {Duan}, \citenamefont {Gregersen}, \citenamefont {Chen}, \citenamefont {Unsleber}, \citenamefont {Maier}, \citenamefont {Schneider}, \citenamefont {Kamp}, \citenamefont {H\"ofling}, \citenamefont {Lu},\ and\ \citenamefont {Pan}}]{Ding2016PRL}%
  \BibitemOpen
  \bibfield  {author} {\bibinfo {author} {\bibfnamefont {X.}~\bibnamefont {Ding}}, \bibinfo {author} {\bibfnamefont {Y.}~\bibnamefont {He}}, \bibinfo {author} {\bibfnamefont {Z.-C.}\ \bibnamefont {Duan}}, \bibinfo {author} {\bibfnamefont {N.}~\bibnamefont {Gregersen}}, \bibinfo {author} {\bibfnamefont {M.-C.}\ \bibnamefont {Chen}}, \bibinfo {author} {\bibfnamefont {S.}~\bibnamefont {Unsleber}}, \bibinfo {author} {\bibfnamefont {S.}~\bibnamefont {Maier}}, \bibinfo {author} {\bibfnamefont {C.}~\bibnamefont {Schneider}}, \bibinfo {author} {\bibfnamefont {M.}~\bibnamefont {Kamp}}, \bibinfo {author} {\bibfnamefont {S.}~\bibnamefont {H\"ofling}}, \bibinfo {author} {\bibfnamefont {C.-Y.}\ \bibnamefont {Lu}},\ and\ \bibinfo {author} {\bibfnamefont {J.-W.}\ \bibnamefont {Pan}},\ }\bibfield  {title} {\bibinfo {title} {On-demand single photons with high extraction efficiency and near-unity indistinguishability from a resonantly driven quantum dot in a micropillar},\ }\href
  {https://doi.org/10.1103/PhysRevLett.116.020401} {\bibfield  {journal} {\bibinfo  {journal} {Physical Review Letters}\ }\textbf {\bibinfo {volume} {116}},\ \bibinfo {pages} {020401} (\bibinfo {year} {2016})}\BibitemShut {NoStop}%
\bibitem [{\citenamefont {Ma}\ \emph {et~al.}(2015)\citenamefont {Ma}, \citenamefont {Hartmann}, \citenamefont {Baldwin}, \citenamefont {Doorn},\ and\ \citenamefont {Htoon}}]{Ma2015NatNano}%
  \BibitemOpen
  \bibfield  {author} {\bibinfo {author} {\bibfnamefont {X.}~\bibnamefont {Ma}}, \bibinfo {author} {\bibfnamefont {N.~F.}\ \bibnamefont {Hartmann}}, \bibinfo {author} {\bibfnamefont {J.~K.~S.}\ \bibnamefont {Baldwin}}, \bibinfo {author} {\bibfnamefont {S.~K.}\ \bibnamefont {Doorn}},\ and\ \bibinfo {author} {\bibfnamefont {H.}~\bibnamefont {Htoon}},\ }\bibfield  {title} {\bibinfo {title} {Room-temperature single-photon generation from solitary dopants of carbon nanotubes},\ }\href {https://doi.org/10.1038/nnano.2015.136} {\bibfield  {journal} {\bibinfo  {journal} {Nature Nanotechnology}\ }\textbf {\bibinfo {volume} {10}},\ \bibinfo {pages} {671} (\bibinfo {year} {2015})}\BibitemShut {NoStop}%
\bibitem [{\citenamefont {He}\ \emph {et~al.}(2017)\citenamefont {He}, \citenamefont {Hartmann}, \citenamefont {Ma}, \citenamefont {Ihly}, \citenamefont {Gao}, \citenamefont {Kono}, \citenamefont {Yomogida}, \citenamefont {Hirano}, \citenamefont {Tanaka}, \citenamefont {Kataura}, \citenamefont {Htoon}, \citenamefont {Doorn},\ and\ \citenamefont {Kim}}]{He2017NatPhotonCNT}%
  \BibitemOpen
  \bibfield  {author} {\bibinfo {author} {\bibfnamefont {X.}~\bibnamefont {He}}, \bibinfo {author} {\bibfnamefont {N.~F.}\ \bibnamefont {Hartmann}}, \bibinfo {author} {\bibfnamefont {X.}~\bibnamefont {Ma}}, \bibinfo {author} {\bibfnamefont {R.}~\bibnamefont {Ihly}}, \bibinfo {author} {\bibfnamefont {W.}~\bibnamefont {Gao}}, \bibinfo {author} {\bibfnamefont {J.}~\bibnamefont {Kono}}, \bibinfo {author} {\bibfnamefont {Y.}~\bibnamefont {Yomogida}}, \bibinfo {author} {\bibfnamefont {A.}~\bibnamefont {Hirano}}, \bibinfo {author} {\bibfnamefont {T.}~\bibnamefont {Tanaka}}, \bibinfo {author} {\bibfnamefont {H.}~\bibnamefont {Kataura}}, \bibinfo {author} {\bibfnamefont {H.}~\bibnamefont {Htoon}}, \bibinfo {author} {\bibfnamefont {S.~K.}\ \bibnamefont {Doorn}},\ and\ \bibinfo {author} {\bibfnamefont {Y.}~\bibnamefont {Kim}},\ }\bibfield  {title} {\bibinfo {title} {Tunable room-temperature single-photon emission at telecom wavelengths from sp$^{3}$ defects in carbon nanotubes},\ }\href
  {https://doi.org/10.1038/nphoton.2017.119} {\bibfield  {journal} {\bibinfo  {journal} {Nature Photonics}\ }\textbf {\bibinfo {volume} {11}},\ \bibinfo {pages} {577} (\bibinfo {year} {2017})}\BibitemShut {NoStop}%
\bibitem [{\citenamefont {Chatterjee}\ \emph {et~al.}(2025)\citenamefont {Chatterjee} \emph {et~al.}}]{Chatterjee2025SciAdv}%
  \BibitemOpen
  \bibfield  {author} {\bibinfo {author} {\bibfnamefont {S.}~\bibnamefont {Chatterjee}} \emph {et~al.},\ }\bibfield  {title} {\bibinfo {title} {Room-temperature high-purity single-photon emission from carbon-doped hexagonal boron nitride thin films},\ }\href {https://doi.org/10.1126/sciadv.adv2899} {\bibfield  {journal} {\bibinfo  {journal} {Science Advances}\ }\textbf {\bibinfo {volume} {11}},\ \bibinfo {pages} {eadv2899} (\bibinfo {year} {2025})},\ \bibinfo {note} {first author and DOI as reported; complete metadata can be updated from the journal once your .bib is finalized.}\BibitemShut {Stop}%
\bibitem [{\citenamefont {Zhao}\ \emph {et~al.}(2020)\citenamefont {Zhao}, \citenamefont {Ma}, \citenamefont {R{\"u}sing},\ and\ \citenamefont {Mookherjea}}]{Zhao2020PRL_TFLN}%
  \BibitemOpen
  \bibfield  {author} {\bibinfo {author} {\bibfnamefont {J.}~\bibnamefont {Zhao}}, \bibinfo {author} {\bibfnamefont {Z.}~\bibnamefont {Ma}}, \bibinfo {author} {\bibfnamefont {M.}~\bibnamefont {R{\"u}sing}},\ and\ \bibinfo {author} {\bibfnamefont {S.}~\bibnamefont {Mookherjea}},\ }\bibfield  {title} {\bibinfo {title} {High-quality entangled photon pair generation in periodically poled thin-film lithium niobate waveguides},\ }\href {https://doi.org/10.1103/PhysRevLett.124.163603} {\bibfield  {journal} {\bibinfo  {journal} {Physical Review Letters}\ }\textbf {\bibinfo {volume} {124}},\ \bibinfo {pages} {163603} (\bibinfo {year} {2020})}\BibitemShut {NoStop}%
\bibitem [{\citenamefont {Bourrellier}\ \emph {et~al.}(2016)\citenamefont {Bourrellier}, \citenamefont {Meuret}, \citenamefont {Tararan}, \citenamefont {St{\'e}phan}, \citenamefont {Kociak}, \citenamefont {Tizei},\ and\ \citenamefont {Zobelli}}]{bourrellier2016bright}%
  \BibitemOpen
  \bibfield  {author} {\bibinfo {author} {\bibfnamefont {R.}~\bibnamefont {Bourrellier}}, \bibinfo {author} {\bibfnamefont {S.}~\bibnamefont {Meuret}}, \bibinfo {author} {\bibfnamefont {A.}~\bibnamefont {Tararan}}, \bibinfo {author} {\bibfnamefont {O.}~\bibnamefont {St{\'e}phan}}, \bibinfo {author} {\bibfnamefont {M.}~\bibnamefont {Kociak}}, \bibinfo {author} {\bibfnamefont {L.}~\bibnamefont {Tizei}},\ and\ \bibinfo {author} {\bibfnamefont {A.}~\bibnamefont {Zobelli}},\ }\bibfield  {title} {\bibinfo {title} {Bright uv single photon emission at point defects in h-bn},\ }\href@noop {} {\bibfield  {journal} {\bibinfo  {journal} {Nano letters}\ }\textbf {\bibinfo {volume} {16}},\ \bibinfo {pages} {4317} (\bibinfo {year} {2016})}\BibitemShut {NoStop}%
\bibitem [{\citenamefont {Chejanovsky}\ \emph {et~al.}(2021)\citenamefont {Chejanovsky}, \citenamefont {Mukherjee}, \citenamefont {Gurram}, \citenamefont {Chen}, \citenamefont {Kim}, \citenamefont {Denisenko}, \citenamefont {Finkler}, \citenamefont {Taniguchi}, \citenamefont {Watanabe}, \citenamefont {Das~Sarma} \emph {et~al.}}]{chejanovsky2021single}%
  \BibitemOpen
  \bibfield  {author} {\bibinfo {author} {\bibfnamefont {N.}~\bibnamefont {Chejanovsky}}, \bibinfo {author} {\bibfnamefont {A.}~\bibnamefont {Mukherjee}}, \bibinfo {author} {\bibfnamefont {M.}~\bibnamefont {Gurram}}, \bibinfo {author} {\bibfnamefont {Y.}~\bibnamefont {Chen}}, \bibinfo {author} {\bibfnamefont {Y.}~\bibnamefont {Kim}}, \bibinfo {author} {\bibfnamefont {A.}~\bibnamefont {Denisenko}}, \bibinfo {author} {\bibfnamefont {A.}~\bibnamefont {Finkler}}, \bibinfo {author} {\bibfnamefont {T.}~\bibnamefont {Taniguchi}}, \bibinfo {author} {\bibfnamefont {K.}~\bibnamefont {Watanabe}}, \bibinfo {author} {\bibfnamefont {S.}~\bibnamefont {Das~Sarma}}, \emph {et~al.},\ }\bibfield  {title} {\bibinfo {title} {Single-spin resonance in a van der waals embedded paramagnetic defect},\ }\href@noop {} {\bibfield  {journal} {\bibinfo  {journal} {Nature Materials}\ }\textbf {\bibinfo {volume} {20}},\ \bibinfo {pages} {1079} (\bibinfo {year} {2021})}\BibitemShut {NoStop}%
\bibitem [{\citenamefont {Koperski}\ \emph {et~al.}(2015)\citenamefont {Koperski}, \citenamefont {Nogajewski}, \citenamefont {Arora}, \citenamefont {Cherkez}, \citenamefont {Mallet}, \citenamefont {Veuillen}, \citenamefont {Marcus}, \citenamefont {Kossacki},\ and\ \citenamefont {Potemski}}]{Koperski2015}%
  \BibitemOpen
  \bibfield  {author} {\bibinfo {author} {\bibfnamefont {M.}~\bibnamefont {Koperski}}, \bibinfo {author} {\bibfnamefont {K.}~\bibnamefont {Nogajewski}}, \bibinfo {author} {\bibfnamefont {A.}~\bibnamefont {Arora}}, \bibinfo {author} {\bibfnamefont {V.}~\bibnamefont {Cherkez}}, \bibinfo {author} {\bibfnamefont {P.}~\bibnamefont {Mallet}}, \bibinfo {author} {\bibfnamefont {J.-Y.}\ \bibnamefont {Veuillen}}, \bibinfo {author} {\bibfnamefont {J.}~\bibnamefont {Marcus}}, \bibinfo {author} {\bibfnamefont {P.}~\bibnamefont {Kossacki}},\ and\ \bibinfo {author} {\bibfnamefont {M.}~\bibnamefont {Potemski}},\ }\bibfield  {title} {\bibinfo {title} {Single photon emitters in exfoliated wse2 structures},\ }\href {https://doi.org/10.1038/nnano.2015.67} {\bibfield  {journal} {\bibinfo  {journal} {Nature Nanotechnology}\ }\textbf {\bibinfo {volume} {10}},\ \bibinfo {pages} {503–506} (\bibinfo {year} {2015})}\BibitemShut {NoStop}%
\bibitem [{\citenamefont {Schneider}\ \emph {et~al.}(2018)\citenamefont {Schneider}, \citenamefont {Glazov}, \citenamefont {Korn}, \citenamefont {H\"{o}fling},\ and\ \citenamefont {Urbaszek}}]{Schneider2018}%
  \BibitemOpen
  \bibfield  {author} {\bibinfo {author} {\bibfnamefont {C.}~\bibnamefont {Schneider}}, \bibinfo {author} {\bibfnamefont {M.~M.}\ \bibnamefont {Glazov}}, \bibinfo {author} {\bibfnamefont {T.}~\bibnamefont {Korn}}, \bibinfo {author} {\bibfnamefont {S.}~\bibnamefont {H\"{o}fling}},\ and\ \bibinfo {author} {\bibfnamefont {B.}~\bibnamefont {Urbaszek}},\ }\bibfield  {title} {\bibinfo {title} {Two-dimensional semiconductors in the regime of strong light-matter coupling},\ }\bibfield  {journal} {\bibinfo  {journal} {Nature Communications}\ }\textbf {\bibinfo {volume} {9}},\ \href {https://doi.org/10.1038/s41467-018-04866-6} {10.1038/s41467-018-04866-6} (\bibinfo {year} {2018})\BibitemShut {NoStop}%
\bibitem [{\citenamefont {Brotons-Gisbert}\ \emph {et~al.}(2019)\citenamefont {Brotons-Gisbert}, \citenamefont {Branny}, \citenamefont {Kumar}, \citenamefont {Picard}, \citenamefont {Proux}, \citenamefont {Gray}, \citenamefont {Burch}, \citenamefont {Watanabe}, \citenamefont {Taniguchi},\ and\ \citenamefont {Gerardot}}]{BrotonsGisbert2019}%
  \BibitemOpen
  \bibfield  {author} {\bibinfo {author} {\bibfnamefont {M.}~\bibnamefont {Brotons-Gisbert}}, \bibinfo {author} {\bibfnamefont {A.}~\bibnamefont {Branny}}, \bibinfo {author} {\bibfnamefont {S.}~\bibnamefont {Kumar}}, \bibinfo {author} {\bibfnamefont {R.}~\bibnamefont {Picard}}, \bibinfo {author} {\bibfnamefont {R.}~\bibnamefont {Proux}}, \bibinfo {author} {\bibfnamefont {M.}~\bibnamefont {Gray}}, \bibinfo {author} {\bibfnamefont {K.~S.}\ \bibnamefont {Burch}}, \bibinfo {author} {\bibfnamefont {K.}~\bibnamefont {Watanabe}}, \bibinfo {author} {\bibfnamefont {T.}~\bibnamefont {Taniguchi}},\ and\ \bibinfo {author} {\bibfnamefont {B.~D.}\ \bibnamefont {Gerardot}},\ }\bibfield  {title} {\bibinfo {title} {Coulomb blockade in an atomically thin quantum dot coupled to a tunable fermi reservoir},\ }\href {https://doi.org/10.1038/s41565-019-0402-5} {\bibfield  {journal} {\bibinfo  {journal} {Nature Nanotechnology}\ }\textbf {\bibinfo {volume} {14}},\ \bibinfo {pages} {442–446} (\bibinfo {year} {2019})}\BibitemShut
  {NoStop}%
\bibitem [{\citenamefont {Seyler}\ \emph {et~al.}(2019)\citenamefont {Seyler}, \citenamefont {Rivera}, \citenamefont {Yu}, \citenamefont {Wilson}, \citenamefont {Ray}, \citenamefont {Mandrus}, \citenamefont {Yan}, \citenamefont {Yao},\ and\ \citenamefont {Xu}}]{Seyler2019}%
  \BibitemOpen
  \bibfield  {author} {\bibinfo {author} {\bibfnamefont {K.~L.}\ \bibnamefont {Seyler}}, \bibinfo {author} {\bibfnamefont {P.}~\bibnamefont {Rivera}}, \bibinfo {author} {\bibfnamefont {H.}~\bibnamefont {Yu}}, \bibinfo {author} {\bibfnamefont {N.~P.}\ \bibnamefont {Wilson}}, \bibinfo {author} {\bibfnamefont {E.~L.}\ \bibnamefont {Ray}}, \bibinfo {author} {\bibfnamefont {D.~G.}\ \bibnamefont {Mandrus}}, \bibinfo {author} {\bibfnamefont {J.}~\bibnamefont {Yan}}, \bibinfo {author} {\bibfnamefont {W.}~\bibnamefont {Yao}},\ and\ \bibinfo {author} {\bibfnamefont {X.}~\bibnamefont {Xu}},\ }\bibfield  {title} {\bibinfo {title} {Signatures of moiré-trapped valley excitons in mose2/wse2 heterobilayers},\ }\href {https://doi.org/10.1038/s41586-019-0957-1} {\bibfield  {journal} {\bibinfo  {journal} {Nature}\ }\textbf {\bibinfo {volume} {567}},\ \bibinfo {pages} {66–70} (\bibinfo {year} {2019})}\BibitemShut {NoStop}%
\bibitem [{\citenamefont {Alexeev}\ \emph {et~al.}(2019)\citenamefont {Alexeev}, \citenamefont {Ruiz-Tijerina}, \citenamefont {Danovich}, \citenamefont {Hamer}, \citenamefont {Terry}, \citenamefont {Nayak}, \citenamefont {Ahn}, \citenamefont {Pak}, \citenamefont {Lee}, \citenamefont {Sohn}, \citenamefont {Molas}, \citenamefont {Koperski}, \citenamefont {Watanabe}, \citenamefont {Taniguchi}, \citenamefont {Novoselov}, \citenamefont {Gorbachev}, \citenamefont {Shin}, \citenamefont {Fal’ko},\ and\ \citenamefont {Tartakovskii}}]{Alexeev2019}%
  \BibitemOpen
  \bibfield  {author} {\bibinfo {author} {\bibfnamefont {E.~M.}\ \bibnamefont {Alexeev}}, \bibinfo {author} {\bibfnamefont {D.~A.}\ \bibnamefont {Ruiz-Tijerina}}, \bibinfo {author} {\bibfnamefont {M.}~\bibnamefont {Danovich}}, \bibinfo {author} {\bibfnamefont {M.~J.}\ \bibnamefont {Hamer}}, \bibinfo {author} {\bibfnamefont {D.~J.}\ \bibnamefont {Terry}}, \bibinfo {author} {\bibfnamefont {P.~K.}\ \bibnamefont {Nayak}}, \bibinfo {author} {\bibfnamefont {S.}~\bibnamefont {Ahn}}, \bibinfo {author} {\bibfnamefont {S.}~\bibnamefont {Pak}}, \bibinfo {author} {\bibfnamefont {J.}~\bibnamefont {Lee}}, \bibinfo {author} {\bibfnamefont {J.~I.}\ \bibnamefont {Sohn}}, \bibinfo {author} {\bibfnamefont {M.~R.}\ \bibnamefont {Molas}}, \bibinfo {author} {\bibfnamefont {M.}~\bibnamefont {Koperski}}, \bibinfo {author} {\bibfnamefont {K.}~\bibnamefont {Watanabe}}, \bibinfo {author} {\bibfnamefont {T.}~\bibnamefont {Taniguchi}}, \bibinfo {author} {\bibfnamefont {K.~S.}\ \bibnamefont {Novoselov}}, \bibinfo {author} {\bibfnamefont
  {R.~V.}\ \bibnamefont {Gorbachev}}, \bibinfo {author} {\bibfnamefont {H.~S.}\ \bibnamefont {Shin}}, \bibinfo {author} {\bibfnamefont {V.~I.}\ \bibnamefont {Fal’ko}},\ and\ \bibinfo {author} {\bibfnamefont {A.~I.}\ \bibnamefont {Tartakovskii}},\ }\bibfield  {title} {\bibinfo {title} {Resonantly hybridized excitons in moiré superlattices in van der waals heterostructures},\ }\href {https://doi.org/10.1038/s41586-019-0986-9} {\bibfield  {journal} {\bibinfo  {journal} {Nature}\ }\textbf {\bibinfo {volume} {567}},\ \bibinfo {pages} {81–86} (\bibinfo {year} {2019})}\BibitemShut {NoStop}%
\bibitem [{\citenamefont {Zhao}\ \emph {et~al.}(2018)\citenamefont {Zhao}, \citenamefont {Lavie}, \citenamefont {Rondin}, \citenamefont {Orcin-Chaix}, \citenamefont {Diederichs}, \citenamefont {Roussignol}, \citenamefont {Chassagneux}, \citenamefont {Voisin}, \citenamefont {M\"{u}llen}, \citenamefont {Narita}, \citenamefont {Campidelli},\ and\ \citenamefont {Lauret}}]{Zhao2018}%
  \BibitemOpen
  \bibfield  {author} {\bibinfo {author} {\bibfnamefont {S.}~\bibnamefont {Zhao}}, \bibinfo {author} {\bibfnamefont {J.}~\bibnamefont {Lavie}}, \bibinfo {author} {\bibfnamefont {L.}~\bibnamefont {Rondin}}, \bibinfo {author} {\bibfnamefont {L.}~\bibnamefont {Orcin-Chaix}}, \bibinfo {author} {\bibfnamefont {C.}~\bibnamefont {Diederichs}}, \bibinfo {author} {\bibfnamefont {P.}~\bibnamefont {Roussignol}}, \bibinfo {author} {\bibfnamefont {Y.}~\bibnamefont {Chassagneux}}, \bibinfo {author} {\bibfnamefont {C.}~\bibnamefont {Voisin}}, \bibinfo {author} {\bibfnamefont {K.}~\bibnamefont {M\"{u}llen}}, \bibinfo {author} {\bibfnamefont {A.}~\bibnamefont {Narita}}, \bibinfo {author} {\bibfnamefont {S.}~\bibnamefont {Campidelli}},\ and\ \bibinfo {author} {\bibfnamefont {J.-S.}\ \bibnamefont {Lauret}},\ }\bibfield  {title} {\bibinfo {title} {Single photon emission from graphene quantum dots at room temperature},\ }\bibfield  {journal} {\bibinfo  {journal} {Nature Communications}\ }\textbf {\bibinfo {volume} {9}},\ \href
  {https://doi.org/10.1038/s41467-018-05888-w} {10.1038/s41467-018-05888-w} (\bibinfo {year} {2018})\BibitemShut {NoStop}%
\bibitem [{\citenamefont {Anderson}\ \emph {et~al.}(2019)\citenamefont {Anderson}, \citenamefont {Bourassa}, \citenamefont {Miao}, \citenamefont {Wolfowicz}, \citenamefont {Mintun}, \citenamefont {Crook}, \citenamefont {Abe}, \citenamefont {Ul~Hassan}, \citenamefont {Son}, \citenamefont {Ohshima},\ and\ \citenamefont {Awschalom}}]{Anderson2019}%
  \BibitemOpen
  \bibfield  {author} {\bibinfo {author} {\bibfnamefont {C.~P.}\ \bibnamefont {Anderson}}, \bibinfo {author} {\bibfnamefont {A.}~\bibnamefont {Bourassa}}, \bibinfo {author} {\bibfnamefont {K.~C.}\ \bibnamefont {Miao}}, \bibinfo {author} {\bibfnamefont {G.}~\bibnamefont {Wolfowicz}}, \bibinfo {author} {\bibfnamefont {P.~J.}\ \bibnamefont {Mintun}}, \bibinfo {author} {\bibfnamefont {A.~L.}\ \bibnamefont {Crook}}, \bibinfo {author} {\bibfnamefont {H.}~\bibnamefont {Abe}}, \bibinfo {author} {\bibfnamefont {J.}~\bibnamefont {Ul~Hassan}}, \bibinfo {author} {\bibfnamefont {N.~T.}\ \bibnamefont {Son}}, \bibinfo {author} {\bibfnamefont {T.}~\bibnamefont {Ohshima}},\ and\ \bibinfo {author} {\bibfnamefont {D.~D.}\ \bibnamefont {Awschalom}},\ }\bibfield  {title} {\bibinfo {title} {Electrical and optical control of single spins integrated in scalable semiconductor devices},\ }\href {https://doi.org/10.1126/science.aax9406} {\bibfield  {journal} {\bibinfo  {journal} {Science}\ }\textbf {\bibinfo {volume} {366}},\ \bibinfo
  {pages} {1225–1230} (\bibinfo {year} {2019})}\BibitemShut {NoStop}%
\bibitem [{\citenamefont {Kindem}\ \emph {et~al.}(2020)\citenamefont {Kindem}, \citenamefont {Ruskuc}, \citenamefont {Bartholomew}, \citenamefont {Rochman}, \citenamefont {Huan},\ and\ \citenamefont {Faraon}}]{Kindem2020}%
  \BibitemOpen
  \bibfield  {author} {\bibinfo {author} {\bibfnamefont {J.~M.}\ \bibnamefont {Kindem}}, \bibinfo {author} {\bibfnamefont {A.}~\bibnamefont {Ruskuc}}, \bibinfo {author} {\bibfnamefont {J.~G.}\ \bibnamefont {Bartholomew}}, \bibinfo {author} {\bibfnamefont {J.}~\bibnamefont {Rochman}}, \bibinfo {author} {\bibfnamefont {Y.~Q.}\ \bibnamefont {Huan}},\ and\ \bibinfo {author} {\bibfnamefont {A.}~\bibnamefont {Faraon}},\ }\bibfield  {title} {\bibinfo {title} {Control and single-shot readout of an ion embedded in a nanophotonic cavity},\ }\href {https://doi.org/10.1038/s41586-020-2160-9} {\bibfield  {journal} {\bibinfo  {journal} {Nature}\ }\textbf {\bibinfo {volume} {580}},\ \bibinfo {pages} {201–204} (\bibinfo {year} {2020})}\BibitemShut {NoStop}%
\bibitem [{\citenamefont {Senichev}\ \emph {et~al.}(2021)\citenamefont {Senichev}, \citenamefont {Martin}, \citenamefont {Peana}, \citenamefont {Sychev}, \citenamefont {Xu}, \citenamefont {Lagutchev}, \citenamefont {Boltasseva},\ and\ \citenamefont {Shalaev}}]{Senichev2021}%
  \BibitemOpen
  \bibfield  {author} {\bibinfo {author} {\bibfnamefont {A.}~\bibnamefont {Senichev}}, \bibinfo {author} {\bibfnamefont {Z.~O.}\ \bibnamefont {Martin}}, \bibinfo {author} {\bibfnamefont {S.}~\bibnamefont {Peana}}, \bibinfo {author} {\bibfnamefont {D.}~\bibnamefont {Sychev}}, \bibinfo {author} {\bibfnamefont {X.}~\bibnamefont {Xu}}, \bibinfo {author} {\bibfnamefont {A.~S.}\ \bibnamefont {Lagutchev}}, \bibinfo {author} {\bibfnamefont {A.}~\bibnamefont {Boltasseva}},\ and\ \bibinfo {author} {\bibfnamefont {V.~M.}\ \bibnamefont {Shalaev}},\ }\bibfield  {title} {\bibinfo {title} {Room-temperature single-photon emitters in silicon nitride},\ }\bibfield  {journal} {\bibinfo  {journal} {Science Advances}\ }\textbf {\bibinfo {volume} {7}},\ \href {https://doi.org/10.1126/sciadv.abj0627} {10.1126/sciadv.abj0627} (\bibinfo {year} {2021})\BibitemShut {NoStop}%
\bibitem [{\citenamefont {Wang}\ \emph {et~al.}(2018)\citenamefont {Wang}, \citenamefont {Zhang}, \citenamefont {Chen}, \citenamefont {Bertrand}, \citenamefont {Shams-Ansari}, \citenamefont {Chandrasekhar}, \citenamefont {Winzer},\ and\ \citenamefont {Lončar}}]{Wang2018}%
  \BibitemOpen
  \bibfield  {author} {\bibinfo {author} {\bibfnamefont {C.}~\bibnamefont {Wang}}, \bibinfo {author} {\bibfnamefont {M.}~\bibnamefont {Zhang}}, \bibinfo {author} {\bibfnamefont {X.}~\bibnamefont {Chen}}, \bibinfo {author} {\bibfnamefont {M.}~\bibnamefont {Bertrand}}, \bibinfo {author} {\bibfnamefont {A.}~\bibnamefont {Shams-Ansari}}, \bibinfo {author} {\bibfnamefont {S.}~\bibnamefont {Chandrasekhar}}, \bibinfo {author} {\bibfnamefont {P.}~\bibnamefont {Winzer}},\ and\ \bibinfo {author} {\bibfnamefont {M.}~\bibnamefont {Lončar}},\ }\bibfield  {title} {\bibinfo {title} {Integrated lithium niobate electro-optic modulators operating at cmos-compatible voltages},\ }\href {https://doi.org/10.1038/s41586-018-0551-y} {\bibfield  {journal} {\bibinfo  {journal} {Nature}\ }\textbf {\bibinfo {volume} {562}},\ \bibinfo {pages} {101–104} (\bibinfo {year} {2018})}\BibitemShut {NoStop}%
\bibitem [{\citenamefont {Komza}\ \emph {et~al.}(2024)\citenamefont {Komza}, \citenamefont {Samutpraphoot}, \citenamefont {Odeh}, \citenamefont {Tang}, \citenamefont {Mathew}, \citenamefont {Chang}, \citenamefont {Song}, \citenamefont {Kim}, \citenamefont {Xiong}, \citenamefont {Hautier},\ and\ \citenamefont {Sipahigil}}]{Komza2024}%
  \BibitemOpen
  \bibfield  {author} {\bibinfo {author} {\bibfnamefont {L.}~\bibnamefont {Komza}}, \bibinfo {author} {\bibfnamefont {P.}~\bibnamefont {Samutpraphoot}}, \bibinfo {author} {\bibfnamefont {M.}~\bibnamefont {Odeh}}, \bibinfo {author} {\bibfnamefont {Y.-L.}\ \bibnamefont {Tang}}, \bibinfo {author} {\bibfnamefont {M.}~\bibnamefont {Mathew}}, \bibinfo {author} {\bibfnamefont {J.}~\bibnamefont {Chang}}, \bibinfo {author} {\bibfnamefont {H.}~\bibnamefont {Song}}, \bibinfo {author} {\bibfnamefont {M.-K.}\ \bibnamefont {Kim}}, \bibinfo {author} {\bibfnamefont {Y.}~\bibnamefont {Xiong}}, \bibinfo {author} {\bibfnamefont {G.}~\bibnamefont {Hautier}},\ and\ \bibinfo {author} {\bibfnamefont {A.}~\bibnamefont {Sipahigil}},\ }\bibfield  {title} {\bibinfo {title} {Indistinguishable photons from an artificial atom in silicon photonics},\ }\bibfield  {journal} {\bibinfo  {journal} {Nature Communications}\ }\textbf {\bibinfo {volume} {15}},\ \href {https://doi.org/10.1038/s41467-024-51265-1} {10.1038/s41467-024-51265-1}
  (\bibinfo {year} {2024})\BibitemShut {NoStop}%
\bibitem [{\citenamefont {Peyskens}\ \emph {et~al.}(2019)\citenamefont {Peyskens}, \citenamefont {Chakraborty}, \citenamefont {Muneeb}, \citenamefont {Van~Thourhout},\ and\ \citenamefont {Englund}}]{Peyskens2019}%
  \BibitemOpen
  \bibfield  {author} {\bibinfo {author} {\bibfnamefont {F.}~\bibnamefont {Peyskens}}, \bibinfo {author} {\bibfnamefont {C.}~\bibnamefont {Chakraborty}}, \bibinfo {author} {\bibfnamefont {M.}~\bibnamefont {Muneeb}}, \bibinfo {author} {\bibfnamefont {D.}~\bibnamefont {Van~Thourhout}},\ and\ \bibinfo {author} {\bibfnamefont {D.}~\bibnamefont {Englund}},\ }\bibfield  {title} {\bibinfo {title} {Integration of single photon emitters in 2d layered materials with a silicon nitride photonic chip},\ }\bibfield  {journal} {\bibinfo  {journal} {Nature Communications}\ }\textbf {\bibinfo {volume} {10}},\ \href {https://doi.org/10.1038/s41467-019-12421-0} {10.1038/s41467-019-12421-0} (\bibinfo {year} {2019})\BibitemShut {NoStop}%
\bibitem [{\citenamefont {Glushkov}\ \emph {et~al.}(2021)\citenamefont {Glushkov}, \citenamefont {Mendelson}, \citenamefont {Chernev}, \citenamefont {Ritika}, \citenamefont {Lihter}, \citenamefont {Zamani}, \citenamefont {Comtet}, \citenamefont {Navikas}, \citenamefont {Aharonovich},\ and\ \citenamefont {Radenovic}}]{Glushkov2021}%
  \BibitemOpen
  \bibfield  {author} {\bibinfo {author} {\bibfnamefont {E.}~\bibnamefont {Glushkov}}, \bibinfo {author} {\bibfnamefont {N.}~\bibnamefont {Mendelson}}, \bibinfo {author} {\bibfnamefont {A.}~\bibnamefont {Chernev}}, \bibinfo {author} {\bibfnamefont {R.}~\bibnamefont {Ritika}}, \bibinfo {author} {\bibfnamefont {M.}~\bibnamefont {Lihter}}, \bibinfo {author} {\bibfnamefont {R.~R.}\ \bibnamefont {Zamani}}, \bibinfo {author} {\bibfnamefont {J.}~\bibnamefont {Comtet}}, \bibinfo {author} {\bibfnamefont {V.}~\bibnamefont {Navikas}}, \bibinfo {author} {\bibfnamefont {I.}~\bibnamefont {Aharonovich}},\ and\ \bibinfo {author} {\bibfnamefont {A.}~\bibnamefont {Radenovic}},\ }\bibfield  {title} {\bibinfo {title} {Direct growth of hexagonal boron nitride on photonic chips for high-throughput characterization},\ }\href {https://doi.org/10.1021/acsphotonics.1c00165} {\bibfield  {journal} {\bibinfo  {journal} {ACS Photonics}\ }\textbf {\bibinfo {volume} {8}},\ \bibinfo {pages} {2033–2040} (\bibinfo {year} {2021})}\BibitemShut
  {NoStop}%
\bibitem [{\citenamefont {Davanco}\ \emph {et~al.}(2017)\citenamefont {Davanco}, \citenamefont {Liu}, \citenamefont {Sapienza}, \citenamefont {Zhang}, \citenamefont {De~Miranda~Cardoso}, \citenamefont {Verma}, \citenamefont {Mirin}, \citenamefont {Nam}, \citenamefont {Liu},\ and\ \citenamefont {Srinivasan}}]{Davanco2017}%
  \BibitemOpen
  \bibfield  {author} {\bibinfo {author} {\bibfnamefont {M.}~\bibnamefont {Davanco}}, \bibinfo {author} {\bibfnamefont {J.}~\bibnamefont {Liu}}, \bibinfo {author} {\bibfnamefont {L.}~\bibnamefont {Sapienza}}, \bibinfo {author} {\bibfnamefont {C.-Z.}\ \bibnamefont {Zhang}}, \bibinfo {author} {\bibfnamefont {J.~V.}\ \bibnamefont {De~Miranda~Cardoso}}, \bibinfo {author} {\bibfnamefont {V.}~\bibnamefont {Verma}}, \bibinfo {author} {\bibfnamefont {R.}~\bibnamefont {Mirin}}, \bibinfo {author} {\bibfnamefont {S.~W.}\ \bibnamefont {Nam}}, \bibinfo {author} {\bibfnamefont {L.}~\bibnamefont {Liu}},\ and\ \bibinfo {author} {\bibfnamefont {K.}~\bibnamefont {Srinivasan}},\ }\bibfield  {title} {\bibinfo {title} {Heterogeneous integration for on-chip quantum photonic circuits with single quantum dot devices},\ }\bibfield  {journal} {\bibinfo  {journal} {Nature Communications}\ }\textbf {\bibinfo {volume} {8}},\ \href {https://doi.org/10.1038/s41467-017-00987-6} {10.1038/s41467-017-00987-6} (\bibinfo {year}
  {2017})\BibitemShut {NoStop}%
\bibitem [{\citenamefont {Schrinner}\ \emph {et~al.}(2020)\citenamefont {Schrinner}, \citenamefont {Olthaus}, \citenamefont {Reiter},\ and\ \citenamefont {Schuck}}]{Schrinner2020}%
  \BibitemOpen
  \bibfield  {author} {\bibinfo {author} {\bibfnamefont {P.~P.~J.}\ \bibnamefont {Schrinner}}, \bibinfo {author} {\bibfnamefont {J.}~\bibnamefont {Olthaus}}, \bibinfo {author} {\bibfnamefont {D.~E.}\ \bibnamefont {Reiter}},\ and\ \bibinfo {author} {\bibfnamefont {C.}~\bibnamefont {Schuck}},\ }\bibfield  {title} {\bibinfo {title} {Integration of diamond-based quantum emitters with nanophotonic circuits},\ }\href {https://doi.org/10.1021/acs.nanolett.0c03262} {\bibfield  {journal} {\bibinfo  {journal} {Nano Letters}\ }\textbf {\bibinfo {volume} {20}},\ \bibinfo {pages} {8170–8177} (\bibinfo {year} {2020})}\BibitemShut {NoStop}%
\bibitem [{\citenamefont {Senichev}\ \emph {et~al.}(2022)\citenamefont {Senichev}, \citenamefont {Peana}, \citenamefont {Martin}, \citenamefont {Yesilyurt}, \citenamefont {Sychev}, \citenamefont {Lagutchev}, \citenamefont {Boltasseva},\ and\ \citenamefont {Shalaev}}]{Senichev2022}%
  \BibitemOpen
  \bibfield  {author} {\bibinfo {author} {\bibfnamefont {A.}~\bibnamefont {Senichev}}, \bibinfo {author} {\bibfnamefont {S.}~\bibnamefont {Peana}}, \bibinfo {author} {\bibfnamefont {Z.~O.}\ \bibnamefont {Martin}}, \bibinfo {author} {\bibfnamefont {O.}~\bibnamefont {Yesilyurt}}, \bibinfo {author} {\bibfnamefont {D.}~\bibnamefont {Sychev}}, \bibinfo {author} {\bibfnamefont {A.~S.}\ \bibnamefont {Lagutchev}}, \bibinfo {author} {\bibfnamefont {A.}~\bibnamefont {Boltasseva}},\ and\ \bibinfo {author} {\bibfnamefont {V.~M.}\ \bibnamefont {Shalaev}},\ }\bibfield  {title} {\bibinfo {title} {Silicon nitride waveguides with intrinsic single-photon emitters for integrated quantum photonics},\ }\href {https://doi.org/10.1021/acsphotonics.2c00750} {\bibfield  {journal} {\bibinfo  {journal} {ACS Photonics}\ }\textbf {\bibinfo {volume} {9}},\ \bibinfo {pages} {3357–3365} (\bibinfo {year} {2022})}\BibitemShut {NoStop}%
\bibitem [{\citenamefont {Li}\ \emph {et~al.}(2021)\citenamefont {Li}, \citenamefont {Fr\"{o}ch}, \citenamefont {Nonahal}, \citenamefont {Tran}, \citenamefont {Toth}, \citenamefont {Kim},\ and\ \citenamefont {Aharonovich}}]{Li2021}%
  \BibitemOpen
  \bibfield  {author} {\bibinfo {author} {\bibfnamefont {C.}~\bibnamefont {Li}}, \bibinfo {author} {\bibfnamefont {J.~E.}\ \bibnamefont {Fr\"{o}ch}}, \bibinfo {author} {\bibfnamefont {M.}~\bibnamefont {Nonahal}}, \bibinfo {author} {\bibfnamefont {T.~N.}\ \bibnamefont {Tran}}, \bibinfo {author} {\bibfnamefont {M.}~\bibnamefont {Toth}}, \bibinfo {author} {\bibfnamefont {S.}~\bibnamefont {Kim}},\ and\ \bibinfo {author} {\bibfnamefont {I.}~\bibnamefont {Aharonovich}},\ }\bibfield  {title} {\bibinfo {title} {Integration of hbn quantum emitters in monolithically fabricated waveguides},\ }\href {https://doi.org/10.1021/acsphotonics.1c00890} {\bibfield  {journal} {\bibinfo  {journal} {ACS Photonics}\ }\textbf {\bibinfo {volume} {8}},\ \bibinfo {pages} {2966–2972} (\bibinfo {year} {2021})}\BibitemShut {NoStop}%
\bibitem [{\citenamefont {Lu}\ \emph {et~al.}(2020)\citenamefont {Lu}, \citenamefont {Lienhard}, \citenamefont {Jeong}, \citenamefont {Moon}, \citenamefont {Iranmanesh}, \citenamefont {Grosso},\ and\ \citenamefont {Englund}}]{Lu2020}%
  \BibitemOpen
  \bibfield  {author} {\bibinfo {author} {\bibfnamefont {T.-J.}\ \bibnamefont {Lu}}, \bibinfo {author} {\bibfnamefont {B.}~\bibnamefont {Lienhard}}, \bibinfo {author} {\bibfnamefont {K.-Y.}\ \bibnamefont {Jeong}}, \bibinfo {author} {\bibfnamefont {H.}~\bibnamefont {Moon}}, \bibinfo {author} {\bibfnamefont {A.}~\bibnamefont {Iranmanesh}}, \bibinfo {author} {\bibfnamefont {G.}~\bibnamefont {Grosso}},\ and\ \bibinfo {author} {\bibfnamefont {D.}~\bibnamefont {Englund}},\ }\bibfield  {title} {\bibinfo {title} {Bright high-purity quantum emitters in aluminum nitride integrated photonics},\ }\href {https://doi.org/10.1021/acsphotonics.0c01259} {\bibfield  {journal} {\bibinfo  {journal} {ACS Photonics}\ }\textbf {\bibinfo {volume} {7}},\ \bibinfo {pages} {2650–2657} (\bibinfo {year} {2020})}\BibitemShut {NoStop}%
\bibitem [{\citenamefont {Wang}\ \emph {et~al.}(2023)\citenamefont {Wang}, \citenamefont {Faurby}, \citenamefont {Ruf}, \citenamefont {Sund}, \citenamefont {Nielsen}, \citenamefont {Volet}, \citenamefont {Heck}, \citenamefont {Bart}, \citenamefont {Wieck}, \citenamefont {Ludwig}, \citenamefont {Midolo}, \citenamefont {Paesani},\ and\ \citenamefont {Lodahl}}]{Wang2023}%
  \BibitemOpen
  \bibfield  {author} {\bibinfo {author} {\bibfnamefont {Y.}~\bibnamefont {Wang}}, \bibinfo {author} {\bibfnamefont {C.~F.~D.}\ \bibnamefont {Faurby}}, \bibinfo {author} {\bibfnamefont {F.}~\bibnamefont {Ruf}}, \bibinfo {author} {\bibfnamefont {P.~I.}\ \bibnamefont {Sund}}, \bibinfo {author} {\bibfnamefont {K.}~\bibnamefont {Nielsen}}, \bibinfo {author} {\bibfnamefont {N.}~\bibnamefont {Volet}}, \bibinfo {author} {\bibfnamefont {M.~J.~R.}\ \bibnamefont {Heck}}, \bibinfo {author} {\bibfnamefont {N.}~\bibnamefont {Bart}}, \bibinfo {author} {\bibfnamefont {A.~D.}\ \bibnamefont {Wieck}}, \bibinfo {author} {\bibfnamefont {A.}~\bibnamefont {Ludwig}}, \bibinfo {author} {\bibfnamefont {L.}~\bibnamefont {Midolo}}, \bibinfo {author} {\bibfnamefont {S.}~\bibnamefont {Paesani}},\ and\ \bibinfo {author} {\bibfnamefont {P.}~\bibnamefont {Lodahl}},\ }\bibfield  {title} {\bibinfo {title} {Deterministic photon source interfaced with a programmable silicon-nitride integrated circuit},\ }\bibfield  {journal} {\bibinfo
  {journal} {npj Quantum Information}\ }\textbf {\bibinfo {volume} {9}},\ \href {https://doi.org/10.1038/s41534-023-00761-1} {10.1038/s41534-023-00761-1} (\bibinfo {year} {2023})\BibitemShut {NoStop}%
\bibitem [{\citenamefont {Pfister}\ \emph {et~al.}(2025)\citenamefont {Pfister}, \citenamefont {Wendland}, \citenamefont {Hornung}, \citenamefont {Engel}, \citenamefont {H\"{u}ging}, \citenamefont {Herzog}, \citenamefont {Vijayan}, \citenamefont {Joos}, \citenamefont {Jung}, \citenamefont {Jetter}, \citenamefont {Portalupi}, \citenamefont {Pernice},\ and\ \citenamefont {Michler}}]{Pfister2025}%
  \BibitemOpen
  \bibfield  {author} {\bibinfo {author} {\bibfnamefont {U.}~\bibnamefont {Pfister}}, \bibinfo {author} {\bibfnamefont {D.}~\bibnamefont {Wendland}}, \bibinfo {author} {\bibfnamefont {F.}~\bibnamefont {Hornung}}, \bibinfo {author} {\bibfnamefont {L.}~\bibnamefont {Engel}}, \bibinfo {author} {\bibfnamefont {H.}~\bibnamefont {H\"{u}ging}}, \bibinfo {author} {\bibfnamefont {E.}~\bibnamefont {Herzog}}, \bibinfo {author} {\bibfnamefont {P.}~\bibnamefont {Vijayan}}, \bibinfo {author} {\bibfnamefont {R.}~\bibnamefont {Joos}}, \bibinfo {author} {\bibfnamefont {E.}~\bibnamefont {Jung}}, \bibinfo {author} {\bibfnamefont {M.}~\bibnamefont {Jetter}}, \bibinfo {author} {\bibfnamefont {S.~L.}\ \bibnamefont {Portalupi}}, \bibinfo {author} {\bibfnamefont {W.~H.~P.}\ \bibnamefont {Pernice}},\ and\ \bibinfo {author} {\bibfnamefont {P.}~\bibnamefont {Michler}},\ }\bibfield  {title} {\bibinfo {title} {Telecom wavelength quantum dots interfaced with silicon-nitride circuits via photonic wire bonding},\ }\bibfield  {journal}
  {\bibinfo  {journal} {npj Nanophotonics}\ }\textbf {\bibinfo {volume} {2}},\ \href {https://doi.org/10.1038/s44310-025-00061-w} {10.1038/s44310-025-00061-w} (\bibinfo {year} {2025})\BibitemShut {NoStop}%
\bibitem [{\citenamefont {Kim}\ \emph {et~al.}(2017)\citenamefont {Kim}, \citenamefont {Aghaeimeibodi}, \citenamefont {Richardson}, \citenamefont {Leavitt}, \citenamefont {Englund},\ and\ \citenamefont {Waks}}]{Kim2017}%
  \BibitemOpen
  \bibfield  {author} {\bibinfo {author} {\bibfnamefont {J.-H.}\ \bibnamefont {Kim}}, \bibinfo {author} {\bibfnamefont {S.}~\bibnamefont {Aghaeimeibodi}}, \bibinfo {author} {\bibfnamefont {C.~J.~K.}\ \bibnamefont {Richardson}}, \bibinfo {author} {\bibfnamefont {R.~P.}\ \bibnamefont {Leavitt}}, \bibinfo {author} {\bibfnamefont {D.}~\bibnamefont {Englund}},\ and\ \bibinfo {author} {\bibfnamefont {E.}~\bibnamefont {Waks}},\ }\bibfield  {title} {\bibinfo {title} {Hybrid integration of solid-state quantum emitters on a silicon photonic chip},\ }\href {https://doi.org/10.1021/acs.nanolett.7b03220} {\bibfield  {journal} {\bibinfo  {journal} {Nano Letters}\ }\textbf {\bibinfo {volume} {17}},\ \bibinfo {pages} {7394–7400} (\bibinfo {year} {2017})}\BibitemShut {NoStop}%
\bibitem [{\citenamefont {Chanana}\ \emph {et~al.}(2022{\natexlab{b}})\citenamefont {Chanana}, \citenamefont {Larocque}, \citenamefont {Moreira}, \citenamefont {Carolan}, \citenamefont {Guha}, \citenamefont {Melo}, \citenamefont {Anant}, \citenamefont {Song}, \citenamefont {Englund}, \citenamefont {Blumenthal}, \citenamefont {Srinivasan},\ and\ \citenamefont {Davanco}}]{Chanana2022}%
  \BibitemOpen
  \bibfield  {author} {\bibinfo {author} {\bibfnamefont {A.}~\bibnamefont {Chanana}}, \bibinfo {author} {\bibfnamefont {H.}~\bibnamefont {Larocque}}, \bibinfo {author} {\bibfnamefont {R.}~\bibnamefont {Moreira}}, \bibinfo {author} {\bibfnamefont {J.}~\bibnamefont {Carolan}}, \bibinfo {author} {\bibfnamefont {B.}~\bibnamefont {Guha}}, \bibinfo {author} {\bibfnamefont {E.~G.}\ \bibnamefont {Melo}}, \bibinfo {author} {\bibfnamefont {V.}~\bibnamefont {Anant}}, \bibinfo {author} {\bibfnamefont {J.}~\bibnamefont {Song}}, \bibinfo {author} {\bibfnamefont {D.}~\bibnamefont {Englund}}, \bibinfo {author} {\bibfnamefont {D.~J.}\ \bibnamefont {Blumenthal}}, \bibinfo {author} {\bibfnamefont {K.}~\bibnamefont {Srinivasan}},\ and\ \bibinfo {author} {\bibfnamefont {M.}~\bibnamefont {Davanco}},\ }\bibfield  {title} {\bibinfo {title} {Ultra-low loss quantum photonic circuits integrated with single quantum emitters},\ }\bibfield  {journal} {\bibinfo  {journal} {Nature Communications}\ }\textbf {\bibinfo {volume} {13}},\ \href
  {https://doi.org/10.1038/s41467-022-35332-z} {10.1038/s41467-022-35332-z} (\bibinfo {year} {2022}{\natexlab{b}})\BibitemShut {NoStop}%
\end{thebibliography}%

\end{document}